\documentclass[12pt]{article}
\pdfoutput=1
\usepackage{amsmath,amssymb,amsfonts, epsfig}
\usepackage{color}
\usepackage{bm}
\usepackage{epsfig}
\usepackage{psfrag}
\usepackage{latexsym}
\usepackage{indentfirst}
\usepackage{fancyhdr}
\usepackage{dsfont}
\usepackage{pifont}
\usepackage{amssymb}
\usepackage{amsmath}
\usepackage{amsfonts}
\usepackage{cite}
\usepackage{bbold}
\usepackage{color}
\usepackage{colordvi}
\usepackage{fancybox}
\usepackage[footnotesize]{caption2}
\usepackage{graphicx}
\usepackage[center,footnotesize,hang]{subfigure}
\usepackage{bbm}
\usepackage{url}
\usepackage{multirow}
\usepackage{tabularx}
\usepackage{array}
\usepackage{arydshln}
\usepackage{enumitem}
\newcommand{\PreserveBackslash}[1]{\let\temp=\\#1\let\\=\temp}
\newcolumntype{C}[1]{>{\PreserveBackslash\centering}p{#1}}
\newcolumntype{R}[1]{>{\PreserveBackslash\raggedleft}p{#1}}
\newcolumntype{L}[1]{>{\PreserveBackslash\raggedright}p{#1}}
\newcommand{\cleqn}{\setcounter{equation}{0}}

\newcommand{\bq}{\begin{eqnarray}}
\newcommand{\nq}{\end{eqnarray}}
\newcommand{\cmark}{\ding{51}}%
\newcommand{\xmark}{\ding{55}}%
\allowdisplaybreaks

\makeatletter
\@addtoreset{equation}{section}
\makeatother

\begin{document}
\title{
\begin{flushright}
\ \\*[-80pt]
\begin{minipage}{0.2\linewidth}
\normalsize
\end{minipage}
\end{flushright}
{\Large \bf
Generalised CP and $\Delta (6n^2)$ Family Symmetry\\ in Semi-Direct Models of Leptons
\\*[20pt]}}

\author{
Gui-Jun~Ding$^{1}$,  \
Stephen~F.~King$^{2}$, \
Thomas~Neder$^{2}$ \
\\*[20pt]
\centerline{
\begin{minipage}{\linewidth}
\begin{center}
$^1${\it \normalsize
Department of Modern Physics, University of Science and Technology of
China,\\
Hefei, Anhui 230026, China}\\
$^2${\it \normalsize
School of Physics and Astronomy,
University of Southampton,
Southampton, SO17 1BJ, U.K.}\\
\end{center}
\end{minipage}}
\\*[50pt]}
\vskip 2 cm
\date{\small
\centerline{ \bf Abstract}
\begin{minipage}{0.9\linewidth}
\medskip
We perform a detailed analysis of $\Delta (6n^2)$ family symmetry
combined with a generalised CP symmetry in the lepton sector,
breaking to different remnant symmetries $G_{\nu}$ in the neutrino and $G_{l}$ in the charged lepton sector,
together with different remnant CP symmetries in each sector.
We discuss the resulting mass and mixing predictions for
$G_{\nu}=Z_2$ with $G_{l}=K_4,Z_p,p>2$ and $G_{\nu}=K_4$ with $G_{l}=Z_2$.  All cases correspond to the preserved symmetry smaller than the full Klein symmetry, as in the semi-direct approach, leading to predictions which depend on a single undetermined real parameter, which mainly determines the reactor angle. We focus on five phenomenologically allowed cases for which we present the resulting predictions for the PMNS parameters as a function of $n$,
as well as the predictions for neutrinoless double beta decay.
\end{minipage}
}

\begin{titlepage}
\maketitle
\thispagestyle{empty}
\end{titlepage}

\section{Introduction}
\cleqn

Following the pioneering measurements of the reactor mixing angle $\theta_{13}$ by the
Daya Bay~\cite{An:2012eh}, RENO~\cite{Ahn:2012nd}, and Double
Chooz~\cite{Abe:2011fz} reactor neutrino experiments, the three lepton
mixing angles $\theta_{12}$, $\theta_{23}$, $\theta_{13}$ and both
mass-squared differences $\Delta m^2_{\rm sol}$ and $\Delta m^2_{\rm atm}$ have by now been
measured to quite good accuracy, as recently summarised, for example, at the Neutrino Oscillation Workshop 2014~\cite{Daya_Bay_NOW_14}.

However the Dirac CP violating oscillation phase has not been measured so far and the neutrino mass squared ordering remains ambiguous. Moreover, if neutrinos are Majorana particles, there are two more unmeasured Majorana CP phases which play a role in determining the neutrinoless double-beta decay mass observable $|m_{ee}|$, which is the $(1,1)$ element of the neutrino mass matrix in the flavour basis. Determining the neutrino mass squared ordering and measuring the Dirac CP violating phase are primary goals of the next generation neutrino oscillation experiments. The CP violation has been firmly established in the quark sector and therefore it is natural to expect that CP violation occurs in the lepton sector as well. Indeed hints of a nonzero $\delta_{\rm{CP}}\sim 3\pi /2$ have begun to show up in
global analyses of neutrino oscillation data~\cite{Tortola:2012te,Capozzi:2013csa,Gonzalez-Garcia:2014bfa}, although we emphasise that such a hint could equally well be due to an upward statistical fluctuation in the
T2K electron appearance measurements~\cite{Daya_Bay_NOW_14} which are largely responsible for the effect.

In recent years, much effort has been devoted to explaining the structure of the lepton mixing angles through the introduction of a non-abelian discrete family symmetry group. For recent reviews of model building and relevant group theory aspects (see for example~\cite{Altarelli:2010gt,King:2013eh}).
Three different model building approaches have been identified,
known as ``direct'', ``semi-direct'' and ``indirect''~\cite{King:2013eh}. In the ``direct'' approach the Klein symmetry $Z_2\times Z_2$ of the Majorana neutrino mass matrix is identified as a subgroup of the family symmetry, while in the ``semi-direct'' approach only part of the Klein symmetry, typically $Z_2$, is contained in the family symmetry group, allowing the reactor angle to be fitted with one free parameter. In the ``indirect approach'' the neutrino mass matrix is constructed from ``vacuum alignments'' which are orthogonal to the symmetry preserving ones, allowing the Klein symmetry to arise as an indirect consequence of the family symmetry (for a recent example of this approach see e.g.~\cite{King:2014iia}).

Inspired by the predictive power of discrete family symmetry, it is conceivable to extend the family symmetry approaches to include a
generalised CP symmetry $H_{\rm{CP}}$ ~\cite{Ecker:1981wv,Grimus:1995zi} in order to allow the prediction of not only the mixing angles but also the CP phases as well. The idea was first explored in the context of CP symmetry combined with $\mu-\tau$ reflection symmetry, resulting in the atmospheric mixing angle $\theta_{23}$ and the Dirac CP phase $\delta_{\rm{CP}}$ both being determined to be maximal, as discussed in~\cite{Harrison:2002kp,Grimus:2003yn,Farzan:2006vj}.
More recently, the phenomenological consequences of imposing both an $S_4$ family symmetry and a generalised CP symmetry have been investigated in a model-independent way~\cite{Feruglio:2012cw,Ding:2013hpa,Li:2013jya,Li:2014eia,Feruglio:2013hia,Luhn:2013lkn}.
Assuming the symmetry breaking of $S_4\rtimes H_{\rm{CP}}$ to $Z_2\times \rm{CP}$ in the neutrino sector and to some abelian subgroup of $S_4$ in the charged lepton sector, all the three lepton mixing angles and CP phases were then given in terms of only one free parameter, where this free parameter results from the fact that only part of the Klein symmetry comes from the discrete family symmetry which is the case for the ``semi-direct'' approach.
Concrete ``semi-direct'' $S_4$ family models with a generalised CP symmetry have been constructed in Refs.~\cite{Feruglio:2012cw,Ding:2013hpa,Li:2013jya,Li:2014eia,Feruglio:2013hia,Luhn:2013lkn} where the spontaneous breaking of the $S_4\rtimes H_{\rm{CP}}$ down to $Z_2\times \rm{CP}$ in the neutrino sector was implemented. A similar generalised analysis has also been considered for $A_4$ family symmetry~\cite{Ding:2013bpa}. The typical prediction for the Dirac CP phase for the ``semi-direct'' approach to $S_4$ and $A_4$
is $|\sin \delta_{\rm{CP}}|=0\ {\rm or} \ 1$ in each case, although other model dependent predictions are also possible.

Other models with a family symmetry and a generalised CP symmetry can also be found in Refs.~\cite{Krishnan:2012me,Mohapatra:2012tb,Nishi:2013jqa,Ding:2013nsa}.
The interplay between flavor symmetries and CP symmetries has been generally discussed in~\cite{Holthausen:2012dk,Chen:2014tpa}. In addition, there are other theoretical approaches involving both family symmetry and CP violation~\cite{Branco:1983tn,Chen:2009gf,Antusch:2011sx,Girardi:2013sza}.

A generalised CP analysis of $\Delta(6n^2)$ has been performed
recently~\cite{King:2014rwa} based on a ``direct'' approach with the full Klein symmetry $Z_2\times Z_2$ preserved in the neutrino sector and a $Z_3$ preserved in the charged lepton sector. However the result of this analysis is that $|\sin \delta_{\rm{CP}}|=0$ corresponding to the Dirac CP phase being either zero or $\pm \pi$. In fact, this result was obtained originally without imposing any CP symmetry~\cite{King:2013vna}, the difference being that in the earlier approach the Majorana phases were not determined.
Clearly this result is rather disappointing from an experimental point of view, given the large effort going into measuring the Dirac CP phase and the difficulty of proving that $|\sin \delta_{\rm{CP}}|=0$. Therefore it is of interest to consider the analogous situation for the ``semi-direct'' approach. Here we shall focus on the infinite series of finite groups
$\Delta (6n^2)$ but relax the requirement of having the full Klein symmetry in the neutrino sector, and in addition consider more general preserved symmetries in the charged lepton sector. In other words we follow the  ``semi-direct'' approach for the infinite series of groups based on $\Delta(6n^2)$.

In this paper, then, we study generalised CP symmetry for all
$\Delta(6n^2)$ family symmetry groups where the CP symmetry is assumed to exist at a high energy scale and the requirement of having the full Klein symmetry is relaxed, as in the so-called ``semi-direct'' approach. The work here follows on from a similar analysis of the semi-direct approach performed by two of us based on the group $\Delta (96)$~\cite{Ding:2014ssa}. Here we investigate the lepton mixing parameters which can be obtained from the original symmetry $\Delta (6n^2)\rtimes H_{\rm{CP}}$ breaking to different remnant symmetries in the neutrino and charged lepton sectors, namely $G_{\nu}$ and $G_l$ subgroups in the neutrino and the charged lepton sector respectively, while the remnant CP symmetries from the breaking of $H_{\rm{CP}}$ are $H^{\nu}_{\rm{CP}}$ and $H^{l}_{\rm{CP}}$, respectively. The generalised CP transformation compatible with an $\Delta(6n^2)$ family symmetry is defined, and a model-independent analysis of the lepton mixing matrix is performed by scanning all the possible remnant subgroups in the neutrino and charged lepton sectors. Relaxing the requirement of having the full Klein symmetry in the neutrino sector given by a subgroup of $\Delta(6n^2)$, as in the semi-direct approach, we are led to a large number of possibilities
where the results depend on a single parameter, expressed as an angle which determines the reactor angle. We systematically discuss all such
possibilities consistent with existing phenomenological data,
then analyse in detail the resulting predictions for mixing parameters. Our results divide into two cases, the first in which the residual symmetry in the neutrino sector is $Z_2\times CP$ and the second in which the same residual symmetry $Z_2\times CP$ is preserved
by the charged lepton sector. More precisely, we discuss the resulting
mass and mixing predictions for all possible cases where the family symmetry $\Delta(6n^2)$ enhanced with generalised CP is broken to $G_{\nu}=Z_2$ with $G_{l}=K_4,Z_p,p>2$ and $G_{\nu}=K_4$ with $G_{l}=Z_2$. We are led to six phenomenologically allowed
mixing patterns and present the resulting predictions for PMNS parameters as a function of $n$,
as well as the predictions for neutrinoless double beta decay.

While this paper was being prepared, a study of generalised CP within the semi-direct approach appeared based on the infinite series of finite groups $\Delta (6n^2)$ and $\Delta (3n^2)$~\cite{Hagedorn:2014wha}. Where the results overlap for $\Delta (6n^2)$ they appear to be broadly in agreement, although the case that the residual symmetry $Z_2\times CP$ is preserved by the charged lepton sector was not considered in~\cite{Hagedorn:2014wha}. The present paper focuses exclusively on $\Delta (6n^2)$, and, apart from considering extra cases not previously considered, presents the numerical results in a quite different and complementary way. Many of the numerical results contained in this paper, for example, the predictions for neutrinoless double beta decay, were not previously considered at all.

The remainder of this paper is organised as follows.
In Section~\ref{sec:GCP} we consider Generalised CP with $\Delta(6n^2)$. In Section~\ref{sec:Z2xCP_neutrino} we consider possible lepton mixing from ``semi-direct'' approach with residual symmetry $Z_2\times CP$ in the neutrino sector. In Section~\ref{sec:Z2xCP_charged_lepton} we consider possible lepton mixing from ``semi-direct'' approach with residual symmetry $Z_2\times CP$ in the charged lepton sector. The phenomenological predictions of the neutrinoless double beta decay for all the viable cases are presented in Section~\ref{sec:nubeta}, Finally Section~\ref{sec:conclusions} concludes the paper.

\section{\label{sec:GCP}Generalised CP with $\Delta(6n^2)$}
\cleqn

Let us consider a theory with both family symmetry $G_f$ and generalized CP
symmetry at high energy scale. A field multiplet $\varphi_{\mathbf{r}}$
embedded into the representation $\mathbf{r}$ of $G_f$ transforms under the action of the family symmetry group $G_f$ as
\begin{equation}
\varphi_{\mathbf{r}}\stackrel{g}{\longmapsto}\rho_{\mathbf{r}}(g)\varphi_{\mathbf{r}},\quad
g\in G_f\,,
\end{equation}
where $\rho_{\mathbf{r}}(g)$ is the representation matrix of $g$ in the
representation $\mathbf{r}$. Furthermore, the most general CP transformations act on the field $\varphi_{\mathbf{r}}$ as:
\begin{equation}
\varphi_{\mathbf{r}}\stackrel{CP}{\longmapsto}X_{\mathbf{r}}\varphi^{*}_{\mathbf{r}}(x_P)\,,
\end{equation}
where $x_{P}=\left(t,-\mathbf{x}\right)$, $X_{\mathbf{r}}$ is a unitary matrix, and it is the so-called generalized CP transformation. The generalized CP symmetry has to be consistent with the family symmetry. It has been firmly established that the generalized CP symmetry can only be compatible with the
family symmetry if the following consistency equation is
satisfied~\cite{Ecker:1981wv,Grimus:1995zi,Holthausen:2012dk,Feruglio:2012cw}:
\begin{equation}
\label{eq:consistency_condition}X_{\mathbf{r}}\rho^{*}_{\mathbf{r}}(g)X^{\dagger}_{\mathbf{r}}=\rho_{\mathbf{r}}(g^{\prime}),\quad
g,g^{\prime}\in G_{f}\,.
\end{equation}
Hence the generalized CP transformation is related to an automorphism which
maps $g$ into $g^{\prime}$. Furthermore, it was recently shown that
physical CP transformations have to be given by class-inverting automorphism
of $G_f$~\cite{Chen:2014tpa}. In this work, we shall investigate the
$\Delta(6n^2)$ series as the family symmetry group. The group theory of $\Delta(6n^2)$ is presented in Appendix~\ref{sec:Appd_group_theory}. With the help of the
computer algebra program system \texttt{GAP}~\cite{GAP4:2011,REPSN:2011,SmallGroups:2011,SONATA:2003}, we
have studied the automorphism group of the $\Delta(6n^2)$  until
$n=19$~\footnote{The $\Delta(6n^2)$ group with $n>19$ are not available in
\texttt{GAP} so far.}. The results are collected in
Table~\ref{tab:automorphism}. We find that the outer
automorphism groups of members of the $\Delta(6n^2)$ series are generally non-trivial except for
$\Delta(6)\cong S_3$ and $\Delta(24)\cong S_4$. However, there is only one
class-inverting outer automorphism for $n\neq 3 \mathbb{Z}$ while no
class-inverting automorphism exists for $n=3 \mathbb{Z}$. In fact, we find an outer automorphism $u$ acting on the generators as:
\begin{equation}
\label{eq:class_inverting_aut}a\stackrel{u}{\longmapsto}a^2,\quad
b\stackrel{u}{\longmapsto}b,\quad c\stackrel{u}{\longmapsto}d,\quad
d\stackrel{u}{\longmapsto}c\,.
\end{equation}
It can be straightforwardly checked that this automorphism $u$ maps each
conjugacy class into the inverse one for $n\neq 3 \mathbb{Z}$. In the case
of $n=3 \mathbb{Z}$, we have
\begin{equation}
\frac{2n^2}{3}C_{2}^{(\tau)}\stackrel{u}{\longmapsto}\frac{2n^2}{3}C_{2}^{(-\tau)},\qquad
\left(\frac{2n^2}{3}C_{2}^{(\tau)}\right)^{-1}=\frac{2n^2}{3}C_{2}^{(\tau)},\quad
\tau=0,1,2\,.
\end{equation}
Hence both $\frac{2n^2}{3}C_{2}^{(1)}$ and $\frac{2n^2}{3}C_{2}^{(2)}$ are
not mapped into their inverse classes although the latter is still true for the remaining classes. As a result, we conjecture that the $\Delta(6n^2)$ group with $n\neq3 \mathbb{Z}$ admits a unique class-inverting automorphism given by Eq.~\eqref{eq:class_inverting_aut} although we can not prove it in a strict
mathematical manner so far. In the following, we shall concentrate on the
$n\neq3 \mathbb{Z}$ case without mention. The generalized CP transformation
corresponding to $u$, which is denoted by $X_{\mathbf{r}}(u)$, would be
physically well-defined, as suggested in Ref.~\cite{Chen:2014tpa}. Its
concrete form is fixed by the consistency equations as follows:
\begin{eqnarray}
\nonumber&&X_{\mathbf{r}}\left(u\right)\rho^{*}_{\mathbf{r}}(a)X^{\dagger}_{\mathbf{r}}\left(u\right)=\rho_{\mathbf{r}}\left(u\left(a\right)\right)=\rho_{\mathbf{r}}(a^2)\,,\\
\nonumber&&X_{\mathbf{r}}\left(u\right)\rho^{*}_{\mathbf{r}}(b)X^{\dagger}_{\mathbf{r}}\left(u\right)=\rho_{\mathbf{r}}\left(u\left(b\right)\right)=\rho_{\mathbf{r}}\left(b\right)\,,\\
\nonumber&&X_{\mathbf{r}}\left(u\right)\rho^{*}_{\mathbf{r}}(c)X^{\dagger}_{\mathbf{r}}\left(u\right)=\rho_{\mathbf{r}}\left(u\left(c\right)\right)=\rho_{\mathbf{r}}\left(d\right)\,,\\
\label{eq:conmsistency_equations}&&X_{\mathbf{r}}\left(u\right)\rho^{*}_{\mathbf{r}}(d)X^{\dagger}_{\mathbf{r}}\left(u\right)=\rho_{\mathbf{r}}\left(u\left(d\right)\right)=\rho_{\mathbf{r}}\left(c\right)\,.
\end{eqnarray}
In our basis, presented in section~\ref{sec:Appd_group_theory}, we can
determine that
\begin{equation}
\label{eq:gcp_trans}X_{\mathbf{r}}\left(u\right)=\rho_{\mathbf{r}}(b)\,.
\end{equation}
Furthermore, including inner automorphisms, the full set of
generalized CP transformations compatible with $\Delta(6n^2)$ family
symmetry is
\begin{equation}
\label{eq:GCP_all}X_{\mathbf{r}}=\rho_{\mathbf{r}}(g),\quad g\in\Delta(6n^2)\,.
\end{equation}
Consequently the generalized CP transformations are of the same form as the
family symmetry transformations in the chosen basis. In particular, we see
that the conventional CP transformation with $\rho_{\mathbf{r}}(1)$=1 is
allowed. As a consequence, all coupling constants would be real in a
$\Delta(6n^2)$ model with imposed CP symmetry since all the CG coefficients
are real, as shown in Appendix~\ref{sec:appendix_CG_coefficients}. In the case of $n=3\mathbb{Z}$, the consistency equations of
Eq.~\eqref{eq:conmsistency_equations} are also satisfied except for $\mathbf{r}$ being the doublet representations $\mathbf{2_2}$, $\mathbf{2_3}$ or $\mathbf{2_4}$. Hence the generalized CP transformations in Eq.~\eqref{eq:gcp_trans} can also be imposed on a model with $n=3\mathbb{Z}$ if the fields transforming as $\mathbf{2_2}$, $\mathbf{2_3}$ or $\mathbf{2_4}$ are absent.

\begin{table}[t!]
\begin{center}
\begin{tabular}{|c|c|c|c|c|c|}
\hline\hline

$n$ &  {\tt $G_f$} &  {\tt GAP-Id}  & {\tt Inn($G_f$)}  &   {\tt Out($G_f$)}
& {\tt Num.} \\  \hline

1  & $\Delta(6)\equiv S_3$   &  [6,1]   &    $S_3$    &    $Z_1$     &  1 \\
\hline

2 &  $\Delta(24)\equiv S_4$  &   [24,12]      &  $S_4$    &    $Z_1$     & 1
\\ \hline

3  &  $\Delta(54)$   & [54,8]   &   $\left(Z_3\times Z_3\right)\rtimes Z_2$
&    $S_4$     & 0 \\ \hline

4  &  $\Delta(96)$   & [96,64]   &  $\Delta(96)$   &    $Z_2$     & 1\\
\hline

5  &  $\Delta(150)$   & [150,5]   &  $\Delta(150)$   &    $Z_4$     & 1\\
\hline

6  &  $\Delta(216)$   & [216,95]   &   $\left(Z_3\times A_4\right)\rtimes
Z_2$    &    $S_3$     & 0 \\ \hline

7  &  $\Delta(294)$   & [294,7]   &  $\Delta(294)$   &    $Z_6$     & 1\\
\hline

8  &  $\Delta(384)$   & [384,568]   &  $\Delta(384)$   &    $K_4$     & 1\\
\hline

9  &  $\Delta(486)$   & [486,61]   &   $\left(\left(Z_9\times
Z_3\right)\rtimes Z_3\right)\rtimes Z_2$    &    $Z_3\times S_3$     & 0 \\
\hline

10  &  $\Delta(600)$   & [600,179]   &  $\Delta(600)$   &    $Z_4$     & 1\\
\hline

11  &  $\Delta(726)$   & [726,5]   &  $\Delta(726)$   &    $Z_{10}$     &
1\\ \hline

12  &  $\Delta(864)$   & [864,701]   &
$\left(Z_3\times\left(\left(Z_4\times Z_4\right)\rtimes
Z_3\right)\right)\rtimes Z_2$    &    $D_{12}$     & 0 \\ \hline

13  &  $\Delta(1014)$   & [1014,7]   &  $\Delta(1014)$   &    $Z_{12}$     &
1\\ \hline

14  &  $\Delta(1176)$   & [1176,243]   &  $\Delta(1176)$   &    $Z_{6}$
& 1\\ \hline

15  &  $\Delta(1350)$   & [1350,46]   &
$\left(Z_3\times\left(\left(Z_5\times Z_5\right)\rtimes
Z_3\right)\right)\rtimes Z_2$    &    $Z_{4}\times S_3$     & 0 \\ \hline

16  &  $\Delta(1536)$   & [1536,408544632]   &  $\Delta(1536)$   &
$Z_{4}\times Z_2$     & 1\\ \hline

17  &  $\Delta(1734)$   & [1734,5]   &  $\Delta(1734)$   &    $Z_{16}$     &
1\\ \hline

18  &  $\Delta(1944)$   & [1944,849]   &   $\left(\left(Z_{18}\times
Z_6\right)\rtimes Z_3\right)\rtimes Z_2$    &    $Z_{3}\times S_3$     & 0
\\ \hline

19  &  $\Delta(2166)$   & [2166,15]   &  $\Delta(2166)$   &    $Z_{18}$
& 1\\ \hline\hline

\end{tabular}
\caption{\label{tab:automorphism}The automorphism groups of the
$\Delta(6n^2)$ group series, where {\tt Inn($G_f$)} and  {\tt Out($G_f$)}
denote inner automorphism group and outer automorphism group of the family
symmetry group {\tt $G_f$} respectively. The last column gives the number of
class-inverting outer automorphisms. Note that the inner automorphism group of $\Delta(6n^2)$ with $n=3\mathbb{Z}$ is
isomorphic to $\Delta(6n^2)/Z_3$ since its center is the $Z_3$ subgroup
generated by $c^{\frac{n}{3}}d^{\frac{2n}{3}}$.}
\end{center}
\end{table}

\section{\label{sec:Z2xCP_neutrino}Possible lepton mixing from ``semi-direct'' approach with residual
symmetry $Z_2\times CP$ in the neutrino sector}
\label{3}
\cleqn

In the following, we shall analyze all possible lepton mixing patterns in the ``semi-direct'' method~\cite{King:2013eh,Ding:2013hpa,Li:2013jya,Li:2014eia,Ding:2013bpa,Ding:2013nsa,Ding:2014ssa}.
In this context, both family symmetry $\Delta(6n^2)$ and the consistent generalized CP symmetry are imposed onto the theory at high energy scales. The full symmetry group is $\Delta(6n^2)\rtimes
H_{CP}$, which is broken down to $G_{l}\rtimes H^{l}_{CP}$ and $Z_2\times
H^{\nu}_{CP}$ residual symmetries in the charged lepton and neutrino sectors
respectively. $G_{l}$ is usually taken to be an abelian subgroup of
$\Delta(6n^2)$ of order larger than 2 to avoid degenerate charged lepton masses. The misalignment between the two residual symmetries generates the PMNS matrix. In this approach, only remnant symmetries are considered and we do not discuss how the required symmetry breaking is dynamically achieved as there are generally more than one mechanism and many possible specific model realizations. It is remarkable that one column of the lepton mixing matrix can be fixed and the resulting lepton mixing parameters are generally constrained to depend on only one free parameter in this approach. We shall consider all the possible remnant symmetries $G_{l}\rtimes H^{l}_{CP}$ and $Z_2\times H^{\nu}_{CP}$, and predictions for the lepton mixing angles and CP phases would be investigated. As usual the three generation of the left-handed lepton doublet fields are assigned to the faithful representation $\mathbf{3}_{1,1}$ which is denoted by $\mathbf{3}$ for simplicity in the following.

\subsection{\label{subsec:charge_lepton_one_column}Charged lepton sector}

The remnant symmetry $G_l$ implies that the charged lepton mass matrix is invariant under the transformation $\ell_{L}\rightarrow
\rho_{\mathbf{3}}(g_{l})\ell_{L}$, where $\ell_{L}$ stands for the three generations of left-handed lepton doublets, $g_{l}$ is the generator of $G_l$, and $\rho_{\mathbf{3}}(g_{l})$ is the representation matrix of $g_{l}$ in the triplet representation $\mathbf{3}$. As a consequence, the charged lepton mass matrix satisfies
\begin{equation}
\label{eq:ch_inv}\rho^{\dagger}_{\mathbf{3}}(g_{l})m^{\dagger}_{l}m_{l}\rho_{\mathbf{3}}(g_{l})=m^{\dagger}_{l}m_{l}\,,
\end{equation}
where the charged lepton mass matrix $m_{l}$ is defined in the convention, $\ell^{c}m_{l}\ell_{L}$. Let us denote the diagonalization
matrix of $m^{\dagger}_{l}m_{l}$ by $U_{l}$, i.e.
\begin{equation}
\label{eq:ch_dia}U^{\dagger}_{l}m^{\dagger}_{l}m_{l}U_{l}=\text{diag}\left(m^2_{e},
m^2_{\mu}, m^2_{\tau}\right)\equiv \widehat{m}^2_{l}\,.
\end{equation}
where $m_{e}$, $m_{\mu}$ and $m_{\tau}$ are the electron, muon and tau
masses respectively. Substituting Eq.~\eqref{eq:ch_dia} into
Eq.~\eqref{eq:ch_inv}, we obtain
\begin{equation}
\widehat{m}^2_{l}\left[U^{\dagger}_{l}\rho_{\mathbf{3}}(g_{l})U_{l}\right]=\left[U^{\dagger}_{l}\rho_{\mathbf{3}}(g_{l})U_{l}\right]\widehat{m}^2_{l}\,.
\end{equation}
One can see that $U^{\dagger}_{l}\rho_{\mathbf{3}}(g_{l})U_{l}$ has to be diagonal. Therefore $U_{l}$ not only diagonalizes $m^{\dagger}_{l}m_{l}$ but also the matrix $\rho_{\mathbf{3}}(g_{l})$. As a result, the unitary diagonalization matrix $U_{l}$ is completely fixed by the remnant flavor symmetry $G_{l}$ once the eigenvalues of $\rho_{\mathbf{3}}(g_{l})$ are non-degenerate. In the present work, we shall focus on the case that $G_{l}$ is a cyclic subgroup of $\Delta(6n^2)$. Hence the generator $g_{l}$ of $G_{l}$ could be of the form $c^{s}d^{t}$, $bc^{s}d^{t}$, $ac^{s}d^{t}$,
$a^2c^{s}d^{t}$, $abc^{s}d^{t}$ or $a^2bc^{s}d^{t}$ with $s,t=0,1,\ldots,
n-1$. If the eigenvalues of $\rho_{\mathbf{3}}(g_{l})$ are degenerate such
that its diagonalization matrix $U_{l}$ can not be fixed uniquely, we could extend $G_{l}$ from a single cyclic subgroup to the product of several cyclic subgroups. This scenario is beyond the scope of this work except that the simplest $K_4$ extension is included. Given the explicit form of the representation matrices listed in Appendix~\ref{sec:Appd_group_theory}, the charged lepton diagonalization matrices $U_{l}$ for different cases of $G_{l}$ can be calculated straightforwardly, and the results are summarized in the following. Since the charged lepton masses can not be constrained at all in the present approach (in other word, the order of the eigenvalues of $\rho_{\mathbf{3}}(g_{l})$ is indeterminate), $U_l$ can undergo rephasing and permutations from the left.
\begin{itemize}

\item{$G_{l}=\left\langle c^{s}d^{t}\right\rangle$}

\begin{equation}
U_{l}=\left(\begin{array}{ccc}
 1  &  0   &  0  \\
 0  &  1   &  0  \\
 0  &  0   &  1
\end{array}
\right)\,.
\end{equation}
Note that the parameters $s$ and $t$ should be subject to the following
constraints
\begin{equation}
\label{eq:cons_s_t}s+t\neq0~\textrm{mod}~n,\qquad
s-2t\neq0~\textrm{mod}~n,\qquad t-2s\neq0~\textrm{mod}~n\,,
\end{equation}
otherwise the eigenvalues of $c^{s}d^{t}$ would be degenerate and consequently $U_{l}$ can not be determined uniquely. For the value of $s=t=n/2$, the remnant symmetry could be chose to be $K^{(c^{n/2},d^{n/2})}_4\equiv$$\big\{1$, $c^{n/2}$, $d^{n/2}$, $c^{n/2}d^{n/2}$$\big\}$ instead, and then corresponding unitary transformation $U_{l}$ is still a unit matrix. The constraints of Eq.~\eqref{eq:cons_s_t} will be assumed for the subgroup $G_{l}=\left\langle c^{s}d^{t}\right\rangle$ in the following.

\item{$G_{l}=\left\langle bc^{s}d^{t}\right\rangle$}

\begin{equation}
\label{eq:Ul_bcsdt}U_{l}=\frac{1}{\sqrt{2}}\left(\begin{array}{ccc}
e^{-i\pi\frac{s+t}{2n}}  &  0~   &~ e^{-i\pi\frac{s+t}{2n}}\\
0                        &  \sqrt{2}~  &~  0  \\
-e^{i\pi\frac{s+t}{2n}}  &  0~   &~ e^{i\pi\frac{s+t}{2n}}
\end{array}
\right)\,.
\end{equation}
To avoid degenerate eigenvalues, we should exclude the values
\begin{equation}
s-t=0, n/3, 2n/3 ~\textrm{mod}~n\,.
\end{equation}
For the case of $s=t$, the order of the element $bc^{s}d^{s}$ is two and one could extend $G_{l}$ from $\langle bc^{s}d^{s}\rangle=\left\{1, bc^{s}d^{s}\right\}$ to the Klein four subgroup $K^{(c^{n/2}d^{n/2},bc^{s}d^{s})}_4\equiv$$\big\{1$, $c^{n/2}d^{n/2}$, $bc^{s}d^{s}$, $bc^{s+n/2}d^{s+n/2}\big\}$. Then the unitary transformation $U_{l}$ is still of the form in Eq.~\eqref{eq:Ul_bcsdt} with $s=t$.

\item{$G_{l}=\left\langle ac^{s}d^{t}\right\rangle$}

\begin{equation}
\label{eq:ul_ac^sd^t}U_{l}=\frac{1}{\sqrt{3}}\left(\begin{array}{ccc}
e^{-2i\pi\frac{s}{n}}   ~&~   \omega^2e^{-2i\pi\frac{s}{n}} ~&~  \omega
e^{-2i\pi\frac{s}{n}} \\
e^{-2i\pi\frac{t}{n}}  ~&~  \omega e^{-2i\pi\frac{t}{n}} ~&~
\omega^2e^{-2i\pi\frac{t}{n}} \\
1    ~&~   1    ~&~  1
\end{array}
\right)\,,
\end{equation}
where $\omega=e^{2i\pi/3}=-\frac{1}{2}+i\frac{\sqrt{3}}{2}$ is the third root of unity. Notice that the order of the element $ac^{s}d^{t}$ is three regardless of the values of $s$ and $t$, and its eigenvalues are $1$, $\omega$ and $\omega^2$.

\item{$G_{l}=\left\langle a^2c^{s}d^{t}\right\rangle$}

\begin{equation}
\label{eq:ul_a^2c^sd^t}U_{l}=\frac{1}{\sqrt{3}}\left(\begin{array}{ccc}
e^{-2i\pi\frac{t}{n}}  ~&~  \omega^2e^{-2i\pi\frac{t}{n}}  ~&~  \omega
e^{-2i\pi\frac{t}{n}} \\
e^{2i\pi\frac{s-t}{n}}   ~&~  \omega e^{2i\pi\frac{s-t}{n}} ~&~
\omega^2e^{2i\pi\frac{s-t}{n}} \\
1   ~&~  1  ~&~  1
\end{array}
\right)\,.
\end{equation}
Note that because $\left(ac^{t}d^{t-s}\right)^2=a^2c^sd^t$ holds, this $U_{l}$ can be obtained from the one in
Eq.~\eqref{eq:ul_ac^sd^t} by the replacement $s\rightarrow t$,
$t\rightarrow t-s$.

\item{$G_{l}=\left\langle abc^{s}d^{t}\right\rangle$}

\begin{equation}
\label{eq:abcsdt}U_{l}=\frac{1}{\sqrt{2}}\left(\begin{array}{ccc}
e^{i\pi\frac{t-2s}{2n}}  ~&~  e^{i\pi\frac{t-2s}{2n}}   ~&~    0  \\
-e^{-i\pi\frac{t-2s}{2n}}  ~&~ e^{-i\pi\frac{t-2s}{2n}}  ~&~  0  \\
0   ~&~  0   ~&~  \sqrt{2}
\end{array}
\right)\,.
\end{equation}
Non-degeneracy of the eigenvalues of $abc^{s}d^{t}$ requires $t\neq0, n/3$, $2n/3$. In the case of $t=0$, the degeneracy can be avoided
by expanding $G_{l}$ to the Klein four subgroup $K^{(c^{n/2},abc^{s})}_4\equiv$$\big\{1$, $c^{n/2}$, $abc^{s}$,
$abc^{s+n/2}\big\}$, whose diagonalization matrix is of the same form as Eq.~\eqref{eq:abcsdt} with $t=0$.

\item{$G_{l}=\left\langle a^2bc^{s}d^{t}\right\rangle$}

\begin{equation}
\label{eq:a2bcsdt}U_{l}=\frac{1}{\sqrt{2}}\left(\begin{array}{ccc}
\sqrt{2}  ~&~  0   ~&~   0  \\
0   ~&~  e^{i\pi\frac{s-2t}{2n}}   ~&~   e^{i\pi\frac{s-2t}{2n}} \\
0   ~&~  -e^{-i\pi\frac{s-2t}{2n}}  ~&~  e^{-i\pi\frac{s-2t}{2n}}
\end{array}
\right)\,.
\end{equation}
Here the parameter $s$ can not be equal to $0$, $n/3$ or $2n/3$, otherwise two eigenvalues of $a^2bc^{s}d^{t}$ would be identical. For the extended residual symmetry $G_{l}=K^{(d^{n/2},a^2bd^{t})}_4\equiv$$\big\{1$,
$d^{n/2}$, $a^2bd^{t}$, $a^2bd^{t+n/2}\big\}$, the corresponding unitary transformation is still given by Eq.~\eqref{eq:a2bcsdt} with $s=0$.

\end{itemize}

\subsection{\label{subsec:neutrino_one_column}Neutrino sector}

In the present work, we assume the light neutrinos are Majorana particles. As a consequence, the remnant flavor symmetry $G_{\nu}$ in the neutrino sector can only be a $K_4$ or $Z_2$ subgroup. The phenomenological consequence of $G_{\nu}=K_4$ has been studied in Refs.~\cite{King:2014rwa} by two of us. Here we shall concentrate on $G_{\nu}=Z_2$ case with generalized CP symmetry which allows us to predict CP phases. The $Z_2$ subgroups of $\Delta(6n^2)$ can be generated by
\begin{equation}
\label{eq:z2_1}bc^xd^x,\quad abc^y,\quad a^2bd^z,\qquad x,y,z=0,1\ldots n-1
\end{equation}
and additionally
\begin{equation}
\label{eq:z2_2}c^{n/2},\quad d^{n/2},\quad c^{n/2}d^{n/2}
\end{equation}
for $n=2\mathbb{Z}$. It is notable that the $Z_2$ elements in Eq.~\eqref{eq:z2_1} and Eq.~\eqref{eq:z2_2} are conjugate to each other respectively:
\begin{subequations}
\begin{eqnarray}
\nonumber&\hskip-0.2in\left(c^{\gamma}d^{\delta}\right)bc^{x}d^{x}\left(c^{\gamma}d^{\delta}\right)^{-1}=bc^{x-\delta-\gamma}d^{x-\delta-\gamma},\quad &\left(bc^{\gamma}d^{\delta}\right)bc^{x}d^{x}\left(bc^{\gamma}d^{\delta}\right)^{-1}=bc^{-x+\delta+\gamma}d^{-x+\delta+\gamma}\,,\\
\nonumber&\hskip-0.2in\left(ac^{\gamma}d^{\delta}\right)bc^{x}d^{x}\left(ac^{\gamma}d^{\delta}\right)^{-1}=a^2bd^{-x+\delta+\gamma},\quad &\left(a^2c^{\gamma}d^{\delta}\right)bc^{x}d^{x}\left(a^2c^{\gamma}d^{\delta}\right)^{-1}=abc^{-x+\delta+\gamma}\,,\\
\label{eq:neutrino_conjugate_1}&\hskip-0.2in\left(abc^{\gamma}d^{\delta}\right)bc^{x}d^{x}\left(abc^{\gamma}d^{\delta}\right)^{-1}=a^2bd^{x-\delta-\gamma},\quad &\left(a^2bc^{\gamma}d^{\delta}\right)bc^{x}d^{x}\left(a^2bc^{\gamma}d^{\delta}\right)^{-1}=abc^{x-\delta-\gamma}\,.\\
\nonumber&&\\
\nonumber&\left(c^{\gamma}d^{\delta}\right)c^{n/2}\left(c^{\gamma}d^{\delta}\right)^{-1}=c^{n/2},\qquad &\left(bc^{\gamma}d^{\delta}\right)c^{n/2}\left(bc^{\gamma}d^{\delta}\right)^{-1}=d^{n/2}\,,\\
\nonumber&\left(ac^{\gamma}d^{\delta}\right)c^{n/2}\left(ac^{\gamma}d^{\delta}\right)^{-1}=c^{n/2}d^{n/2},\qquad
&\left(a^2c^{\gamma}d^{\delta}\right)c^{n/2}\left(a^2c^{\gamma}d^{\delta}\right)^{-1}=d^{n/2}\,,\\
\label{eq:neutrino_conjugate_4}&\left(abc^{\gamma}d^{\delta}\right)c^{n/2}\left(abc^{\gamma}d^{\delta}\right)^{-1}=c^{n/2},\qquad
&\left(a^2bc^{\gamma}d^{\delta}\right)c^{n/2}\left(a^2bc^{\gamma}d^{\delta}\right)^{-1}=c^{n/2}d^{n/2}\,.
\end{eqnarray}
\end{subequations}
The remnant generalized CP symmetry should be compatible with the residual
$Z_2$ symmetry in the neutrino sector, and therefore the corresponding consistency equation should be satisfied, i.e.,
\begin{equation}
\label{eq:consistency_neutrino}X_{\nu\mathbf{r}}\rho^{*}_{\mathbf{r}}(g)X^{-1}_{\nu\mathbf{r}}=\rho_{\mathbf{r}}(g),\quad
g\in Z_2\,,
\end{equation}
which means that the residual CP and residual flavor transformations are commutable with each other~\cite{Feruglio:2012cw,Ding:2013hpa} in the neutrino sector. For a given solution $X_{\nu\mathbf{r}}$ of Eq.~\eqref{eq:consistency_neutrino}, one can check that $\rho_{\mathbf{r}}(g)X_{\nu\mathbf{r}}$ is also a solution. The remnant CP symmetries consistent with the $Z_2$ elements in Eq.~\eqref{eq:z2_1} and Eq.~\eqref{eq:z2_2} are summarized as follows.
\begin{itemize}
\item{$g=bc^{x}d^{x},~~x=0,1,2\ldots n-1$}
\begin{equation}
X_{\nu\mathbf{r}}=\rho_{\mathbf{r}}(c^{\gamma}d^{-2x-\gamma}),\quad
\rho_{\mathbf{r}}(bc^{\gamma}d^{-\gamma}),~~\gamma=0,1,2\ldots n-1\,.
\end{equation}

\item{$g=abc^{y},~~y=0,1,2\ldots n-1$}
\begin{equation}
X_{\nu\mathbf{r}}=\rho_{\mathbf{r}}(c^{\gamma}d^{2y+2\gamma}),~~
\rho_{\mathbf{r}}(abc^{\gamma}d^{2\gamma}),\quad \gamma=0,1,2\ldots n-1\,.
\end{equation}

\item{$g=a^2bd^{z},~~z=0,1,2\ldots n-1$}
\begin{equation}
X_{\nu\mathbf{r}}=\rho_{\mathbf{r}}(c^{2z+2\delta}d^{\delta}),~~\rho_{\mathbf{r}}(a^2bc^{2\delta}d^{\delta}),\quad
\delta=0,1,2\ldots n-1\,.
\end{equation}

\item{$g=c^{n/2}$}
\begin{equation}
X_{\nu\mathbf{r}}=\rho_{\mathbf{r}}(c^{\gamma}d^{\delta}),~~\rho_{\mathbf{r}}(abc^{\gamma}d^{\delta}),\quad
\gamma,\delta=0,1,2\ldots n-1\,.
\end{equation}

\item{$g=d^{n/2}$}
\begin{equation}
X_{\nu\mathbf{r}}=\rho_{\mathbf{r}}(c^{\gamma}d^{\delta}),~~\rho_{\mathbf{r}}(a^2bc^{\gamma}d^{\delta}),\quad
\gamma,\delta=0,1,2\ldots n-1\,.
\end{equation}

\item{$g=c^{n/2}d^{n/2}$}
\begin{equation}
X_{\nu\mathbf{r}}=\rho_{\mathbf{r}}(c^{\gamma}d^{\delta}),~~\rho_{\mathbf{r}}(bc^{\gamma}d^{\delta}),\quad
\gamma,\delta=0,1,2\ldots n-1\,.
\end{equation}
\end{itemize}
As we shall demonstrate in the following, the remnant CP symmetry should be symmetric to avoid degenerate lepton masses. Then the viable CP transformations would be constrained to be $\rho_{\mathbf{r}}(abc^{\gamma}d^{2\gamma})$, $\rho_{\mathbf{r}}(a^2bc^{2\delta}d^{\delta})$ and $\rho_{\mathbf{r}}(bc^{\gamma}d^{-\gamma})$ together with $\rho_{\mathbf{r}}(c^{\gamma}d^{\delta})$ for $g=c^{n/2}$, $d^{n/2}$ and $c^{n/2}d^{n/2}$ respectively. The full symmetry $\Delta(6n^2)\rtimes H_{CP}$ is broken down to $Z_2\times
H^{\nu}_{CP}$ in the neutrino sector. The invariance of the light neutrino mass matrix $m_{\nu}$ under the remnant family
symmetry $G_{\nu}=Z_2$ and the remnant CP symmetry $H^{\nu}_{CP}$ leads to
\begin{eqnarray}
\nonumber&&\rho^{T}_{\mathbf{3}}(g_{\nu})m_{\nu}\rho_{\mathbf{3}}(g_{\nu})=m_{\nu},\quad
g_{\nu}\in Z^{\nu}_2\,,\\
\label{eq:invariance_neutrino}&&X^{T}_{\nu\mathbf{3}}m_{\nu}X_{\nu\mathbf{3}}=m^{*}_{\nu},\quad X_{\nu}\in
H^{\nu}_{CP}\,,
\end{eqnarray}
from which we can construct the explicit form of $m_{\nu}$ and then diagonalize it.
\begin{description}[labelindent=-0.8em, leftmargin=0.3em]
\item[~~(\romannumeral1)]{$G_{\nu}=Z^{bc^xd^x}_2\equiv\left\{1,bc^xd^x\right\}$,
      $X_{\nu\mathbf{r}}=\left\{\rho_{\mathbf{r}}(c^{\gamma}d^{-2x-\gamma}),
      \rho_{\mathbf{r}}(bc^{x+\gamma}d^{-x-\gamma})\right\}$ }

The light neutrino mass matrix satisfying Eq.~\eqref{eq:invariance_neutrino} is of the following form
\begin{equation}
m_{\nu}=
\left(
\begin{array}{ccc}
 m_{11}e^{-2i\pi\frac{\gamma}{n}}  & m_{12}e^{i\pi\frac{2x+\gamma}{n}} &
 m_{13}e^{-2i\pi\frac{x+\gamma}{n}} \\
 m_{12}e^{i\pi\frac{2x+\gamma}{n}} &  m_{22}e^{4i\pi\frac{x+\gamma}{n}} &
 m_{12}e^{i \pi\frac{\gamma}{n}} \\
m_{13}e^{-2i\pi\frac{x+\gamma}{n}}  &   m_{12}e^{i\pi\frac{\gamma}{n}} &
m_{11}e^{-2i\pi\frac{2x+\gamma}{n}}
\end{array}
\right)\,,
\end{equation}
where $m_{11}$, $m_{12}$, $m_{13}$ and $m_{22}$ are real parameters. This
neutrino mass matrix is diagonalized by the unitary transformation
$U_{\nu}$ via
\begin{equation}
U^{T}_{\nu}m_{\nu}U_{\nu}=\text{diag}\left(m_1,m_2,m_3\right)\,,
\end{equation}
where $U_{\nu}$ is
\begin{equation}
\label{eq:unu_bcxdx}U_{\nu}=\frac{1}{\sqrt{2}}
\left(
\begin{array}{ccc}
 e^{i\pi\frac{\gamma}{n}} & -e^{i\pi\frac{\gamma}{n}}\sin\theta &
 e^{i\pi\frac{\gamma}{n}}\cos\theta \\
 0 &  e^{-2i\pi\frac{x+\gamma}{n}}
   \sqrt{2}\cos\theta~ & ~e^{-2i\pi\frac{x+\gamma}{n}} \sqrt{2}\sin\theta
   \\
 -e^{i\pi\frac{2x+\gamma}{n}} & -e^{i\pi\frac{2x+\gamma}{n}}\sin\theta &
 e^{i\pi\frac{2x+\gamma}{n}}\cos\theta
\end{array}
\right)K_{\nu}\,,
\end{equation}
where $K_{\nu}$ is a diagonal unitary matrix with entries $\pm1$ and $\pm
i$ which encode the CP parity of the neutrino states and renders the light
neutrino masses positive. We shall omit the factor $K_{\nu}$ in the
following cases for simplicity of notation. The angle $\theta$ is given by
\begin{equation}
\tan2\theta=\frac{2\sqrt{2}m_{12}}{m_{11}+m_{13}-m_{22}}\,.
\end{equation}
The light neutrino masses are
\begin{eqnarray}
\nonumber&&m_1=\left|m_{11}-m_{13}\right|,\\
\nonumber&&m_2=\frac{1}{2}\left|m_{11}+m_{13}+m_{22}-\text{sign}\left((m_{11}+m_{13}-m_{22})\cos2\theta\right)\sqrt{(m_{11}+m_{13}-m_{22})^2+8m^2_{12}}\right|,\\
\nonumber&&m_3=\frac{1}{2}\left|m_{11}+m_{13}+m_{22}+\text{sign}\left((m_{11}+m_{13}-m_{22})\cos2\theta\right)\sqrt{(m_{11}+m_{13}-m_{22})^2+8m^2_{12}}\right|\,.
\end{eqnarray}
Here the order of the three eigenvalues $m_1$, $m_2$ and $m_3$ can not be pinned down, consequently the unitary matrix $U_{\nu}$ is determined up to permutations of the columns (the same turns out to be true in the following cases), and the neutrino mass spectrum can be either normal ordering or inverted ordering. Moreover, as four parameters $m_{11}$, $m_{12}$, $m_{13}$ and $m_{22}$ are involved in the neutrino masses, the measured mass squared splitting can be accounted for easily.

\item[~~(\romannumeral2)] {$G_{\nu}=Z^{abc^y}_2\equiv\left\{1,abc^y\right\}$,
    $X_{\nu\mathbf{r}}=\left\{\rho_{\mathbf{r}}(c^{\gamma}d^{2y+2\gamma}),~~
    \rho_{\mathbf{r}}(abc^{y+\gamma}d^{2y+2\gamma})\right\}$}

In this case, the light neutrino mass matrix takes the form:
\begin{equation}
m_{\nu}=
\left(
\begin{array}{ccc}
m_{11}e^{-2i\pi\frac{\gamma}{n}}   &  m_{12}e^{-2i\pi\frac{y+\gamma}{n}}
    & m_{13}e^{i\pi\frac{2y+\gamma}{n}} \\
 m_{12}e^{-2i\pi\frac{y+\gamma}{n}}  & m_{11}e^{-2i\pi\frac{2y+\gamma}{n}}
 & m_{13}e^{i\pi\frac{\gamma}{n}}   \\
 m_{13}e^{i\pi\frac{2y+\gamma}{n}}  & m_{13}e^{i\pi\frac{\gamma}{n}}  &
 m_{33}e^{4i\pi\frac{y+\gamma}{n}}  \\
\end{array}
\right)\,,
\end{equation}
where $m_{11}$, $m_{12}$, $m_{13}$ and $m_{33}$ are real. The unitary
matrix $U_{\nu}$ which diagonalizes the above neutrino mass matrix is
given by
\begin{equation}
U_{\nu}=\frac{1}{\sqrt{2}}
\left(
\begin{array}{ccc}
e^{i\pi\frac{\gamma}{n}} & e^{i\pi\frac{\gamma}{n}}\cos\theta &
e^{i\pi\frac{\gamma}{n}}\sin\theta \\
-e^{i\pi\frac{2y+\gamma}{n}} & e^{i\pi\frac{2y+\gamma}{n}} \cos\theta &
e^{i\pi\frac{2y+\gamma}{n}}\sin\theta \\
0 & -e^{-2i\pi\frac{y+\gamma}{n}}\sqrt{2}\sin\theta~ &
~e^{-2i\pi\frac{y+\gamma}{n}}\sqrt{2}\cos\theta \\
\end{array}
\right)\,,
\end{equation}
with
\begin{equation}
\tan2\theta=\frac{2\sqrt{2}\,m_{13}}{m_{33}-m_{11}-m_{12}}\,.
\end{equation}
The light neutrino mass eigenvalues are determined to be
\begin{eqnarray}
\nonumber&&m_1=\left|m_{11}-m_{12}\right|,\\
\nonumber&&m_2=\frac{1}{2}\left|m_{11}+m_{12}+m_{33}+\text{sign}\left((m_{11}+m_{12}-m_{33})\cos2\theta\right)\sqrt{(m_{11}+m_{12}-m_{33})^2+8m^2_{13}}\right|,\\
\nonumber&&m_2=\frac{1}{2}\left|m_{11}+m_{12}+m_{33}-\text{sign}\left((m_{11}+m_{12}-m_{33})\cos2\theta\right)\sqrt{(m_{11}+m_{12}-m_{33})^2+8m^2_{13}}\right|\,.
\end{eqnarray}

\item[~~(\romannumeral3)]{$G_{\nu}=Z^{a^2bd^z}_2\equiv\left\{1,a^2bd^z\right\}$,
    $X_{\nu\mathbf{r}}=\left\{\rho_{\mathbf{r}}(c^{2z+2\delta}d^{\delta}),
    \rho_{\mathbf{r}}(a^2bc^{2z+2\delta}d^{z+\delta})\right\}$}

The light neutrino mass matrix, which is invariant under both remnant family symmetry and remnant CP symmetry, is of the form:
\begin{equation}
m_{\nu}=
\left(
\begin{array}{ccc}
m_{11}e^{-4i\pi\frac{z+\delta}{n}}   & m_{12}e^{-i\pi\frac{\delta}{n}} &
m_{12}e^{-i\pi\frac{2z+\delta}{n}} \\
m_{12}e^{-i\pi\frac{\delta}{n}}  & m_{22}e^{2i\pi\frac{2z+\delta}{n}} &
m_{23}e^{2i\pi\frac{z+\delta}{n}} \\
m_{12}e^{-i\pi\frac{2z+\delta}{n}}~  & ~m_{23}e^{2i\pi\frac{z+\delta}{n}}~
&
~m_{22}e^{2i\pi\frac{\delta}{n}}
\end{array}
\right)\,,
\end{equation}
where $m_{11}$, $m_{12}$, $m_{22}$ and $m_{23}$ are real parameters. The neutrino diagonalization matrix $U_{\nu}$ is given by
\begin{equation}
U_{\nu}=\frac{1}{\sqrt{2}}
\left(
\begin{array}{ccc}
 0 & ~-e^{2i\pi\frac{z+\delta}{n}}\sqrt{2}\sin\theta~ &
    ~e^{2i\pi\frac{z+\delta}{n}}\sqrt{2}\cos\theta \\
 e^{-i\pi\frac{2z+\delta}{n}} & e^{-i\pi\frac{2z+\delta}{n}}\cos\theta &
 e^{-i\pi\frac{2z+\delta}{n}}\sin\theta \\
 -e^{-i\pi\frac{\delta}{n}} & e^{-i\pi\frac{\delta}{n}}\cos\theta &
 e^{-i\pi\frac{\delta}{n}}\sin\theta
\end{array}
\right)\,,
\end{equation}
where the angle $\theta$ fulfills
\begin{equation}
\tan2\theta=\frac{2\sqrt{2}m_{12}}{m_{11}-m_{22}-m_{23}}\,.
\end{equation}
Finally the light neutrino masses are
\begin{eqnarray}
\nonumber&&m_1=\left|m_{22}-m_{23}\right|,\\
\nonumber&&m_2=\frac{1}{2}\left|m_{11}+m_{22}+m_{23}-\text{sign}\left((m_{11}-m_{22}-m_{23})\cos2\theta\right)\sqrt{(m_{11}-m_{22}-m_{23})^2+8m^2_{12}}\right|,\\
\nonumber&&m_3=\frac{1}{2}\left|m_{11}+m_{22}+m_{23}+\text{sign}\left((m_{11}-m_{22}-m_{23})\cos2\theta\right)\sqrt{(m_{11}-m_{22}-m_{23})^2+8m^2_{12}}\right|\,.
\end{eqnarray}

\item[~~(\romannumeral4)]{$G_{\nu}=Z^{c^{n/2}}_2\equiv\left\{1,c^{n/2}\right\}$,
    $X_{\nu\mathbf{r}}=\left\{\rho_{\mathbf{r}}(c^{\gamma}d^{\delta}),
    \rho_{\mathbf{r}}(abc^{\gamma}d^{\delta})\right\}$}

\begin{itemize}

\item{$X_{\nu\mathbf{r}}=\rho_{\mathbf{r}}(c^{\gamma}d^{\delta})$}

The light neutrino mass matrix is constrained to be of the following
form
\begin{equation}
m_{\nu}=\left(
\begin{array}{ccc}
m_{11}e^{-2i\pi\frac{\gamma}{n}}  & m_{12}e^{-i\pi\frac{\delta}{n}} &
0 \\
m_{12}e^{-i\pi\frac{\delta}{n}}  &
m_{22}e^{-2i\pi\frac{\delta-\gamma}{n}}  & 0 \\
 0 & 0 & m_{33}e^{2i\pi\frac{\delta}{n}}
\end{array}
\right)\,,
\end{equation}
where $m_{11}$, $m_{12}$, $m_{22}$ and $m_{33}$ are real. The unitary transformation $U_{\nu}$ is
\begin{equation}
U_{\nu}=
\left(
\begin{array}{ccc}
e^{i\pi\frac{\gamma}{n}}\cos\theta &  e^{i\pi\frac{\gamma}{n}}\sin\theta
& 0 \\
-e^{i\pi\frac{\delta-\gamma}{n}}\sin\theta~ &
~e^{i\pi\frac{\delta-\gamma}{n}}\cos\theta & 0 \\
 0 & 0 & e^{-i\pi\frac{\delta}{n}}
\end{array}
\right)\,,
\end{equation}
where
\begin{equation}
\tan2\theta=\frac{2m_{12}}{m_{22}-m_{11}}\,.
\end{equation}
The light neutrino masses are determined to be
\begin{eqnarray}
\nonumber&&
m_1=\frac{1}{2}\left|m_{11}+m_{22}-\text{sign}\left((m_{22}-m_{11})\cos2\theta\right)\sqrt{(m_{22}-m_{11})^2+4m^2_{12}}\right|,\\
\nonumber&&
m_2=\frac{1}{2}\left|m_{11}+m_{22}+\text{sign}\left((m_{22}-m_{11})\cos2\theta\right)\sqrt{(m_{22}-m_{11})^2+4m^2_{12}}\right|,\\
&&m_3=\left|m_{33}\right|\,.
\end{eqnarray}

\item{$X_{\nu\mathbf{r}}=\rho_{\mathbf{r}}(abc^{\gamma}d^{\delta})$}

For the case of $\delta\neq2\gamma~\text{mod}~n$, the light neutrino masses would be partially degenerate. This is obviously unviable. The reason is that the corresponding generalized CP transformation matrix is not symmetric~\footnote{In the basis in which the neutrino mass matrix is diagonal with $m_{\nu}=\textrm{diag}(m_1, m_2, m_3)$, the general CP transformation $\widehat{X}$ which leaves $m_{\nu}$ invariant: $\widehat{X}^{T}m_{mu}\widehat{X}=m^{*}_{\nu}$, should be of the form $\widehat{X}=\textrm{diag}(\pm1,\pm1,\pm1)$. One can go to an arbitrary basis and define the corresponding CP symmetry transformation $X=V^{\dagger}\widehat{X}V^{*}$ as a symmetry of the general neutrino mass matrix, where $V$ is the basis transformation. As a result, the remnant CP symmetry $X$ in the neutrino sector should be symmetric. The same conclusion has been obtained in Ref.~\cite{Feruglio:2012cw}.}. Therefore we shall concentrate on the case of $\delta=2\gamma~\text{mod}~ n$ in the following. The neutrino mass matrix is given by
\begin{equation}
m_{\nu}=
\left(
\begin{array}{ccc}
m_{11}e^{i\phi} & m_{12}e^{-2i\pi\frac{\gamma}{n}}  & 0 \\
m_{12}e^{-2i\pi\frac{\gamma}{n}}~  &
~m_{11}e^{-i(4\pi\frac{\gamma}{n}+\phi)}  & 0 \\
 0 & 0 & m_{33}e^{4i\pi\frac{\gamma}{n}}
\end{array}
\right)\,,
\end{equation}
where $m_{11}$, $m_{12}$, $m_{33}$ and $\phi$ are real free parameters. The neutrino diagonalization matrix is
\begin{equation}
U_{\nu}=\frac{1}{\sqrt{2}}\left(
\begin{array}{ccc}
 e^{-i\frac{\phi}{2}} &   e^{-i\frac{\phi}{2}} & 0   \\
 -e^{i(\frac{\phi}{2}+2\pi\frac{\gamma}{n})}~ &
~e^{i(\frac{\phi}{2}+2\pi\frac{\gamma}{n})} &   0 \\
 0 & 0 & \sqrt{2}\,e^{-2i\pi\frac{\gamma}{n}}
\end{array}
\right)\,.
\end{equation}
The light neutrino mass eigenvalues are
\begin{equation}
m_1=\left|m_{11}-m_{12}\right|,\quad m_2=\left|m_{11}+m_{12}\right|,\quad m_{3}=\left|m_{33}\right|\,.
\end{equation}
The ordering of the neutrino masses can not be determined as well.

\end{itemize}

\item[~~(\romannumeral5)]{$G_{\nu}=Z^{d^{n/2}}_2\equiv\left\{1,d^{n/2}\right\}$,
    $X_{\nu\mathbf{r}}=\left\{\rho_{\mathbf{r}}(c^{\gamma}d^{\delta}),
    \rho_{\mathbf{r}}(a^2bc^{\gamma}d^{\delta})\right\}$}

\begin{itemize}

\item{$X_{\nu\mathbf{r}}=\rho_{\mathbf{r}}(c^{\gamma}d^{\delta})$}

The light neutrino mass matrix is constrained by residual family and residual CP symmetries to be
\begin{equation}
m_{\nu}=\left(
\begin{array}{ccc}
 m_{11}e^{-2i\pi\frac{\gamma}{n}}  & 0 & 0 \\
 0 & m_{22}e^{-2i\pi\frac{\delta-\gamma}{n}} &
 m_{23}e^{i\pi\frac{\gamma}{n}}    \\
 0 & m_{23}e^{i\pi\frac{\gamma}{n}}  & m_{33}e^{2i\pi\frac{\delta}{n}}
\end{array}
\right)\,,
\end{equation}
where $m_{11}$, $m_{22}$, $m_{23}$ and $m_{33}$ are real. The neutrino diagonalization matrix is
\begin{equation}
U_{\nu}=\left(
\begin{array}{ccc}
 e^{i\pi\frac{\gamma}{n}}   & 0 & 0 \\
 0 & e^{i\pi\frac{\delta-\gamma}{n}}\cos\theta~ &
 ~e^{i\pi\frac{\delta-\gamma}{n}}\sin\theta \\
 0 & -e^{-i\pi\frac{\delta}{n}}\sin\theta~ &
 ~e^{-i\pi\frac{\delta}{n}}\cos\theta
\end{array}
\right)\,,
\end{equation}
with
\begin{equation}
\tan2\theta=\frac{2m_{23}}{m_{33}-m_{22}}\,.
\end{equation}
The light neutrino masses take the form
\begin{eqnarray}
\nonumber&&\hskip-0.3in m_1=\left|m_{11}\right|,\\
\nonumber&&\hskip-0.3in
m_2=\frac{1}{2}\left|m_{22}+m_{33}-\text{sign}\left((m_{33}-m_{22})\cos2\theta\right)\sqrt{(m_{33}-m_{22})^2+4m^2_{23}}\right|,\\
&&\hskip-0.3in
m_3=\frac{1}{2}\left|m_{22}+m_{33}+\text{sign}\left((m_{33}-m_{22})\cos2\theta\right)\sqrt{(m_{33}-m_{22})^2+4m^2_{23}}\right|\,.
\end{eqnarray}

\item{$X_{\nu\mathbf{r}}=\rho_{\mathbf{r}}(a^2bc^{\gamma}d^{\delta})$}

As has been shown above, $X_{\nu\mathbf{r}}$ has to be symmetric. Then the requirement $\gamma=2\delta~\textrm{mod}~n$ follows immediately, otherwise the light neutrino masses would be partially degenerate. In this case, the neutrino mass matrix takes the form:
\begin{equation}
m_{\nu}=\left(
\begin{array}{ccc}
 m_{11}e^{-4i\pi\frac{\delta}{n}}  & 0 & 0 \\
 0 & m_{22}e^{i\phi}  & m_{23}e^{2i\pi\frac{\delta}{n}}  \\
 0 & m_{23}e^{2i\pi\frac{\delta}{n}}~ &
 ~m_{22}e^{i(4\pi\frac{\delta}{n}-\phi)}
\end{array}
\right)\,,
\end{equation}
where $m_{11}$, $m_{22}$, $m_{23}$ and $\phi$ are real. It is
diagonalized by the unitary matrix
\begin{equation}
U_{\nu}=\frac{1}{\sqrt{2}}
\left(
\begin{array}{ccc}
 \sqrt{2} e^{2i\pi\frac{\delta}{n}} & 0 & 0   \\
 0 & e^{-i\frac{\phi}{2}} &   e^{-i\frac{\phi}{2}} \\
 0 & -e^{i\left(\frac{\phi}{2}-2\pi\frac{\delta}{n}\right)}~ &
   ~e^{i\left(\frac{\phi}{2}-2\pi\frac{\delta}{n}\right)}
\end{array}
\right)\,.
\end{equation}
The light neutrino masses are
\begin{equation}
m_1=\left|m_{11}\right|,\quad m_2=\left|m_{22}-m_{23}\right|,\quad m_3=\left|m_{22}+m_{23}\right|\,.
\end{equation}

\end{itemize}

\item[~~(\romannumeral6)]{$G_{\nu}=Z^{c^{n/2}d^{n/2}}_2\equiv\left\{1,c^{n/2}d^{n/2}\right\}$,
    $X_{\nu\mathbf{r}}=\left\{\rho_{\mathbf{r}}(c^{\gamma}d^{\delta}),
    \rho_{\mathbf{r}}(bc^{\gamma}d^{\delta})\right\}$}

\begin{itemize}

\item{$X_{\nu\mathbf{r}}=\rho_{\mathbf{r}}(c^{\gamma}d^{\delta})$}

The light neutrino mass matrix invariant under both the residual family and residual CP symmetries is
\begin{equation}
m_{\nu}=\left(
\begin{array}{ccc}
m_{11}e^{-2i\pi\frac{\gamma}{n}}  & 0 &
m_{13}e^{-i\pi\frac{\gamma-\delta}{n}} \\
0 & m_{22}e^{-2i\pi\frac{\delta-\gamma}{n}} & 0 \\
m_{13}e^{-i\pi\frac{\gamma-\delta}{n}} & 0 &
m_{33}e^{2i\pi\frac{\delta}{n}} \\
\end{array}
\right)\,,
\end{equation}
where $m_{11}$, $m_{13}$, $m_{22}$ and $m_{33}$ are real parameters. The unitary transformation $U_{\nu}$ is given by
\begin{equation}
U_{\nu}=\left(
\begin{array}{ccc}
 e^{i\pi\frac{\gamma}{n}}\cos\theta & 0 &
 e^{i\pi\frac{\gamma}{n}}\sin\theta \\
 0 & e^{i\pi\frac{\delta-\gamma}{n}} & 0   \\
 -e^{-i\pi\frac{\delta}{n}}\sin\theta & 0 &
   e^{-i\pi\frac{\delta}{n}}\cos\theta \\
\end{array}
\right)\,,
\end{equation}
with
\begin{equation}
\tan2\theta=\frac{2m_{13}}{m_{33}-m_{11}}\,.
\end{equation}
The light neutrino mass eigenvalues are
\begin{eqnarray}
\nonumber&&\hskip-0.3in
m_1=\frac{1}{2}\left|m_{11}+m_{33}-\text{sign}\left((m_{33}-m_{11})\cos2\theta\right)\sqrt{(m_{33}-m_{11})^2+4m^2_{13}}\right|,\\
\nonumber&&\hskip-0.3in m_{2}=\left|m_{22}\right|,\\
&&\hskip-0.3in
m_3=\frac{1}{2}\left|m_{11}+m_{33}+\text{sign}\left((m_{33}-m_{11})\cos2\theta\right)\sqrt{(m_{33}-m_{11})^2+4m^2_{13}}\right|\,.
\end{eqnarray}

\item{$X_{\nu\mathbf{r}}=\rho_{\mathbf{r}}(bc^{\gamma}d^{\delta})$}

In the case of $\gamma+\delta\neq0~\text{mod}~n$, the generalized CP transformation $\rho_{\mathbf{r}}(bc^{\gamma}d^{\delta})$ is not symmetric. As a consequence, the light neutrino masses are partially degenerate. In the following, we shall focus on the case of
$\gamma+\delta=0~\text{mod}~n$. The neutrino mass matrix is determined to be of the following form:
\begin{equation}
m_{\nu}=\left(
\begin{array}{ccc}
 m_{11}e^{i\phi} & 0 & m_{13}e^{-2i\pi\frac{\gamma}{n}}  \\
 0 & m_{22}e^{4i\pi\frac{\gamma}{n}}  &   0 \\
 m_{13}e^{-2i\pi\frac{\gamma}{n}}  & 0 &
 m_{11}e^{-i(\phi+4\pi\frac{\gamma}{n})}  \\
\end{array}
\right)\,,
\end{equation}
where $m_{11}$, $m_{13}$, $m_{22}$ and $\phi$ are real. The neutrino diagonalization matrix is
\begin{equation}
U_{\nu}=\frac{1}{\sqrt{2}}\left(
\begin{array}{ccc}
 e^{-i\frac{\phi}{2}} & 0 &   e^{-i\frac{\phi}{2}} \\
 0 & \sqrt{2} e^{-2i\pi\frac{\gamma}{n}} & 0 \\
-e^{i(\frac{\phi}{2}+2\pi\frac{\gamma}{n})} &   0 &
e^{i(\frac{\phi}{2}+2\pi\frac{\gamma}{n})} \\
\end{array}
\right)\,.
\end{equation}
Finally the light neutrino masses are given by
\begin{equation}
m_1=\left|m_{11}-m_{13}\right|,\quad m_2=\left|m_{22}\right|,\quad m_3=\left|m_{11}+m_{13}\right|\,.
\end{equation}

\end{itemize}

\end{description}

\subsection{\label{subsec:PMNS_one_column}Predictions for lepton flavor mixing}

Since the possible forms of the neutrino and charged lepton mass matrices and their diagonalization matrices have been worked out in previous sections, the lepton flavor mixing matrix can be pinned down immediately
\begin{equation}
U_{PMNS}=U^{\dagger}_{l}U_{\nu}\,.
\end{equation}
Because the order of the charged-lepton and neutrino masses is indeterminate in the present framework, the PMNS matrix $U_{PMNS}$ is determined up to independent permutations of rows and columns. From
Eqs.(\ref{eq:neutrino_conjugate_1},\ref{eq:neutrino_conjugate_4}),
we know that the remnant $Z_2$ symmetries generated by $bc^xd^x$, $abc^y$, $a^2bd^z$ are conjugate to each other, and the same is true for the $Z_2$ symmetry generated by $c^{n/2}$, $d^{n/2}$ and
$c^{n/2}d^{n/2}$. If a pair of residual flavor symmetries $(G^{\prime}_{\nu}, G^{\prime}_{l})$ is conjugated to the pair of groups $(G_{\nu}, G_{l})$ under the group element $g\in\Delta(6n^2)$, it has been established that both pairs lead to the same result for $U_{PMNS}$ even after the generalized CP symmetry is included~\cite{Ding:2013bpa}. As a consequence, we only need to need to consider the representative residual symmetry $G_{\nu}=Z^{bc^xd^x}_2$, $Z^{c^{n/2}}_2$ and $G_{l}=\left\langle c^{s}d^{t}\right\rangle$, $\left\langle
bc^{s}d^{t}\right\rangle$, $\left\langle ac^{s}d^{t}\right\rangle$,
$\left\langle abc^{s}d^{t}\right\rangle$ and $\left\langle
a^2bc^{s}d^{t}\right\rangle$. Because the remnant flavor symmetry in the neutrino sector is taken to be $Z_2$ instead of $K_{4}$ subgroup, only one column of the PMNS matrix can be fixed up to permutation and rephasing of the elements in this 	scenario. The concrete form of the determined columns for different choices of the remnant flavor symmetry is summarized in Table~\ref{tab:PMNS_column}. The present $3\sigma$ confidence level ranges for the magnitude of the elements of the leptonic mixing matrix are fitted to be~\cite{Capozzi:2013csa}:
\begin{equation}
\label{eq:3sigma_ranges}||U_{PMNS}||_{3\sigma}=\left(
\begin{array}{ccc}
0.789\rightarrow0.853 ~&~ 0.501\rightarrow0.594 ~&~
0.133\rightarrow0.172\\
0.194\rightarrow0.558 ~&~ 0.408\rightarrow0.735 ~&~
0.602\rightarrow0.784 \\
0.194\rightarrow0.558 ~&~ 0.408\rightarrow0.735 ~&~
0.602\rightarrow0.784 \\
\end{array}
\right)\,,
\end{equation}
for normal ordering neutrino mass spectrum, and a very similar result is obtained for inverted ordering spectrum. We see that no entry of the PMNS matrix can be zero. As a result, the mixing patterns with a zero element have been ruled out by experimental data of neutrino mixing. In the following, we shall concentrate on the viable case in which no element of the fixed column is zero, and the predictions for the lepton flavor mixing parameters will be investigated for various remnant CP symmetries compatible with remnant flavor symmetry.

\begin{table}[t!]
\renewcommand{\tabcolsep}{2.0mm}
\centering
\begin{tabular}{|c||c|c|}
\hline \hline
 &  &     \\ [-0.16in]
 &  $G_{\nu}=Z^{bc^xd^x}_2$  &  $G_{\nu}=Z^{c^{n/2}}_2$  \\

  &   &      \\ [-0.16in]\hline
 &   &       \\ [-0.16in]

$G_l=\langle c^{s}d^{t}\rangle$  &   $\frac{1}{\sqrt{2}}\left(\begin{array}{c}
0 \\
-1\\
1
\end{array}\right)$\xmark  &  $\left(\begin{array}{c}
0\\
0\\
1
\end{array}\right)$ \xmark  \\
 &   &         \\ [-0.16in]\hline
 &   &        \\ [-0.16in]

$G_l=\langle bc^{s}d^{t}\rangle$ &   $\left(\begin{array}{c}
0\\
\cos\left(\frac{s+t-2x}{2n}\pi\right)\\
\sin\left(\frac{s+t-2x}{2n}\pi\right)
\end{array}
\right)$ \xmark   &  $\frac{1}{\sqrt{2}}\left(\begin{array}{c}
0\\
-1\\
1
\end{array}
\right)$\xmark \\
 &   &         \\ [-0.16in]\hline
 &   &        \\ [-0.16in]

$G_l=\langle ac^{s}d^{t}\rangle$  &  $\sqrt{\frac{2}{3}}\left(
\begin{array}{c}
\sin\left(\frac{s-x}{n}\pi\right) \\
\cos\left(\frac{\pi}{6}-\frac{s-x}{n}\pi\right) \\
\cos\left(\frac{\pi}{6}+\frac{s-x}{n}\pi\right)
\end{array}
\right)$ \cmark  &  $\frac{1}{\sqrt{3}}\left(\begin{array}{c}
1\\
1\\
1
\end{array}
\right)$ \cmark \\
 &   &         \\ [-0.16in]\hline
 &   &        \\ [-0.16in]

$G_l=\langle abc^{s}d^{t}\rangle$  &  $\frac{1}{2}\left(\begin{array}{c}
1\\
1\\
-\sqrt{2}
\end{array}
\right)$ \cmark  & $\left(\begin{array}{c}
0\\
0\\
1
\end{array}
\right)$ \xmark   \\
 &   &         \\ [-0.16in]\hline
 &   &        \\ [-0.16in]

$G_l=\langle a^2bc^{s}d^{t}\rangle$ &  $\frac{1}{2}\left(\begin{array}{c}
1\\
1\\
-\sqrt{2}
\end{array}
\right)$ \cmark  & $\frac{1}{\sqrt{2}}\left(\begin{array}{c}
0\\
-1\\
1
\end{array}
\right)$ \xmark   \\

 &   &      \\ [-0.16in]\hline\hline

\end{tabular}
\caption{\label{tab:PMNS_column} The determined form of one column of the PMNS matrix for different remnant symmetries $G_{\nu}$ and $G_{l}$ which are $Z_2$ and abelian subgroups of $\Delta(6n^2)$ family symmetry groups respectively. The symbol ``\xmark'' denotes that the resulting lepton mixing is ruled out since there is at least one zero element in the fixed column, and the symbol ``\cmark'' denotes that the resulting mixing is viable. Note that for $G_{\nu}=Z^{bc^xd^x}_2$, the cases of $G_l=\langle abc^{s}d^{t}\rangle$ and $G_l=\langle a^2bc^{s}d^{t}\rangle$ are not independent as we have $b\left(abc^{s}d^{t}\right)b=a^2bc^{-t}d^{-s}$ and
$b\left(bc^xd^x\right)b=bc^{-x}d^{-x}$.  }
\end{table}

\begin{description}[labelindent=-0.8em, leftmargin=0.3em]

\item[~~(\uppercase\expandafter{\romannumeral1})]

$G_{l}=\left\langle ac^{s}d^{t}\right\rangle$,
$G_{\nu}=Z^{bc^xd^x}_2$,
$X_{\nu\mathbf{r}}=\left\{\rho_{\mathbf{r}}(c^{\gamma}d^{-2x-\gamma}),
\rho_{\mathbf{r}}(bc^{x+\gamma}d^{-x-\gamma})\right\}$

The PMNS matrix is determined to be
{\footnotesize \begin{eqnarray*}
U^{I}_{PMNS}=\frac{1}{\sqrt{3}}\left(
\begin{array}{ccc}
\sqrt{2}\sin\varphi_1 ~&~
e^{i\varphi_2}\cos\theta-\sqrt{2}\sin\theta\cos\varphi_1 ~&~
e^{i\varphi_2}\sin\theta+\sqrt{2}\cos\theta\cos\varphi_1 \\
\sqrt{2}\cos\left(\frac{\pi}{6}-\varphi_1\right) ~&~
-e^{i\varphi_2}\cos\theta-\sqrt{2}\sin\theta\sin
\left(\frac{\pi}{6}-\varphi_1\right)  ~&~
-e^{i\varphi_2}\sin\theta+\sqrt{2}\cos\theta\sin\left(\frac{\pi}{6}-\varphi_1\right)\\
\sqrt{2}\cos\left(\frac{\pi}{6}+\varphi_1\right) ~&~
e^{i\varphi_2}\cos\theta+\sqrt{2}\sin\theta\sin\left(\frac{\pi}{6}+\varphi_1\right)
~&~
e^{i\varphi_2}\sin\theta-\sqrt{2}\cos\theta\sin\left(\frac{\pi}{6}+\varphi_1\right)
\\
\end{array}
\right)\,,
\end{eqnarray*}}
where
\begin{equation}
\varphi_1=\frac{s-x}{n}\pi,\qquad
\varphi_2=\frac{2t-s-3(\gamma+x)}{n}\pi\,.
\end{equation}
These two parameters $\varphi_1$ and $\varphi_2$ are interdependent of each other, and they can take the discrete values
\begin{eqnarray}
\nonumber&&\varphi_1=0, \pm\frac{1}{n}\pi, \pm\frac{2}{n}\pi, \ldots \pm\frac{n-1}{n}\pi\,,\\
&&\varphi_2~\textrm{mod}~2\pi=0, \frac{1}{n}\pi, \frac{2}{n}\pi, \ldots \frac{2n-1}{n}\pi\,.
\end{eqnarray}
Obviously one column of the PMNS matrix is fixed to be
\begin{equation}
\sqrt{\frac{2}{3}}\left(\begin{array}{c}
\sin\varphi_1\\
\cos\left(\pi/6-\varphi_1\right)\\
\cos\left(\pi/6+\varphi_1\right)
\end{array}
\right)\equiv\mathcal{C}\,.
\end{equation}
For the group $\Delta(24)\equiv S_4$, we have
\begin{equation}
\varphi_1=0~:~\mathcal{C}=\frac{1}{\sqrt{2}}\left(\begin{array}{c}
0\\
1\\
1
\end{array}\right),\qquad
\varphi_1=\pm\frac{\pi}{2}~:~\mathcal{C}=\pm\frac{1}{\sqrt{6}}\left(\begin{array}{c}
2\\
1\\
-1
\end{array}\right)\,.
\end{equation}
Note that the first column of the tri-bimaximal mixing matrix is reproduced for
$\varphi_1=\pm\pi/2$. For the group $\Delta(96)$, we have
\begin{eqnarray}
\nonumber&&\varphi_1=\frac{3\pi}{4},
-\frac{\pi}{4}~:~\mathcal{C}=\pm\frac{1}{6}\left(\begin{array}{c}
2\sqrt{3}\\
-3+\sqrt{3}\\
-3-\sqrt{3}
\end{array}\right), \qquad
\varphi_1=\pm\frac{\pi}{2}~:~\mathcal{C}=\pm\frac{1}{\sqrt{6}}\left(\begin{array}{c}
2\\
1\\
-1
\end{array}\right),\\
&&\varphi_1=\frac{\pi}{4},-\frac{3\pi}{4}~:~\mathcal{C}=\pm\frac{1}{6}\left(\begin{array}{c}
2\sqrt{3}\\
3+\sqrt{3}\\
3-\sqrt{3}
\end{array}\right),~~\qquad
\varphi_1=0~:~\mathcal{C}=\frac{1}{\sqrt{2}}\left(\begin{array}{c}
0\\
1\\
1
\end{array}\right)\,.
\end{eqnarray}
It is interesting that the first and the third columns (up to
permutations) of the Toorop-Feruglio-Hagedorn mixing~\cite{Toorop:2011jn,Ding:2012xx,King:2012in} can be obtained in the
case of $\varphi_1=\pm\pi/4, \pm3\pi/4$. Now we consider the permutations of the rows and the columns, i.e. The PMNS matrix can be multiplied by a $3\times3$ permutation matrix from both the left-hand side and the right-hand side. There are six permutation matrices corresponding to six possible orderings of rows (or columns):
\begin{eqnarray}
\nonumber&&P_{123}=\left(\begin{array}{ccc}
1  & 0  &  0 \\
0  & 1  &  0\\
0  & 0  &  1
\end{array}\right),\quad P_{132}=\left(\begin{array}{ccc}
1  &  0 &  0 \\
0  &  0 &  1 \\
0  &  1 &  0
\end{array}\right),\quad P_{213}=\left(\begin{array}{ccc}
0  &  1  &  0 \\
1  &  0  &  0 \\
0  &  0  &  1
\end{array}\right),\\
\label{eq:per_matrix}&& P_{231}=\left(\begin{array}{ccc}
0   &  1   &  0 \\
0   &  0   &  1  \\
1   &  0   &  0
\end{array}\right),\quad P_{312}=\left(\begin{array}{ccc}
0   &  0  &   1  \\
1   &  0  &   0 \\
0   &  1  &  0
\end{array}\right),\quad P_{321}=\left(\begin{array}{ccc}
0    &   0    &   1  \\
0    &   1    &   0  \\
1    &   0    &   0
\end{array}\right)\,.
\end{eqnarray}
It is well-known that the atmospheric mixing angle $\theta_{23}$ becomes $\pi/2-\theta_{23}$, the Dirac CP phases $\delta_{CP}$ becomes $\pi+\delta_{CP}$ and the other mixing parameters are kept intact if the second and third row of the PMNS matrix are exchanged. Hence the permutation of the second and third row will not be explored explicitly in the following. First we study the situation that the
constant vector $\sqrt{2/3}\left(\sin\varphi_1,
\cos\left(\pi/6-\varphi_1\right), \cos\left(\pi/6+\varphi_1\right)\right)^{T}$ is in the first column. Then the PMNS matrix can be arranged as follows:
\begin{equation}
\label{eq:arrangement_castI_13}U^{I,1st}_{PMNS}=U^{I}_{PMNS},\qquad U^{I,2nd}_{PMNS}=P_{231}U^{I}_{PMNS},\qquad U^{I,3rd}_{PMNS}=P_{312}U^{I}_{PMNS}\,.
\end{equation}
Note that exchanging its second and third row does not lead to new mixing patterns. Moreover, we see that the above three arrangements are related with each other:
\begin{eqnarray}
\nonumber&&U^{I,2nd}_{PMNS}(\theta,\varphi_{1},\varphi_{2})=\textrm{diag}(1,1,-1)U^{I,1st}_{PMNS}(\pi-\theta,\frac{\pi}{3}+\varphi_1,\varphi_2)\textrm{diag}(1,1,-1),\\
\label{eq:PMNS_relation_castI_13}&&U^{I,3rd}_{PMNS}(\theta,\varphi_{1},\varphi_{2})=\textrm{diag}(-1,1,1)U^{I,1st}_{PMNS}(-\theta,-\frac{\pi}{3}+\varphi_1,\varphi_2)\textrm{diag}(1,-1,1)\,,
\end{eqnarray}
where the phase factor $\textrm{diag}\left(\pm1, \pm1, \pm1\right)$ can be absorbed by the lepton fields. Hence it is sufficient to only discuss the first PMNS matrix $U^{I,1st}_{PMNS}$ in detail, the phenomenological predictions for the other two can be easily obtained by variable substitution. In this case, the lepton mixing parameters are predicted to be
\begin{eqnarray}
\nonumber&&\sin^2\theta_{13}=\frac{1}{3}\left(1+\cos^2\theta\cos2\varphi_1+\sqrt{2}\sin2\theta\cos\varphi_2\cos\varphi_1\right),\\
\nonumber&&\sin^2\theta_{12}=\frac{1+\sin^2\theta\cos2\varphi_1-\sqrt{2}\sin2\theta\cos\varphi_2\cos\varphi_1}
{2-\cos^2\theta\cos2\varphi_1-\sqrt{2}\sin2\theta\cos\varphi_2\cos\varphi_1},\\
\nonumber&&\sin^2\theta_{23}=\frac{1-\cos^2\theta\sin\left(\pi/6+2\varphi_1\right)-\sqrt{2}\sin2\theta\cos\varphi_2\sin\left(\pi/6-\varphi_1\right)}
{2-\cos^2\theta\cos2\varphi_1-\sqrt{2}\sin2\theta\cos\varphi_2\cos\varphi_1}\,,\\
\nonumber&&\left|\tan\delta_{CP}\right|=\Big|2\sqrt{2}\sin2\theta\sin\varphi_2(1+2\cos2\varphi_1)\left(2-\cos^2\theta\cos2\varphi_1-\sqrt{2}\sin2\theta\cos\varphi_2\cos\varphi_1\right)\Big/\\
\nonumber&&\quad\Big\{2\sin^22\theta\cos2\varphi_2(\cos3\varphi_1-2\cos\varphi_1)+\cos\varphi_1\left(9-4\cos2\theta+3\cos4\theta-16\cos^2\theta\cos2\varphi_1\right)\\
\nonumber&&\quad-2\sqrt{2}\sin2\theta\cos\varphi_2\left[2-\cos^2\theta(5+\cos2\varphi_1+\cos4\varphi_1)\right]
\Big\}\Big|\,\\
\nonumber&&\left|J_{CP}\right|=\frac{1}{6\sqrt{6}}\left|\sin2\theta\sin\varphi_2\sin3\varphi_1\right|\,,\\
\nonumber&&\left|\tan\alpha_{21}\right|=\left|\frac{2\sin\varphi_2\left(\cos\varphi_2-\sqrt{2}\cos\varphi_1\tan\theta\right)}{\cos2\varphi_2-2\cos\varphi_1\tan\theta\left(\sqrt{2}\cos\varphi_2-\cos\varphi_1\tan\theta\right)}\right|\,,\\
\label{eq:mixing_parameter_III_1st}&&\left|\tan\alpha^{\prime}_{31}\right|=\left|\frac{2\sin\varphi_2\left(\cos\varphi_2+\sqrt{2}\cos\varphi_1\cot\theta\right)}
{\cos2\varphi_2+2\cos\varphi_1\cot\theta\left(\sqrt{2}\cos\varphi_2+\cos\varphi_1\cot\theta\right)}\right|\,,
\end{eqnarray}
where $\alpha^{\prime}_{31}=\alpha_{31}-2\delta_{CP}$, $\delta_{CP}$ is the Dirac CP phase, $\alpha_{21}$ and $\alpha_{31}$ are the Majorana CP phases in the standard parameterization~\cite{pdg}. If we embed the three generations of left-handed lepton doublets into the triplet $\mathbf{3}_{1,n-1}$ which is the complex conjugate representation
of $\mathbf{3}_{1,1}$, all three CP phases $\delta_{CP}$,
$\alpha_{21}$ and $\alpha_{31}$ would become their opposite numbers modulo $2\pi$. Furthermore, the overall sign of $\tan\alpha_{21}$ and $\tan\alpha^{\prime}_{31}$ depends on the CP parity of the neutrino states which is encoded in the matrix $K_{\nu}$ (please see Eq.~\eqref{eq:unu_bcxdx}), and the sign of the Jarlskog invariant $J_{CP}$ depends on the ordering of rows and columns. As a result, all these quantities are presented in terms of absolute values here.

It is notable that all three CP phases depend on both the free continuous parameter $\theta$ and the discrete parameters $\varphi_1$ and $\varphi_2$ associated with flavor and CP symmetries. Obviously both Dirac CP and Majorana CP are conserved for $\varphi_2=0$. Furthermore, the solar mixing angle $\theta_{12}$ and reactor angle $\theta_{13}$ are related by
\begin{equation}
3\cos^2\theta_{12}\cos^2\theta_{13}=2\sin^2\varphi_{1}\,,
\end{equation}
which is independent of the free parameter $\theta$. Taking into account measured values of $\theta_{12}$ and $\theta_{13}$~\cite{Capozzi:2013csa}, we obtain the constraint on $\varphi_1$ as
\begin{equation}
0.417\pi\leq\varphi_1\leq0.583\pi,\qquad \text{or}\qquad
-0.583\pi\leq\varphi_1\leq-0.417\pi\,,
\end{equation}
which indicates that $\varphi_1$ is around $\pm\pi/2$. This mixing pattern is very interesting, and it can accommodate the present neutrino oscillation data very well. After the measured $3\sigma$ ranges of the three lepton mixing angles are imposed, the allowed values of the lepton mixing parameters for $n=2, 3,\ldots, 100$ are displayed in Fig.~\ref{fig:caseIII_1st_angles_ranges} and Fig.~\ref{fig:caseIII_1st_phases_ranges}. In the case that $n$ is divisible by 3, the doublet representations $\mathbf{2_2}$, $\mathbf{2_3}$ and $\mathbf{2_4}$ are assumed to be absent such that generalized CP symmetry in Eq.~\eqref{eq:GCP_all} is consistently defined. Notice that here $n$ should start from $n=2$ since $\Delta(6)\equiv S_3$ does not have three dimensional irreducible representations. If $n$ is divisible by 3, the three permutations $U^{I,1st}_{PMNS}$, $U^{I,2nd}_{PMNS}$ and $U^{I,3rd}_{PMNS}$ give rise to the same predictions for the mixing parameters. The observed values of the three lepton mixing angles can not be achieved simultaneously for $n=3$. In case of $n=2$ and $n=4$, both the atmospheric mixing angle $\theta_{23}$ and the Dirac CP phase $\delta_{CP}$ are maximal while the Majorana phases are zero. It is remarkable that the three CP phases can take any values for sufficiently large $n$, while $\theta_{12}$ is always constrained to be in the range of $0.313\leq\sin^2\theta_{12}\leq0.344$. Hence this mixing pattern can be tested by precisely measuring the solar mixing angle $\theta_{12}$. Notice that $\theta_{12}$ can be measured with rather good accuracy by JUNO experiment~\cite{JUNO}. In the end, the correlations between mixing parameters for $n\rightarrow\infty$ and $n=8$ are displayed in Fig.~\ref{fig:caseIII_1st} where we only show the phenomenologically viable cases for which the measured values of $\theta_{12}$, $\theta_{13}$ and $\theta_{23}$ can be accommodated for certain values of the parameter $\theta$.

\begin{figure}[t!]
\begin{center}
\includegraphics[width=0.99\textwidth]{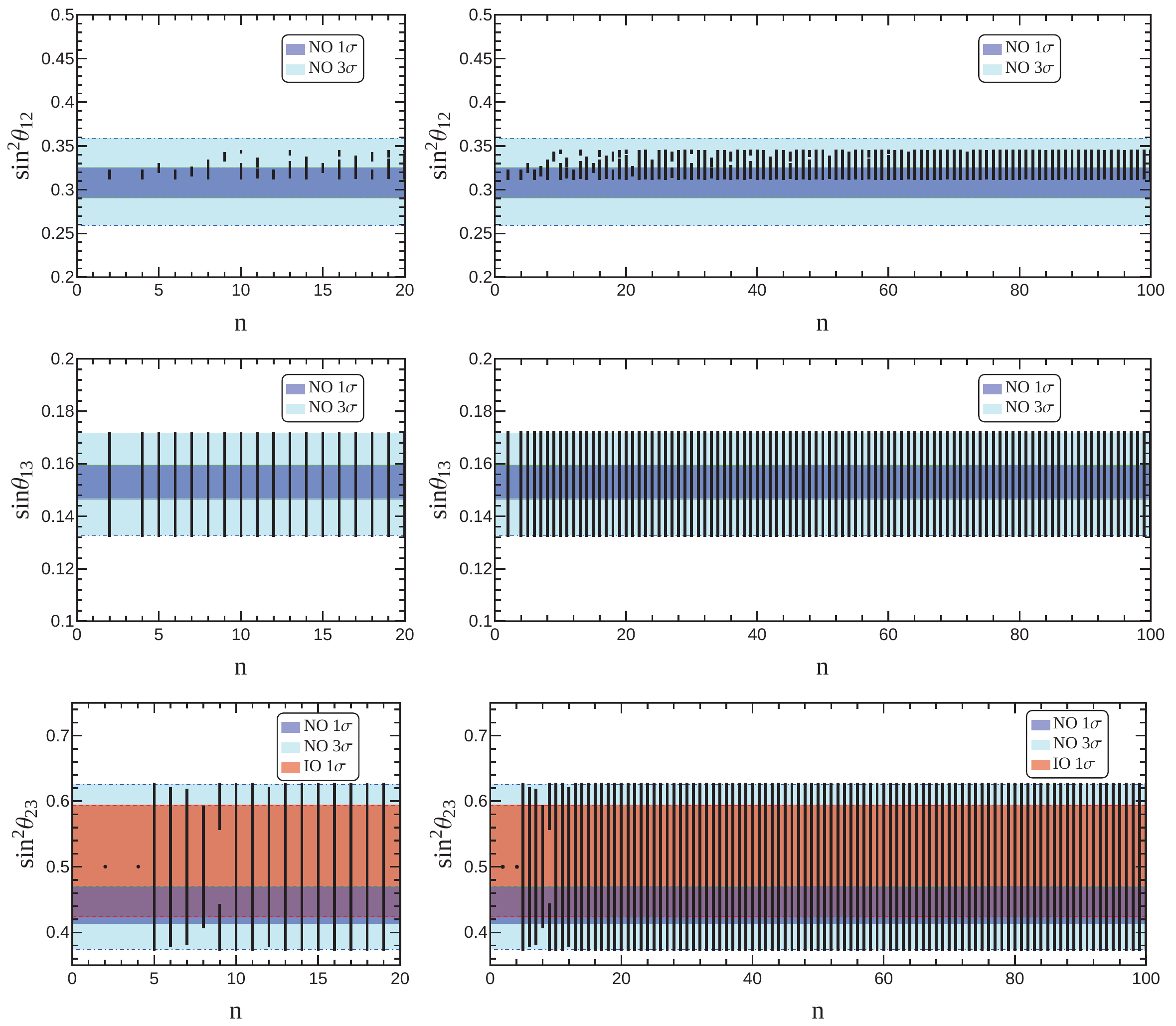}
\caption{\label{fig:caseIII_1st_angles_ranges}Numerical results in case I for the 1st-3rd ordering with the PMNS matrices given in Eq.~\eqref{eq:arrangement_castI_13}: the allowed values of $\sin^2\theta_{12}$, $\sin\theta_{13}$ and $\sin^2\theta_{23}$ for different $n$, where the three lepton mixing angles are required to lie in their $3\sigma$ ranges. The $1\sigma$ and $3\sigma$ bounds of the mixing parameters are taken from Ref.~\cite{Capozzi:2013csa}.}
\end{center}
\end{figure}

\begin{figure}[t!]
\begin{center}
\includegraphics[width=0.99\textwidth]{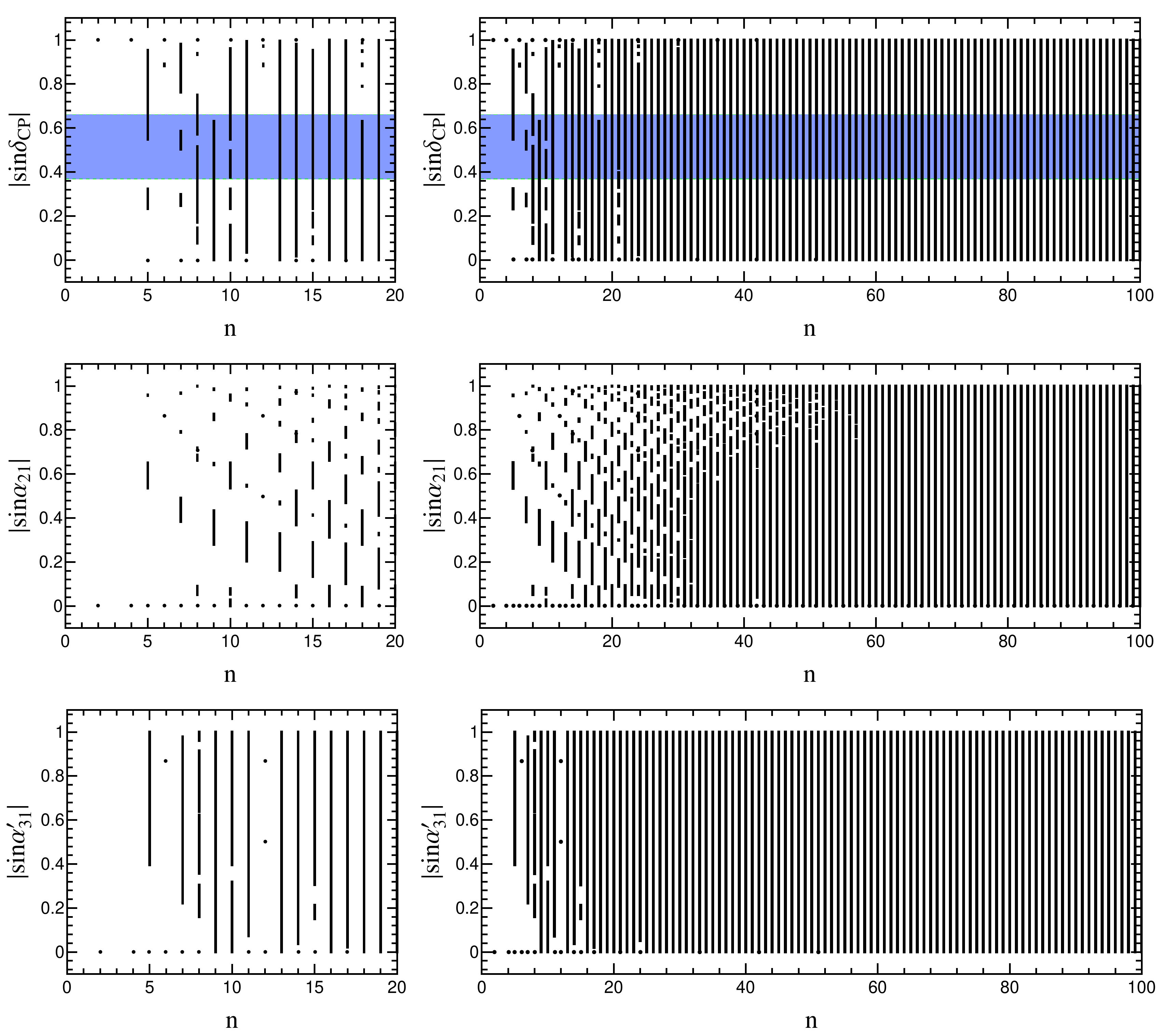}
\caption{\label{fig:caseIII_1st_phases_ranges} Numerical results in case I for the 1st-3rd ordering with the PMNS matrices given in
Eq.~\eqref{eq:arrangement_castI_13}: the possible values of $\left|\sin\delta_{CP}\right|$, $\left|\sin\alpha_{21}\right|$ and $\left|\sin\alpha^{\prime}_{31}\right|$ for different $n$, where the three lepton mixing angles are required to lie in the $3\sigma$ ranges. The $1\sigma$ and $3\sigma$ bounds of the mixing parameters are taken from Ref.~\cite{Capozzi:2013csa}. }
\end{center}
\end{figure}

\begin{figure}[hptb!]
\begin{center}
\includegraphics[width=0.90\textwidth]{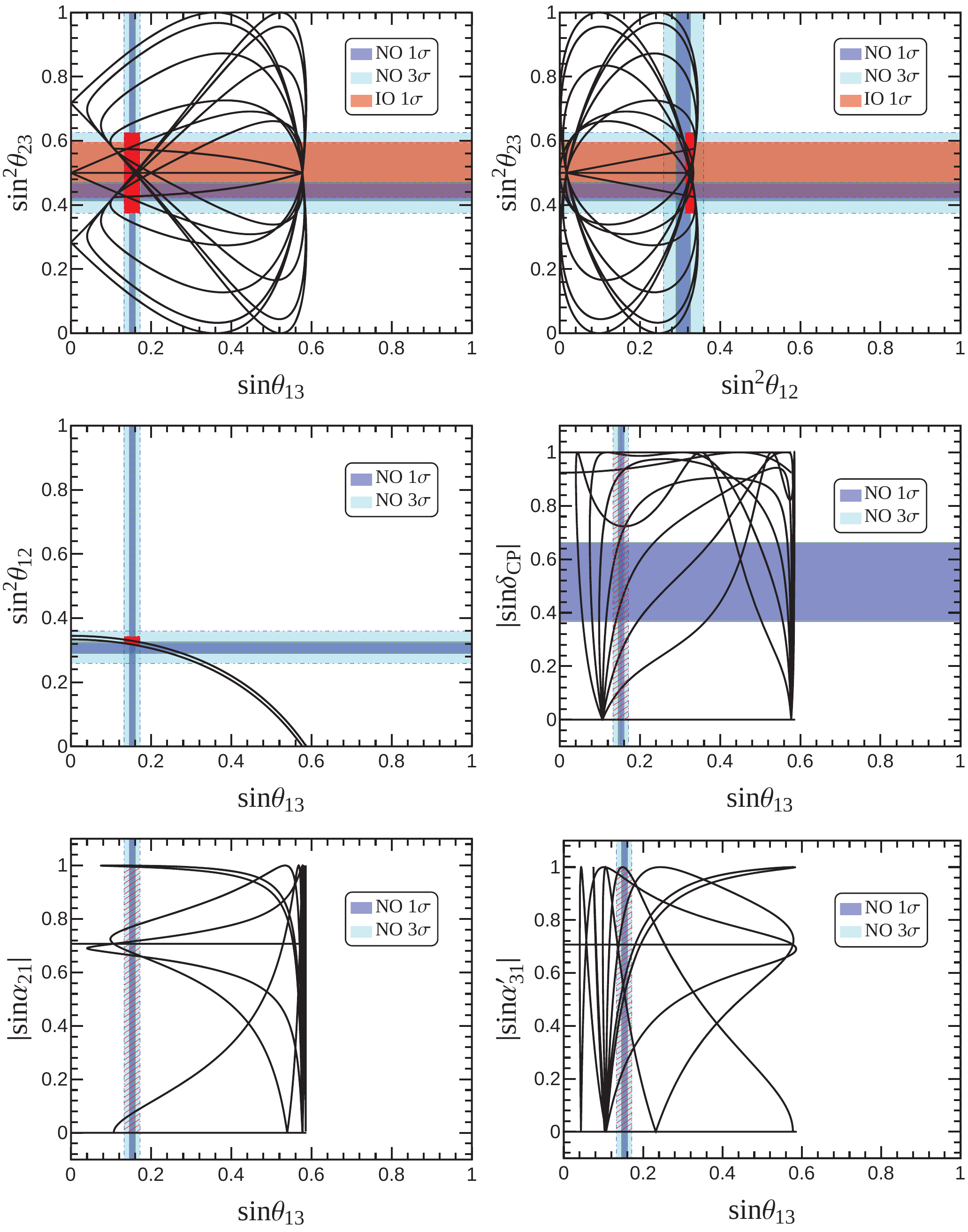}
\caption{\label{fig:caseIII_1st} The correlations among mixing parameters in case I for the 1st-3rd ordering with the PMNS matrices given in Eq.~\eqref{eq:arrangement_castI_13}. The red filled regions denote the allowed values of the mixing parameters if we take the parameters $\varphi_1$ and $\varphi_2$ to be continuous (which is equivalent to taking the limit $n\rightarrow\infty$) and the three mixing angles are required to lie in their $3\sigma$ regions. Note that the three CP phases $\delta_{CP}$, $\alpha_{21}$ and $\alpha^{\prime}_{31}$ are not constrained in this limit. The black curves represent the phenomenologically viable correlations for $n=8$. The $1\sigma$ and $3\sigma$ bounds of the mixing parameters are taken from Ref.~\cite{Capozzi:2013csa}.}
\end{center}
\end{figure}

The vector $\sqrt{2/3}\left(\sin\varphi_1, \cos\left(\pi/6-\varphi_1\right), \cos\left(\pi/6+\varphi_1\right)\right)^{T}$ enforced by the remnant $Z_2$ symmetry can also be the second column of the PMNS matrix. Disregarding the exchange of the second and the third rows, three arrangements are possible as well,
\begin{equation}
\label{eq:arrangement_caseI_46}U^{I,4th}_{PMNS}=U^{I}_{PMNS}P_{213},~~ U^{I,5th}_{PMNS}=P_{231}U^{I}_{PMNS}P_{213},~~ U^{I,6th}_{PMNS}=P_{312}U^{I}_{PMNS}P_{213}\,.
\end{equation}
Analogous to that in Eq.~\eqref{eq:PMNS_relation_castI_13}, these three forms of the PMNS matrix are related by parameter redefinition as follows
\begin{eqnarray}
\nonumber&&U^{I,5th}_{PMNS}(\theta,\varphi_1,\varphi_2)=\textrm{diag}(1,1,-1)U^{I,4th}_{PMNS}(\pi-\theta,\frac{\pi}{3}+\varphi_1,\varphi_2)\textrm{diag}(1,1,-1),\\
&&U^{6th}_{PMNS}(\theta,\varphi_1,\varphi_2)=\textrm{diag}(-1,1,1)U^{I,4th}_{PMNS}(-\theta,-\frac{\pi}{3}+\varphi_1,\varphi_2)\textrm{diag}(-1,1,1)\,.
\end{eqnarray}
For the lepton flavor mixing matrix $U^{4th}_{PMNS}$, one can straightforwardly extract the flavor mixing parameters:
\begin{eqnarray}
\nonumber&&\sin^2\theta_{13}=\frac{1}{3}\left(1+\cos^2\theta\cos2\varphi_1+\sqrt{2}\sin2\theta\cos\varphi_2\cos\varphi_1\right),\\
\nonumber&&\sin^2\theta_{12}=\frac{2\sin^2\varphi_1}{2-\cos^2\theta\cos2\varphi_1-\sqrt{2}\sin2\theta\cos\varphi_2\cos\varphi_1}\,,\\
\nonumber&&\sin^2\theta_{23}=\frac{1-\cos^2\theta\sin\left(\pi/6+2\varphi_1\right)-\sqrt{2}\sin2\theta\cos\varphi_2\sin\left(\pi/6-\varphi_1\right)}
{2-\cos^2\theta\cos2\varphi_1-\sqrt{2}\sin2\theta\cos\varphi_2\cos\varphi_1}\,,\\
\nonumber&&\left|J_{CP}\right|=\frac{1}{6\sqrt{6}}\left|\sin2\theta\sin\varphi_2\sin3\varphi_1\right|\,,\\
\nonumber&&\left|\tan\delta_{CP}\right|=\Big|4\sqrt{2}\sin2\theta\sin\varphi_2\sin3\varphi_1\csc\varphi_1\left(2-\cos2\varphi_1\cos^2\theta-\sqrt{2}\cos\varphi_2\cos\varphi_1\sin2\theta\right)\Big/\\
\nonumber&&\qquad\Big\{-16\cos3\varphi_1\cos^2\theta+8(1-3\cos2\theta)\cos\varphi_1\sin^2\theta+4\cos2\varphi_2(\cos3\varphi_1-2\cos\varphi_1)\sin^22\theta\\
\nonumber&&\qquad+\sqrt{2}\cos\varphi_2\left[8(\cos2\varphi_1+\cos4\varphi_1)\sin\theta
\cos^3\theta+2\sin2\theta+5\sin4\theta\right]\Big\}\Big|\,,\\
\nonumber&&\left|\tan\alpha_{21}\right|=\left|\frac{2\sin\varphi_2\left(\cos\varphi_2-\sqrt{2}\cos\varphi_1\tan\theta\right)}{\cos2\varphi_2-2\cos\varphi_1\tan\theta\left(\sqrt{2}\cos\varphi_2-\cos\varphi_1\tan\theta\right)}\right|\,,\\
\nonumber&&\left|\tan\alpha^{\prime}_{31}\right|=\Big|8\cos\varphi_1\left(\sqrt{2}\cos2\varphi_1\sin2\theta\sin\varphi_2-2\cos2\theta\cos\varphi_1\sin2\varphi_2\right)\Big/\Big\{4(3+\cos4\theta)\\
\label{eq:mixing_parameter_III_4th}&&\times\cos2\varphi_2\cos^2\varphi_1-4\sqrt{2}\cos\varphi_2\cos\varphi_1\cos2\varphi_1\sin4\theta-(3-\cos4\varphi_1+4\cos2\varphi_1)\sin^22\theta\Big\}\Big|\,.
\end{eqnarray}
In this case, we have the following relation
\begin{equation}
3\sin^2\theta_{12}\cos^2\theta_{13}=2\sin^2\varphi_1\,,
\end{equation}
which yields $0.614\leq|\sin\varphi_1|\leq0.727$ at $3\sigma$ confidence level, and therefore the parameter $\varphi_1$ is determined to be in the range
\begin{equation}
\varphi_{1}\in\pm\left(\left[0.210\pi,0.259\pi\right])\cup\left[0.741\pi,0.790\pi\right]\right)\,.
\end{equation}
For the representative values $\pm\pi/4$ and $\pm3\pi/4$ of $\varphi_{1}$, the relatively small $\theta_{13}$ leads to
\begin{equation}
\left(\varphi_1, \varphi_2, \theta\right)\simeq(\pm\frac{\pi}{4},0, \frac{3\pi}{4}),\quad (\pm\frac{\pi}{4},\pi,\frac{\pi}{4}),\quad (\pm\frac{3\pi}{4},0, \frac{\pi}{4}),\quad (\pm\frac{3\pi}{4},\pi,\frac{3\pi}{4})\,.
\end{equation}
Accordingly the atmospheric mixing angle $\theta_{23}$ would be
\begin{equation}
\sin^2\theta_{23}\simeq\frac{1}{4}(2-\sqrt{3})\simeq0.067, \qquad \text{or}\qquad  \sin^2\theta_{23}\simeq\frac{1}{4}(2+\sqrt{3})\simeq0.933\,,
\end{equation}
which is clearly not compatible with the global analysis of neutrino oscillation data~\cite{Capozzi:2013csa}. As a result, the three lepton mixing angles can not be accommodated simultaneously in this
case, and this mixing pattern is not viable. The detailed numerical results are presented in Fig.~\ref{fig:caseIII_4th}. We see that the correct values of the atmospheric mixing angle really can not be achieved for realistic $\theta_{12}$ and $\theta_{13}$.

\begin{figure}[t!]
\begin{center}
\includegraphics[width=0.90\textwidth]{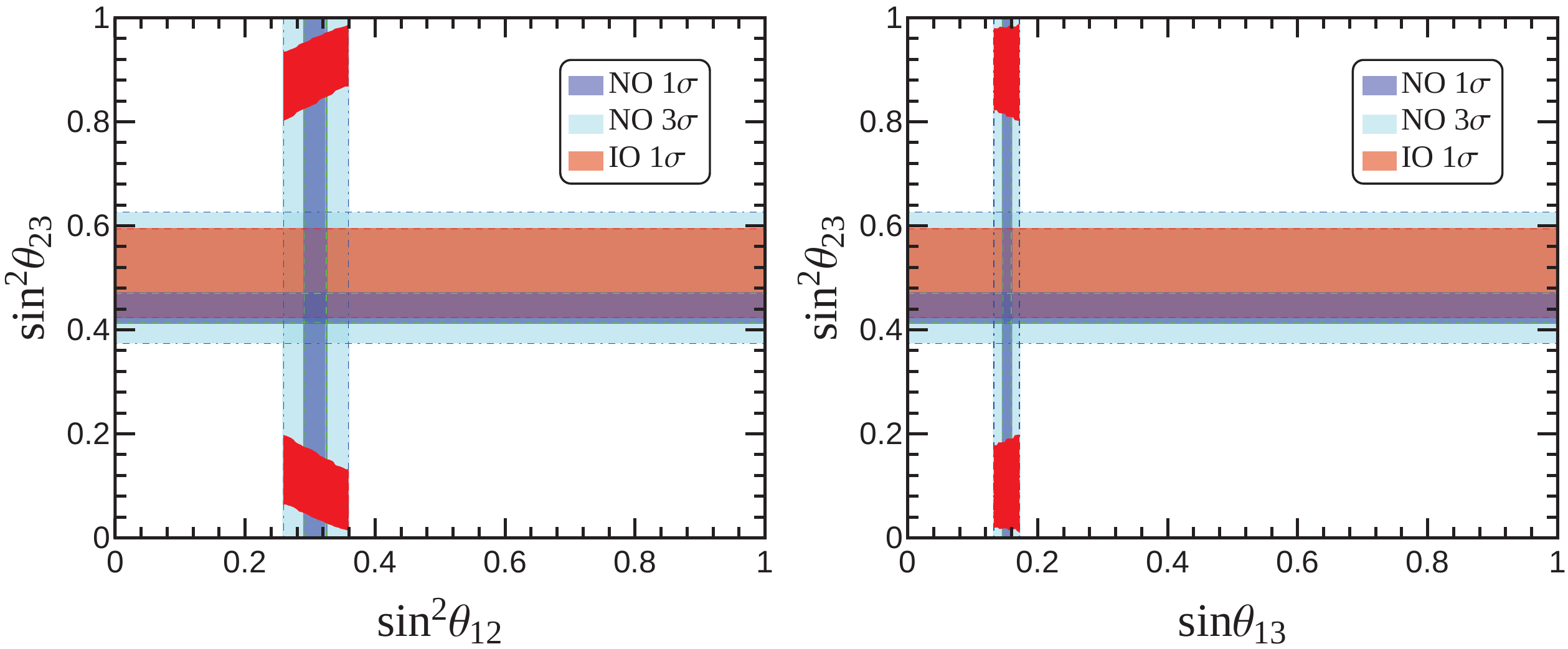}
\caption{\label{fig:caseIII_4th}Numerical results in case I for the 4th-6th ordering with the PMNS matrices given in Eq.~\eqref{eq:arrangement_caseI_46}. The red filled regions denote the allowed values of the mixing parameters if we take the parameters $\varphi_1$ and $\varphi_2$ to be continuous (which is equivalent to taking the limit $n\rightarrow\infty $), where $\theta_{12}$ and $\theta_{13}$ are required to lie in their $3\sigma$ ranges. Obviously the resulting predictions for $\theta_{23}$ are far beyond its $3\sigma$ range. The $1\sigma$ and $3\sigma$ bounds of the mixing parameters are taken from Ref.~\cite{Capozzi:2013csa}.}
\end{center}
\end{figure}

Finally the predicted vector $\sqrt{2/3}\left(\sin\varphi_1, \cos\left(\pi/6-\varphi_1\right),
\cos\left(\pi/6+\varphi_1\right)\right)^{T}$ can be placed in the third column. Using the freedom of exchanging the rows of the PMNS matrix, three configurations are found as well,
\begin{equation}
\label{eq:arrangement_caseI_79}U^{I,7th}_{PMNS}=U^{I}_{PMNS}P_{321},~~ U^{I,8th}_{PMNS}=P_{231}U^{I}_{PMNS}P_{321},~~ U^{I,9th}_{PMNS}=P_{312}U^{I}_{PMNS}P_{321}\,,
\end{equation}
which are related by
\begin{eqnarray}
\nonumber&&U^{I,8th}_{PMNS}(\theta,\varphi_1,\varphi_2)=\textrm{diag}(1,1,-1)U^{I,7th}_{PMNS}(\pi-\theta,\frac{\pi}{3}+\varphi_1,\varphi_2)\textrm{diag}(-1,1,1),\\
&&U^{I,9th}_{PMNS}(\theta,\varphi_1,\varphi_2)=\textrm{diag}(-1,1,1)U^{I,7th}_{PMNS}(-\theta,-\frac{\pi}{3}+\varphi_1,\varphi_2)\textrm{diag}(1,-1,1)\,.
\end{eqnarray}
The lepton mixing parameters for $U^{I,7th}_{PMNS}$ are determined to be
\begin{eqnarray}
\nonumber&&\sin^2\theta_{13}=\frac{2}{3}\sin^2\varphi_1,\qquad\quad
\sin^2\theta_{12}=\frac{1+\sin^2\theta\cos2\varphi_1-\sqrt{2}\sin2\theta\cos\varphi_2\cos\varphi_1}{2+\cos2\varphi_1}\,,\\
\nonumber&&\sin^2\theta_{23}=\frac{1+\sin\left(\pi/6+2\varphi_1\right)}{2+\cos2\varphi_1},\qquad \left|J_{CP}\right|=\frac{1}{6\sqrt{6}}\left|\sin2\theta\sin\varphi_2\sin3\varphi_1\right|\,,\\
\nonumber&&\left|\tan\delta_{CP}\right|=\left|\frac{\sin\varphi_2(2+\cos2\varphi_1)}{\cos\varphi_2\cos2\varphi_1-2\sqrt{2}\cot2\theta\cos\varphi_1}\right|\,,\\
\nonumber&&\left|\tan\alpha_{21}\right|=\Big|4\sqrt{2}\cos\varphi_1\left(\cos2\varphi_1\sin2\theta\sin\varphi_2-\sqrt{2}\cos2\theta\cos\varphi_1\sin2\varphi_2\right)\Big/
\Big\{2(\cos4\theta+3)\\
\nonumber&&\qquad\times\cos2\varphi_2\cos^2\varphi_1-2\sqrt{2}\cos\varphi_2\cos\varphi_1\cos2\varphi_1\sin4\theta+\left(\cos^22\varphi_1-4\cos^2\varphi_1\right)\sin^22\theta\Big\}\Big|\,,\\
\label{eq:mixing_parameter_III_7th}&&\left|\tan\alpha^{\prime}_{31}\right|=\left|\frac{2\sin\varphi_2\sin\theta\left(\sqrt{2}\cos\theta\cos\varphi_1+\cos\varphi_2\sin\theta\right)}
{2\cos^2\theta\cos^2\varphi_1+\sqrt{2}\cos\varphi_2\cos\varphi_1\sin2\theta+\cos2\varphi_2\sin^2\theta}\right|\,.
\end{eqnarray}
The lepton mixing parameters for $U^{I,8th}_{PMNS}$ and $U^{I,9th}_{PMNS}$ can be obtained from Eq.~\eqref{eq:mixing_parameter_III_7th} by the replacement $\theta\rightarrow\pi-\theta$, $\varphi_1\rightarrow\frac{\pi}{3}+\varphi_1$ and $\theta\rightarrow-\theta$, $\varphi_1\rightarrow-\frac{\pi}{3}+\varphi_1$ respectively. We see that both $\theta_{13}$ and $\theta_{23}$ are only determined by the discrete group parameter $\varphi_1$, and they are related by
\begin{equation}
\sin^2\theta_{23}=\frac{1}{2}\pm\frac{1}{2}\tan\theta_{13}\sqrt{2-\tan^2\theta_{13}}\,,
\end{equation}
which yields
\begin{equation}
\theta_{23}\simeq\frac{\pi}{4}\pm\frac{\theta_{13}}{\sqrt{2}}\,.
\end{equation}
For the $3\sigma$ interval $1.76\times10^{-2}\leq\sin^2\theta_{13}\leq2.95\times10^{-2}$~\cite{Capozzi:2013csa}, we have
\begin{equation}
0.378\leq\sin^2\theta_{23}\leq0.406,\qquad \text{or}\qquad 0.594\leq\sin^2\theta_{23}\leq0.622\,.
\end{equation}
This mixing pattern can be directly tested by future atmospheric neutrino oscillation experiments or long baseline neutrino oscillation experiments. If $\theta_{23}$ is found to be nearly maximal, this mixing would be ruled out. Furthermore, the precisely measured $\theta_{13}$ leads to $0.162\leq\left|\sin\varphi_{1}\right|\leq0.210$, and therefore $\varphi_1$ has to be in the following range
\begin{equation}
\varphi_1\in\pm\left(\left[0.0519\pi,0.0675\pi\right]\cup\left[0.933\pi,0.948\pi\right]\right)\,,
\end{equation}
which implies that $\varphi_1$ should be rather close to $0$ or $\pi$. To reproduce the observed value of the reactor mixing angle, the two smallest values for $n$ are 5 and 10, i.e. at least $\Delta(150)$ or $\Delta(600)$ is needed to produce viable mixing in this case. The admissible values of $\sin^2\theta_{23}$ and $\sin\theta_{13}$ for $n=5$, 10, 20 and 30 are  plotted in Fig.~\ref{fig:caseIII_7th_v2}. Furthermore, the variation of the allowed values of the lepton mixing parameters with respect to $n$ are shown in Fig.~\ref{fig:caseIII_7th_angles_ranges} and Fig.~\ref{fig:caseIII_7th_phases_ranges}. Compared with previous cases, both $\theta_{23}$ and $\theta_{13}$ are predicted to take several discrete values until $n=100$ in this case. It is interesting that the Majorana phase $\alpha^{\prime}_{31}$ is constrained to be in the range of $0\leq\left|\sin\alpha^{\prime}_{31}\right|\leq0.91$ while both $\delta_{CP}$ and $\alpha_{21}$ can take any values between $0$ and $2\pi$ for large $n$.

\begin{figure}[t!]
\begin{center}
\includegraphics[width=0.90\textwidth]{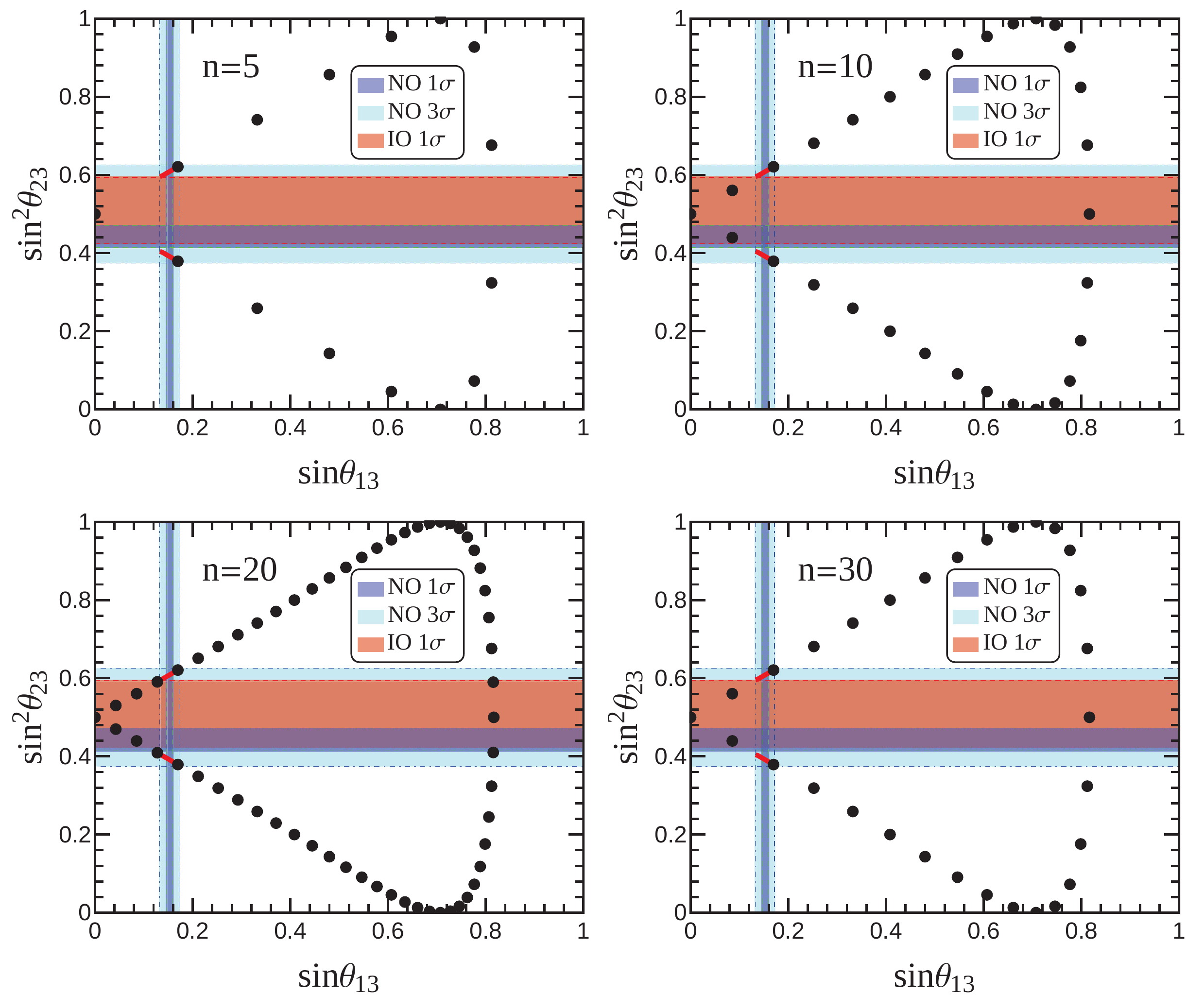}
\caption{\label{fig:caseIII_7th_v2} The possible values of $\sin^2\theta_{23}$ and $\sin\theta_{13}$ for the 7th-9th ordering with the PMNS matrices shown in Eq.~\eqref{eq:arrangement_caseI_79} in case I. The $1\sigma$ and $3\sigma$ bounds of the mixing angles are taken from Ref.~\cite{Capozzi:2013csa}. }
\end{center}
\end{figure}

\begin{figure}[hptb!]
\begin{center}
\includegraphics[width=0.99\textwidth]{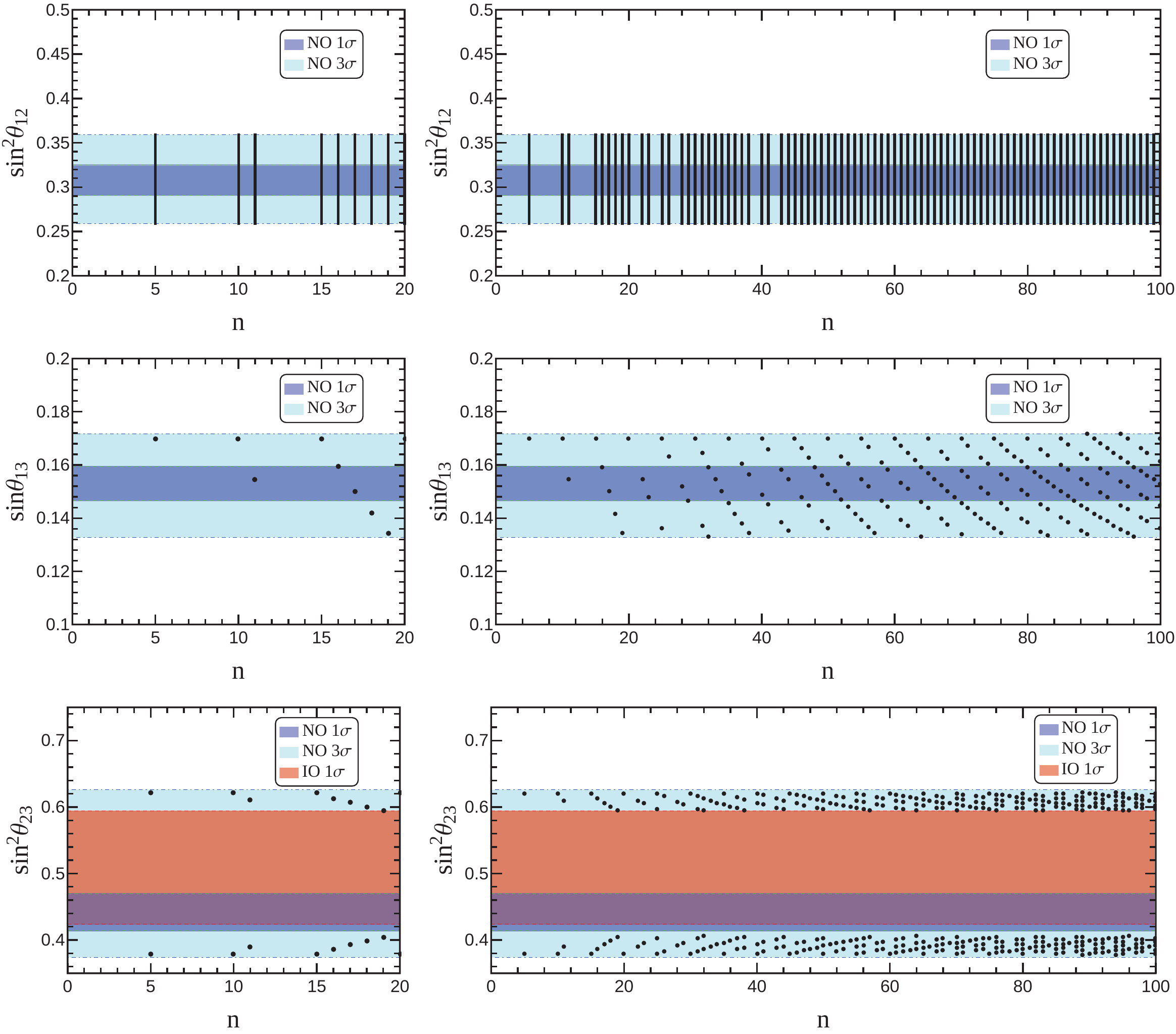}
\caption{\label{fig:caseIII_7th_angles_ranges}
Numerical results in case I for the 7th-9th ordering with the PMNS matrices given in Eq.~\eqref{eq:arrangement_caseI_79}: the allowed values of $\sin^2\theta_{12}$, $\sin\theta_{13}$ and $\sin^2\theta_{23}$ for different $n$, where the three lepton mixing angles are required to lie in their $3\sigma$ ranges. The $1\sigma$ and $3\sigma$ bounds of the mixing parameters are taken from Ref.~\cite{Capozzi:2013csa}.}
\end{center}
\end{figure}

\begin{figure}[hptb!]
\begin{center}
\includegraphics[width=0.99\textwidth]{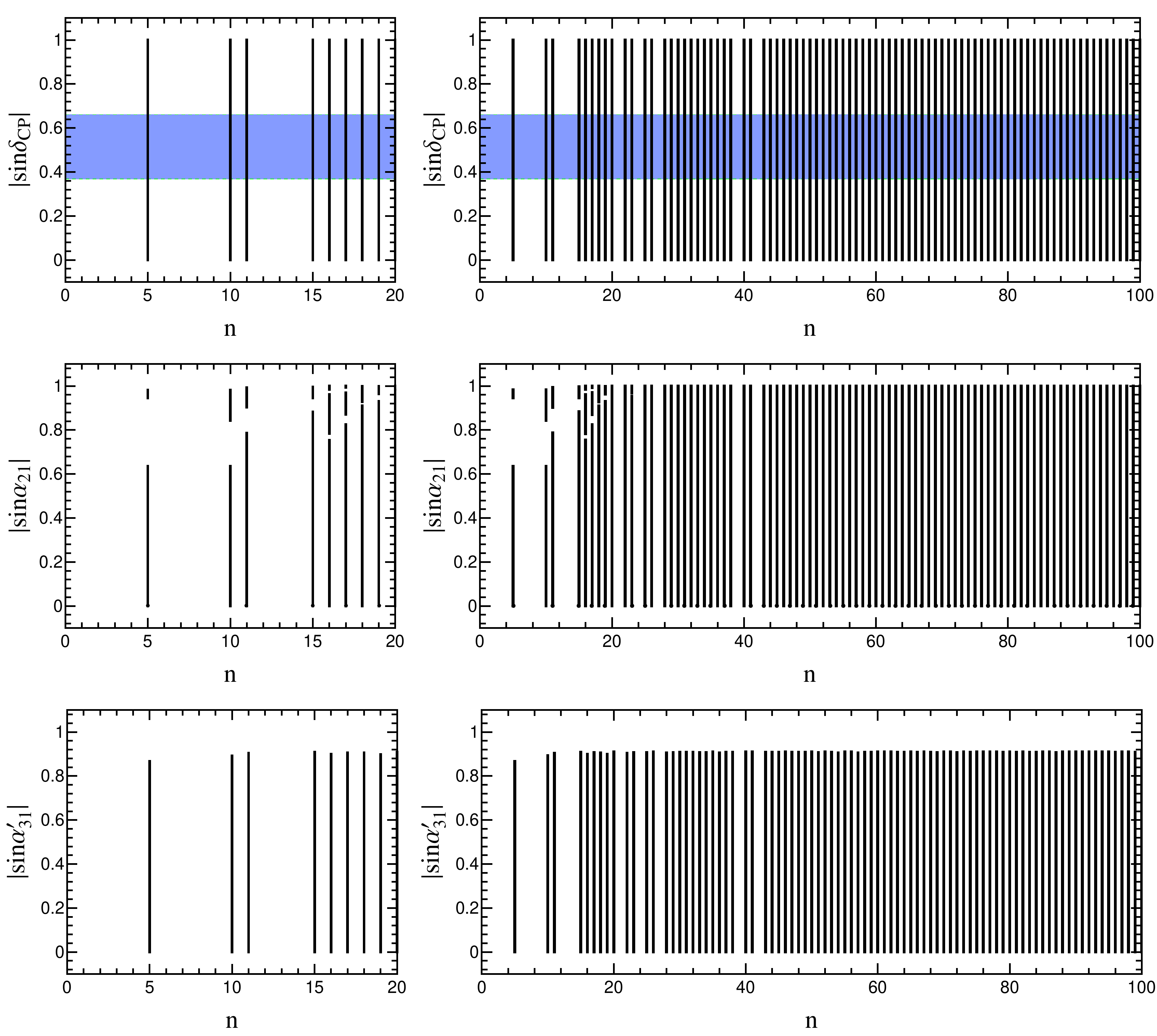}
\caption{\label{fig:caseIII_7th_phases_ranges}
Numerical results in case I for the 7th-9th ordering with the PMNS matrices given in Eq.~\eqref{eq:arrangement_caseI_79}: the allowed values of $\left|\sin\delta_{CP}\right|$, $\left|\sin\alpha_{21}\right|$ and $\left|\sin\alpha^{\prime}_{31}\right|$ for different $n$, where the three lepton mixing angles are required to lie in their $3\sigma$ ranges. The $1\sigma$ and $3\sigma$ bounds of the mixing angles are taken from Ref.~\cite{Capozzi:2013csa}.}
\end{center}
\end{figure}

\item[~~(\uppercase\expandafter{\romannumeral2})]

$G_{l}=\left\langle abc^{s}d^{t}\right\rangle$,
$G_{\nu}=Z^{bc^xd^x}_{2}$,
$X_{\nu\mathbf{r}}=\left\{\rho_{\mathbf{r}}(c^{\gamma}d^{-2x-\gamma}),
\rho_{\mathbf{r}}(bc^{x+\gamma}d^{-x-\gamma})\right\}$

In this case, the PMNS matrix is determined to be
\begin{equation}
\label{eq:PMNS_IV}U^{II}_{PMNS}=\frac{1}{2}\left(
\begin{array}{ccc}
 -\sin\theta-\sqrt{2}e^{i\varphi_3}\cos\theta ~&~ 1 ~&~
 \cos\theta-\sqrt{2} e^{i \varphi_3}
   \sin \theta \\
-\sin\theta+\sqrt{2} e^{i \varphi_3}\cos\theta ~&~ 1  ~&~
\cos\theta+\sqrt{2}e^{i \varphi_3}\sin\theta \\
 -\sqrt{2} \sin \theta  & -\sqrt{2} &
   \sqrt{2} \cos \theta
\end{array}
\right)\,,
\end{equation}
or the one obtained by exchanging the second and the third rows, where the parameter $\varphi_3$ is
\begin{equation}
\varphi_3=-\frac{3 \gamma +2s-t+2x}{n}\pi\,.
\end{equation}
It can take $2n$ discrete values:
\begin{equation}
\varphi_3~~\textrm{mod}~~2\pi=0, \frac{1}{n}\pi,\frac{2}{n}\pi,\ldots, \frac{2n-1}{n}\pi\,.
\end{equation}
The eigenvalues of $abc^{s}d^{t}$ would be degenerate for $t=0$ such that the unitary transformation $U_{l}$ can be be pinned down uniquely. If that is the case, we could choose the residual symmetry to be $G_{l}=K^{(c^{n/2},abc^{s})}_4$ which leads to same PMNS matrix shown in Eq.~\eqref{eq:PMNS_IV} with $t=0$. Obviously this mixing pattern has one column $\left(1/2, 1/2, -1/\sqrt{2}\right)^{T}$ which is the same as the first (second) column of the Bimaximal mixing up to permutations. In order to in accordance with the experimental data, the fixed vector
$\left(1/2, 1/2, -1/\sqrt{2}\right)^{T}$ can only be the second column of
the PMNS matrix. We can straightforwardly read out the predictions for the lepton mixing angles:
\begin{eqnarray}
\nonumber&&\sin^2\theta_{13}=\frac{1}{8}
\left(3-\cos2\theta-2\sqrt{2}\sin2\theta\cos\varphi_3\right),\quad \sin^2\theta_{12}=\frac{2}{5+\cos2\theta+2\sqrt{2}\sin2\theta\cos\varphi_3},\\
\nonumber&&\sin^2\theta_{23}=\frac{3-\cos2\theta+2\sqrt{2}\sin2\theta\cos\varphi_3}{5+\cos2\theta+2\sqrt{2}\sin2\theta\cos\varphi_3},\qquad \left|J_{CP}\right|=\frac{1}{8\sqrt{2}}\left|\sin2\theta\sin\varphi_3\right|,\\
\nonumber&&\left|\tan\delta_{CP}\right|=\left|\frac{8\cos\theta\sin^2\theta\sin2\varphi_3+\sqrt{2}(9\sin\theta+\sin3\theta)\sin\varphi_3}
{4\cos3\theta+\cos\theta\left(4-8\sin^2\theta\cos2\varphi_3\right)+\sqrt{2}(3\sin3\theta-5\sin\theta)\cos\varphi_3}\right|,\\
\nonumber&&\left|\tan\alpha_{21}\right|=\left|
\frac{2\cos^2\theta\sin2\varphi_3+\sqrt{2}\sin2\theta\sin\varphi_3}{\sin^2\theta+2\cos^2\theta\cos2\varphi_3+\sqrt{2}\sin2\theta\cos\varphi_3}\right|,\\
\label{eq:mixing_angles_caseIV}&&\left|\tan\alpha^{\prime}_{31}\right|=\left|
\frac{16\cos2\theta\sin2\varphi_3-8\sqrt{2}\sin2\theta\sin\varphi_3}
{6\sin^22\theta+4\sqrt{2}\sin4\theta\cos\varphi_3-4(3+\cos4\theta)\cos2\varphi_3}\right|\,.
\end{eqnarray}
The following correlation is satisfied:
\begin{equation}
4\sin^2\theta_{12}\cos^2\theta_{13}=1\,,
\end{equation}
which leads to $0.254\leq\sin^2\theta_{12}\leq0.258$ for the measured value of the reactor mixing angle~\cite{Capozzi:2013csa}. Therefore $\sin^2\theta_{12}$ is predicted to be very close to its $3\sigma$ lower bound 0.259~\cite{Capozzi:2013csa} in this case. As a consequence, we suggest that this mixing pattern is a good leading order approximation to the present neutrino mixing data. The reason is that the subleading contributions could easily pull $\theta_{12}$ into a experimentally more favored range. Furthermore, the expression for $\sin^2\theta_{13}$ in Eq.~\eqref{eq:mixing_angles_caseIV} yields
\begin{equation}
\frac{1}{8}\left(3-\sqrt{1+8\cos^2\varphi_3}\right)\leq\sin^2\theta_{13}\leq\frac{1}{8}\left(3+\sqrt{1+8\cos^2\varphi_3}\right)\,.
\end{equation}
In order to be in accordance with experimental data, the parameter $\varphi_3$ has to be in the range
\begin{equation}
\varphi_3\in\left[0,0.135\pi\right]\cup\left[0.865\pi, 1.135\pi\right]\cup\left[1.865\pi,2\pi\right]\,.
\end{equation}
The allowed values of the mixing parameters with respect to $n$ are shown in Fig.~\ref{fig:caseIV_angles_ranges} and Fig.~\ref{fig:caseIV_phases_ranges}, and the correlations between them are plotted in Fig.~\ref{fig:caseIV}, where the $3\sigma$ lower bound of $\sin^2\theta_{12}$ is chosen to be 0.254 instead of 0.259 given in Ref.~\cite{Capozzi:2013csa}. The values of $\varphi_3=0, \pi$ are always acceptable, and the corresponding Dirac and Majorana CP phases are conserved. Note that only the CP conserved cases are viable for $n=2, 3, \ldots,7$. Moreover the CP violating phases $\delta_{CP}$ and $\alpha_{21}$ are predicted to fulfill $\left|\sin\delta_{CP}\right|\leq0.895$ and $\left|\sin\alpha_{21}\right|\leq0.545$ while $\alpha^{\prime}_{31}$ is not constrained at all for large $n$.

\begin{figure}[hptb!]
\begin{center}
\includegraphics[width=0.99\textwidth]{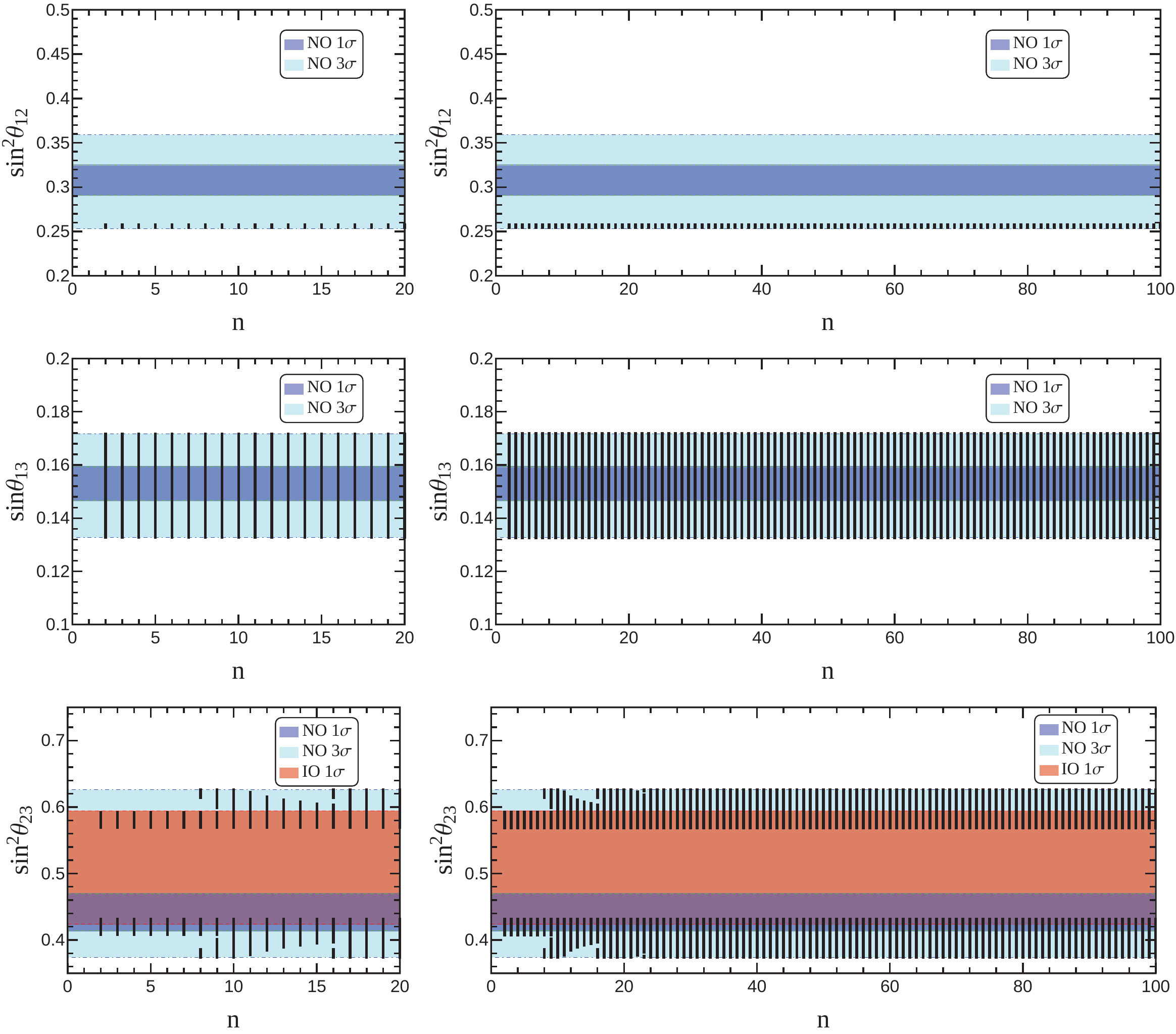}
\caption{\label{fig:caseIV_angles_ranges}
Numerical results in case II: the allowed values of $\sin^2\theta_{12}$, $\sin\theta_{13}$ and $\sin^2\theta_{23}$ for different $n$, where the three lepton mixing angles are required to lie in their $3\sigma$ ranges (the $3\sigma$ lower bound of $\sin^2\theta_{12}$ is chosen to be 0.254 instead of 0.259 given in Ref.~\cite{Capozzi:2013csa}). The $1\sigma$ and $3\sigma$ bounds of the mixing parameters are taken from Ref.~\cite{Capozzi:2013csa}.}
\end{center}
\end{figure}

\begin{figure}[hptb!]
\begin{center}
\includegraphics[width=0.99\textwidth]{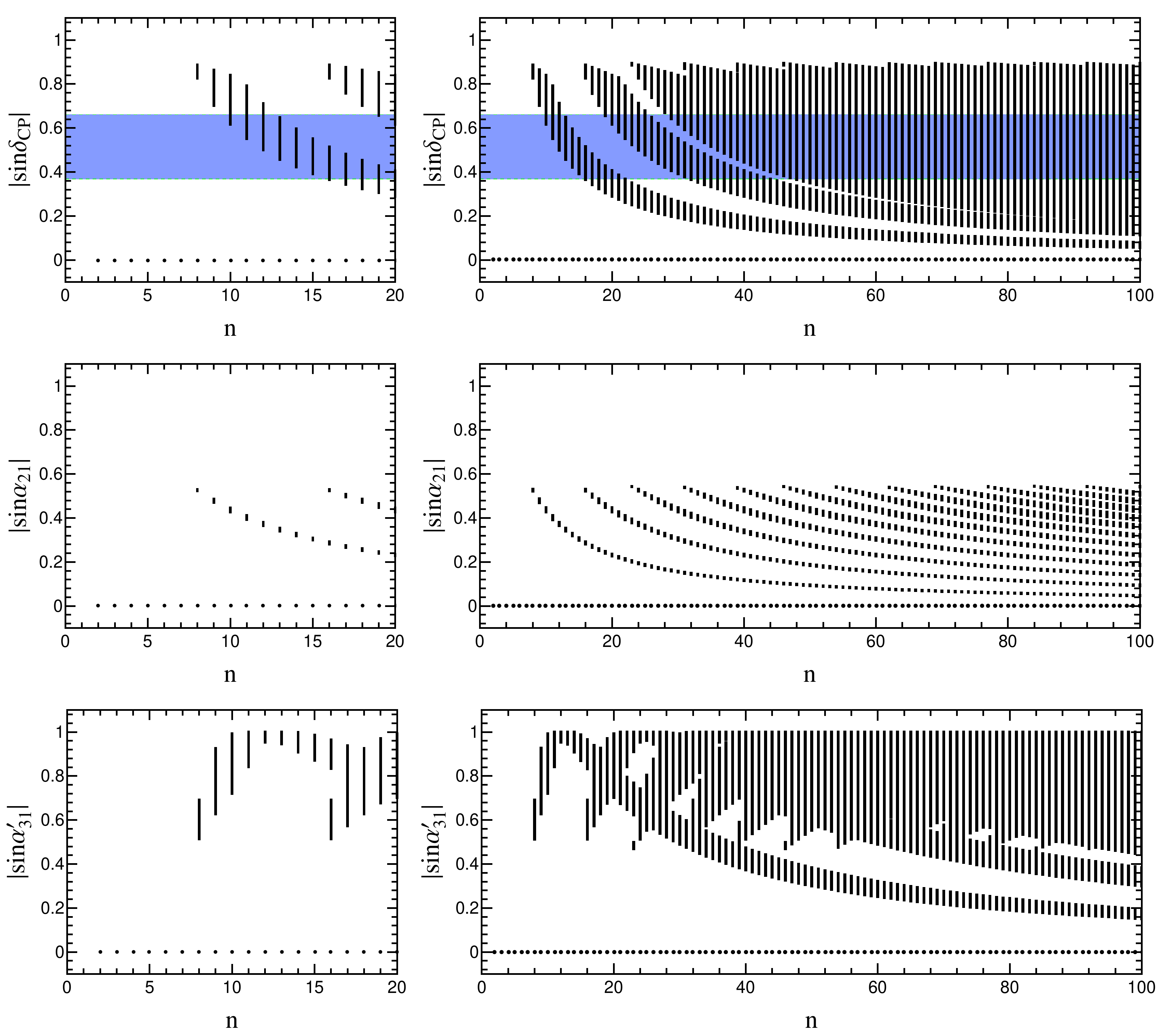}
\caption{\label{fig:caseIV_phases_ranges}
Numerical results in case II: the allowed values of $\left|\sin\delta_{CP}\right|$, $\left|\sin\alpha_{21}\right|$ and $\left|\sin\alpha^{\prime}_{31}\right|$ for different $n$, where the three lepton mixing angles are required to lie in their $3\sigma$ ranges (the $3\sigma$ lower bound of $\sin^2\theta_{12}$ is chosen to be 0.254 instead of 0.259 given in Ref.~\cite{Capozzi:2013csa}). The $1\sigma$ and $3\sigma$ bounds of the mixing parameters are taken from Ref.~\cite{Capozzi:2013csa}. }
\end{center}
\end{figure}

\begin{figure}[hptb!]
\begin{center}
\includegraphics[width=0.90\textwidth]{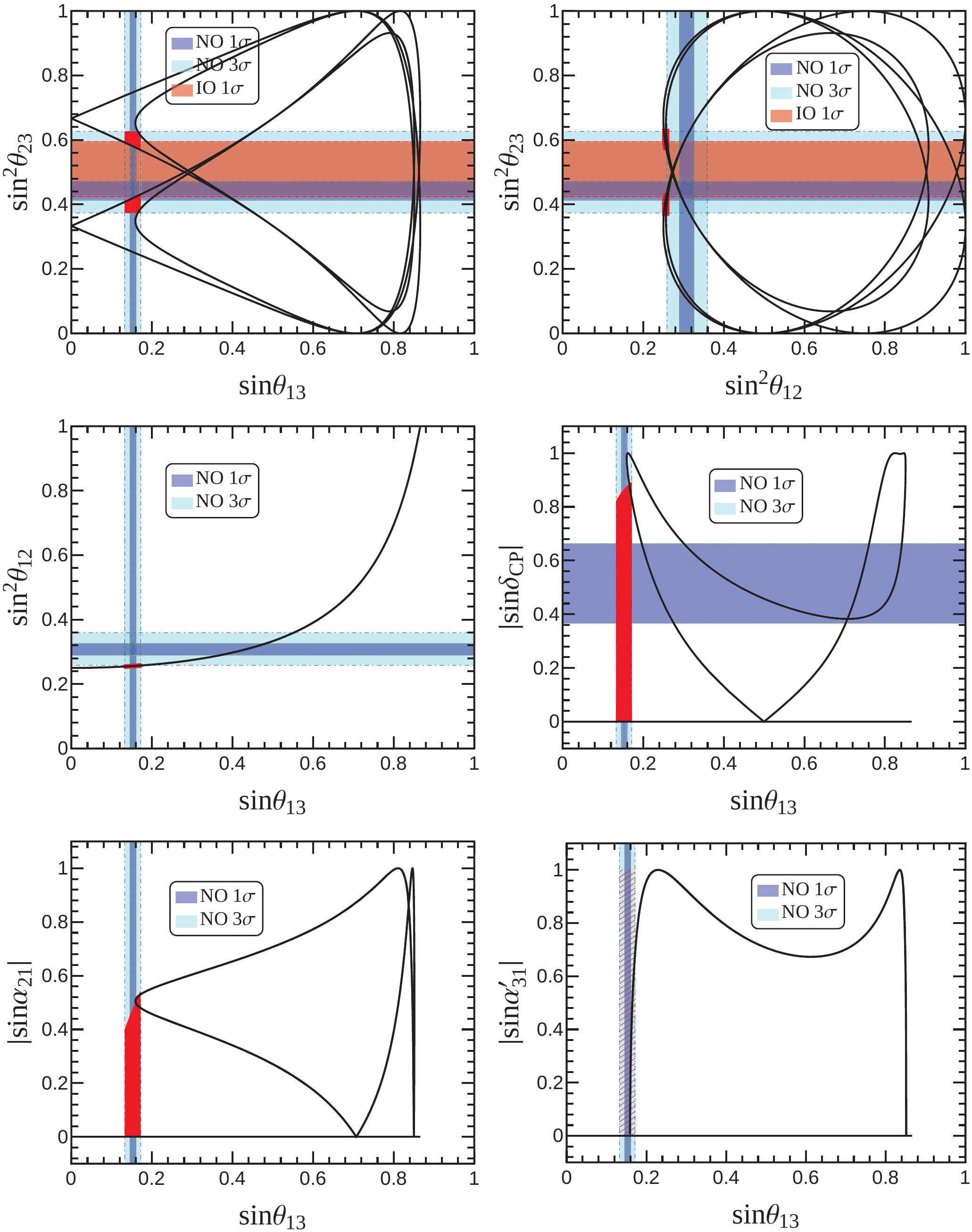}
\caption{\label{fig:caseIV}The correlations among mixing parameters in case II. The red filled regions denote the allowed values of the mixing parameters if we take the parameter $\varphi_3$ to be continuous (which is equivalent to taking the limit $n\rightarrow\infty $) and the three mixing angles are required to lie in their $3\sigma$ ranges (the $3\sigma$ lower bound of $\sin^2\theta_{12}$ is chosen to be 0.254 instead of 0.259 given in Ref.~\cite{Capozzi:2013csa}). Note that the Majorana phase $\alpha^{\prime}_{31}$ is not constrained in this limit. The black curves represent the phenomenologically viable correlations for $n=8$. The $1\sigma$ and $3\sigma$ bounds of the mixing parameters are taken from Ref.~\cite{Capozzi:2013csa}}
\end{center}
\end{figure}

\item[~~(\uppercase\expandafter{\romannumeral3})]

$G_{l}=\left\langle ac^{s}d^{t}\right\rangle$,
$G_{\nu}=Z^{c^{n/2}}_{2}$,
$X_{\nu\mathbf{r}}=\rho_{\mathbf{r}}(c^{\gamma}d^{\delta})$

This case can be realizable if $n$ is divisible by 2, and the PMNS matrix takes the form
\begin{equation}
U^{III}_{PMNS}=\frac{1}{\sqrt{3}}\left(
\begin{array}{ccc}
e^{i\varphi_4}\cos\theta-e^{i\varphi_5}\sin\theta  ~&~ 1 ~&~
e^{i\varphi_4}\sin\theta+e^{i\varphi_5}\cos\theta  \\
\omega e^{i\varphi_4}\cos\theta-\omega^2e^{i\varphi_5}\sin\theta ~&~  1
~&~ \omega e^{i\varphi_4}\sin\theta+\omega^2e^{i\varphi_5}\cos\theta\\
\omega^2e^{i\varphi_4}\cos\theta-\omega e^{i\varphi_5}\sin\theta   ~&~
1~&~ \omega^2e^{i\varphi_4}\sin\theta+\omega e^{i\varphi_5}\cos\theta \\
\end{array}
\right)\,,
\end{equation}
where
\begin{equation}
\varphi_{4}=\frac{\gamma+\delta+2s}{n}\pi,\qquad
\varphi_{5}=\frac{2\delta-\gamma+2t}{n}\pi\,,
\end{equation}
which can take the values
\begin{equation}
\varphi_4, \varphi_5~\textrm{mod}~2\pi=0, \frac{1}{n}\pi, \frac{2}{n}\pi,\ldots,\frac{2n-1}{n}\pi\,.
\end{equation}
Agreement with experimental data can be achieved only if the vector $\left(1/\sqrt{3}, 1/\sqrt{3}, 1/\sqrt{3}\right)^{T}$ is placed in the second column. It is the so-called $\text{TM}_2$ mixing~\cite{Albright:2008rp}. There are three independent arrangements up to the exchange of the second and the third row,
\begin{equation}
U^{III,1st}_{PMNS}=U^{III}_{PMNS},\qquad U^{III,2nd}_{PMNS}=P_{231}U^{III}_{PMNS},\qquad U^{III,3rd}_{PMNS}=P_{312}U^{III}_{PMNS}\,.
\end{equation}
Once can check that they are related as follows,
\begin{eqnarray}
\nonumber&&U^{III,2nd}_{PMNS}(\theta, \varphi_4, \varphi_5)=U^{III,1st}_{PMNS}(\theta, \varphi_4+\frac{2\pi}{3}, \varphi_5-\frac{2\pi}{3}),\\
&&U^{III,3rd}_{PMNS}(\theta, \varphi_4, \varphi_5)=U^{III,1st}_{PMNS}(\theta, \varphi_4-\frac{2\pi}{3}, \varphi_5+\frac{2\pi}{3})\,.
\end{eqnarray}
It is enough to study the phenomenological predictions of $U^{III,1st}_{PMNS}$. The lepton mixing parameters are given by
\begin{eqnarray}
\nonumber&&\sin^2\theta_{13}=\frac{1}{3}\left[1+\sin2\theta\cos(\varphi_5-\varphi_4)\right],\quad
\sin^2\theta_{12}=\frac{1}{2-\sin2\theta\cos(\varphi_5-\varphi_4)}\,,\\
\nonumber&&\sin^2\theta_{23}=\frac{1-\sin2\theta\sin\left(\varphi_5-\varphi_4+\pi/6\right)}{2-\sin2\theta\cos(\varphi_5-\varphi_4)},\qquad \left|J_{CP}\right|=\frac{1}{6 \sqrt{3}}\left|\cos2\theta\right|\,,\\
\nonumber&&\left|\tan\delta_{CP}\right|=\left|\frac{\cot2\theta\left[2-\sin2\theta\cos(\varphi_5-\varphi_4)\right]}{\sin(\varphi_5-\varphi_4)-\sin2\theta\sin(2\varphi_5-2\varphi_4)}\right|,\\
\nonumber&&\left|\tan\alpha_{21}\right|=\left|\frac{\cos^2\theta\sin2\varphi_4+\sin^2\theta\sin2\varphi_5-\sin2\theta\sin(\varphi_5+\varphi_4)}
{\cos^2\theta\cos2\varphi_4+\sin^2\theta\cos2\varphi_5-\sin2\theta\cos(\varphi_5+\varphi_4)}\right|,\\
\label{eq:mixing_parameters_VIII_1}&&\left|\tan\alpha^{\prime}_{31}\right|=\left|\frac{4\cos2\theta\sin(2\varphi_5-2\varphi_4)}{1-3\cos(2\varphi_5-2\varphi_4)-2\cos4\theta\cos^2(\varphi_5-\varphi_4)}\right|\,.
\end{eqnarray}
We see that all the mixing parameters depend on the combination $\varphi_{5}-\varphi_4$ except $\left|\tan\alpha_{21}\right|$. Common to all $\text{TM}_2$ mixing, $\theta_{13}$ and $\theta_{12}$ are related with each other via:
\begin{equation}
3\cos^2\theta_{13}\sin^2\theta_{12}=1.
\end{equation}
Therefore $\theta_{12}$ admits a lower bound $\sin^2\theta_{12}>1/3$. Given the $3\sigma$ interval of $\theta_{13}$~\cite{Capozzi:2013csa}, we find $0.339\leq\sin^2\theta_{12}\leq0.343$. This prediction can be tested at JUNO in near future. In addition, $\theta_{13}$ and $\theta_{23}$ are correlated as follows
\begin{equation}
\frac{3\cos^2\theta_{13}\sin^2\theta_{23}-1}{1-3\sin^2\theta_{13}}=\frac{1}{2}+\frac{\sqrt{3}}{2}\tan\left(\varphi_5-\varphi_4\right)\,.
\end{equation}
The expression for $\theta_{13}$ in Eq.~\eqref{eq:mixing_parameters_VIII_1} implies that
\begin{eqnarray}
\nonumber&\frac{1}{3}(1-|\sin2\theta|)\leq\sin^2\theta_{13}\leq\frac{1}{3}(1+|\sin2\theta|),\\ &\frac{1}{3}(1-\left|\cos(\varphi_{5}-\varphi_4)\right|)\leq\sin^2\theta_{13}\leq\frac{1}{3}(1+\left|\cos(\varphi_{5}-\varphi_4)\right|)\,,
\end{eqnarray}
which yields
\begin{eqnarray}
\nonumber&\theta\in\left[0.183\pi,0.317\pi\right]\cup\left[0.683\pi,0.817\pi\right],\\ &\varphi_{5}-\varphi_4\in\left[-0.135\pi,0.135\pi\right]\cup\left[0.865\pi,1.135\pi\right]\,.
\end{eqnarray}
The allowed values of the mixing parameters for different $n$ are shown in Fig~\ref{fig:caseVIII_angles_ranges} and Fig.~\ref{fig:caseVIII_phases_ranges}. The case of $\varphi_4=\varphi_5$ is always viable for any $n$, and the resulting $\theta_{23}$ and $\delta_{CP}$ are maximal while the Majorana phase $\alpha^{\prime}_{31}$ is trivial. Correlations among the mixing parameters are plotted in Fig.~\ref{fig:caseVIII}. The three CP phases can take any values for large $n$.

\begin{figure}[t!]
\begin{center}
\includegraphics[width=0.99\textwidth]{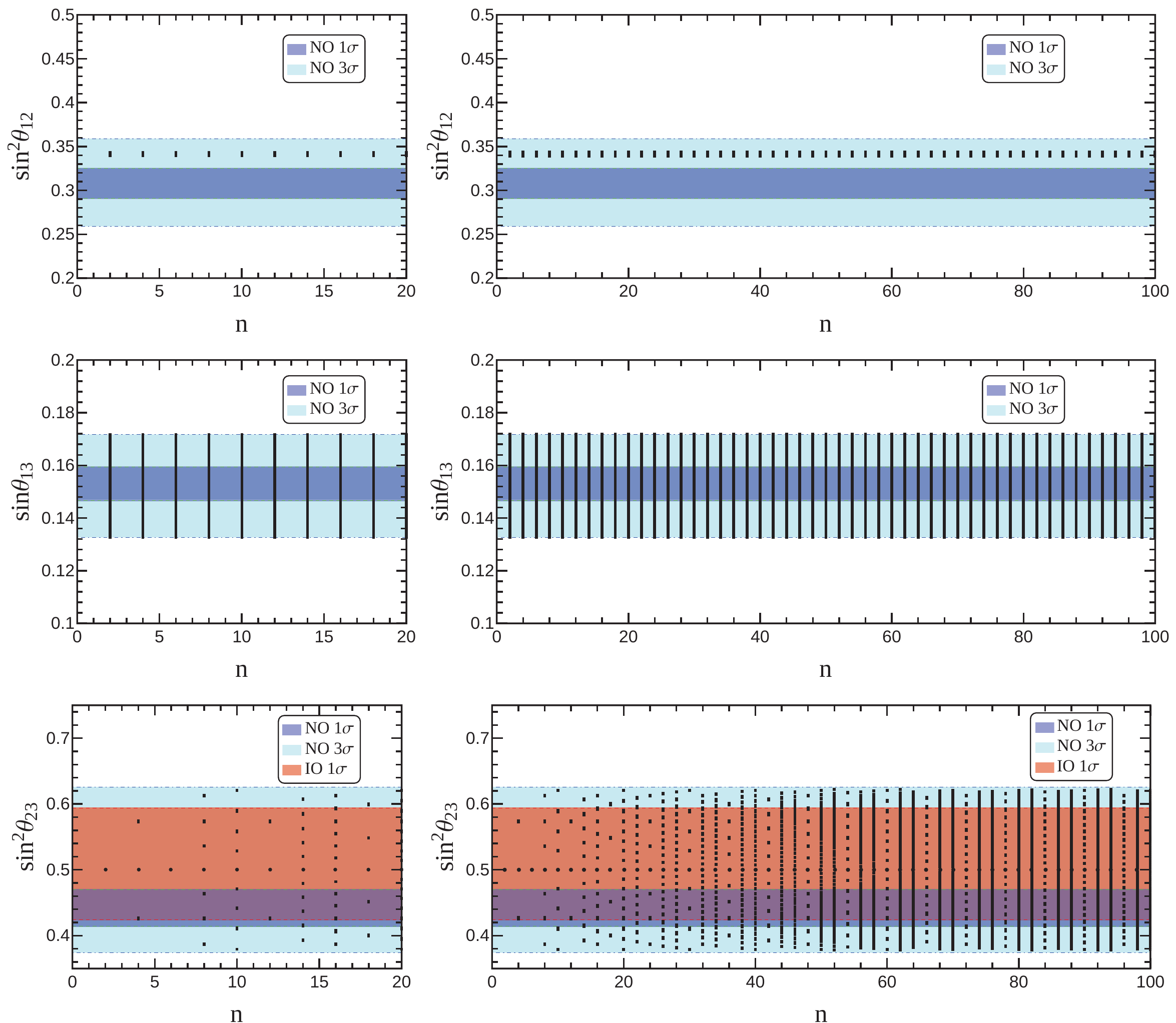}
\caption{\label{fig:caseVIII_angles_ranges}Numerical results in case III: the allowed values of $\sin^2\theta_{12}$, $\sin\theta_{13}$ and $\sin^2\theta_{23}$ for different $n$, where the three lepton mixing angles are required to lie in their $3\sigma$ regions. The $1\sigma$ and $3\sigma$ bounds of the mixing parameters are taken from Ref.~\cite{Capozzi:2013csa}. Note that $n$ should be even in this case.}
\end{center}
\end{figure}

\begin{figure}[hptb!]
\begin{center}
\includegraphics[width=0.99\textwidth]{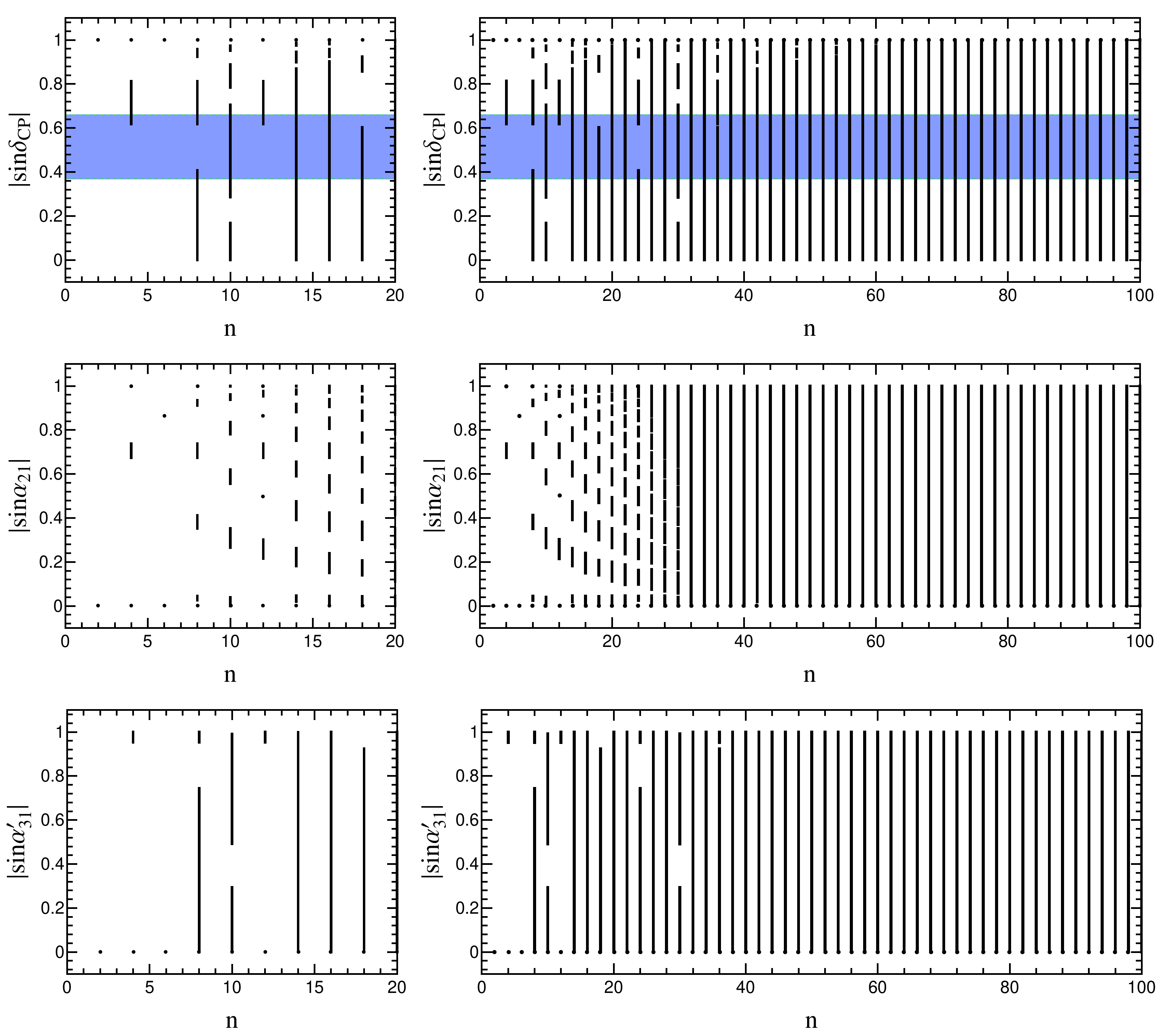}
\caption{\label{fig:caseVIII_phases_ranges}Numerical results in case III:
the allowed values of $\left|\sin\delta_{CP}\right|$, $\left|\sin\alpha_{21}\right|$ and $\left|\sin\alpha^{\prime}_{31}\right|$
for different $n$, where the three lepton mixing angles are required to lie in their $3\sigma$ ranges. The $1\sigma$ and $3\sigma$ bounds of the mixing angles are taken from Ref.~\cite{Capozzi:2013csa}. Note that $n$ should be even in this case. }
\end{center}
\end{figure}

\begin{figure}[hptb!]
\begin{center}
\includegraphics[width=0.90\textwidth]{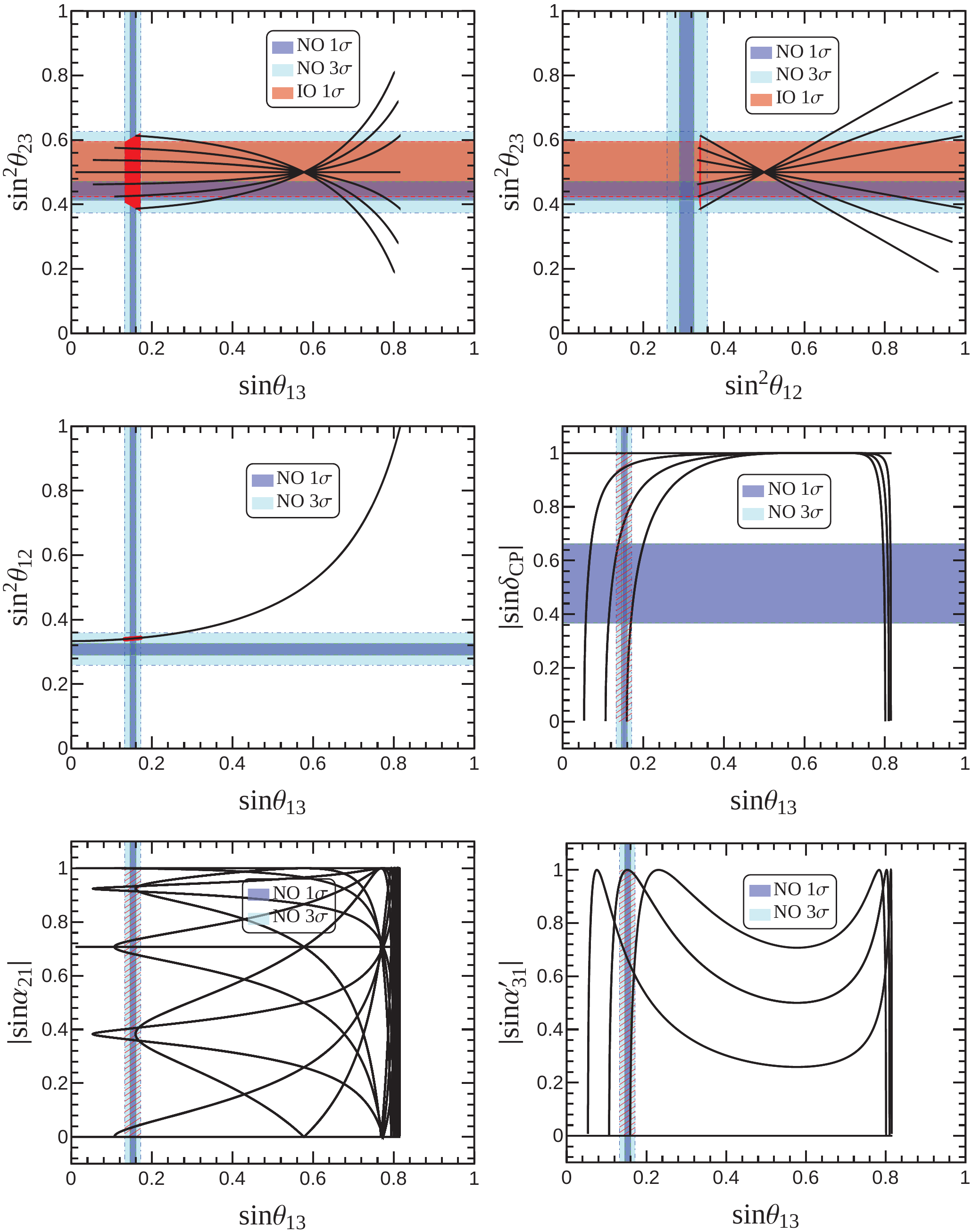}
\caption{\label{fig:caseVIII}The correlations among mixing parameters in case III. The red filled regions denote the allowed values of the mixing parameters if we take the parameters $\varphi_4$ and $\varphi_5$ to be continuous (which is equivalent to taking the limit $n\rightarrow\infty $) and the three mixing angles are required to lie in their $3\sigma$ ranges. Note that the three CP phases $\delta_{CP}$, $\alpha_{21}$ and $\alpha^{\prime}_{31}$ are not constrained in this limit. The black curves represent the phenomenologically viable correlations for $n=8$. The $1\sigma$ and $3\sigma$ bounds of the mixing parameters are taken from Ref.~\cite{Capozzi:2013csa}. }
\end{center}
\end{figure}

\item[~~(\uppercase\expandafter{\romannumeral4})]

$G_{l}=\left\langle ac^{s}d^{t}\right\rangle$,
$G_{\nu}=Z^{c^{n/2}}_{2}$,
$X_{\nu\mathbf{r}}=\rho_{\mathbf{r}}(abc^{\gamma}d^{\delta})$

In this case, the PMNS matrix is of the form
\begin{equation}
\label{eq:PMNS_case_XIII}U^{IV}_{PMNS}=\frac{1}{\sqrt{3}}\left(
\begin{array}{ccc}
i \sqrt{2}e^{i\varphi_7}\sin\left(\varphi_6-\frac{\phi}{2}\right) ~&~ 1
~&~ \sqrt{2}e^{i\varphi_7}\cos\left(\varphi_6-\frac{\phi}{2}\right) \\
i\sqrt{2}e^{i\varphi_7}\cos\left(\varphi_6-\frac{\phi}{2}+\frac{\pi}{6}\right)
~&~ 1
~&~-\sqrt{2}e^{i\varphi_7}\sin\left(\varphi_6-\frac{\phi}{2}+\frac{\pi}{6}\right)
\\
-i\sqrt{2}e^{i\varphi_7}\cos\left(\varphi_6-\frac{\phi}{2}-\frac{\pi}{6}\right)
~&~ 1
~&~\sqrt{2}e^{i\varphi_7}\sin\left(\varphi_6-\frac{\phi}{2}-\frac{\pi}{6}\right)
\end{array}
\right)\,,
\end{equation}
with
\begin{equation}
\varphi_6=\frac{s-t-\gamma}{n}\pi ,\qquad
\varphi_7=\frac{s+t+3\gamma}{n}\pi\,.
\end{equation}
The constant vector $\left(1/\sqrt{3}, 1/\sqrt{3}, 1/\sqrt{3}\right)^{T}$
must be the second column to account for the measured values of the lepton
mixing angles. The PMNS matrices corresponding to other ordering of rows and columns are related to the above one through redefinition of the free parameter $\phi$. This case differs from case III in the remnant CP symmetry, and the resulting PMNS matrix in Eq.~\eqref{eq:PMNS_case_XIII} is still of $\textrm{TM}_2$ form. The associated lepton mixing parameters read as:
\begin{eqnarray}
\nonumber&&\sin^2\theta_{13}=\frac{1}{3}\left[1+\cos(\phi-2\varphi_6)\right],\quad
\sin^2\theta_{12}=\frac{1}{2-\cos(\phi-2\varphi_6)}\,,\\
\nonumber&&\sin^2\theta_{23}=\frac{1-\sin\left(\phi-2\varphi_6+\pi/6\right)}{2-\cos(\phi-2\varphi_6)}\,,\\
\label{eq:mixing_parameters_case_XIII}&&\tan\delta_{CP}=\tan\alpha^{\prime}_{31}=J_{CP}=0,\qquad
\left|\tan\alpha_{21}\right|=|\tan(2\varphi_7)|\,.
\end{eqnarray}
It is remarkable that the contribution of $\varphi_6$ can be absorbed into the free parameter $\phi$ via redefinition $\phi\rightarrow\phi+2\varphi_6$, the reason is that the PMNS matrix in Eq.~\eqref{eq:PMNS_case_XIII} and the resulting mixing parameters in Eq.~\eqref{eq:mixing_parameters_case_XIII} depend on the combination $\phi-2\varphi_6$. Regarding to the CP violating phases, both $\delta_{CP}$ and $\alpha^{\prime}_{31}$ are always conserved while $\alpha_{21}$ can be any value of $0$, $\frac{1}{n}\pi$, $\frac{2}{n}\pi$, $\ldots$, $\frac{2n-1}{n}\pi$ in this scenario. Furthermore, the three mixing angles are strongly related with each other as follows:
\begin{equation}
3\cos^2\theta_{13}\sin^2\theta_{12}=1,\qquad
\sin^2\theta_{23}=\frac{1}{2}\pm\frac{1}{2}\tan\theta_{13}\sqrt{2-\tan^2\theta_{13}}\,.
\end{equation}
For the best fitting value of $\sin^2\theta_{13}=0.0234$~\cite{Capozzi:2013csa}, the solar and atmospheric angles are determined to be
\begin{equation}
\sin^2\theta_{12}\simeq0.341,\qquad \sin^2\theta_{23}\simeq0.391~~ \text{or}~~ 0.609\,,
\end{equation}
which are compatible with the experimentally allowed regions.
These correlations between the three mixing angles are shown in Fig.~\ref{fig:caseXIII}. We see that both $\theta_{12}$ and $\theta_{23}$ are constrained to be in a narrow range. The deviation of $\theta_{23}$ from maximal mixing is somewhat large. Hence this mixing pattern can be checked or ruled by precisely measuring $\theta_{12}$ and $\theta_{23}$ in next generation neutrino oscillation experiments.

\begin{figure}[t!]
\begin{center}
\includegraphics[width=0.99\textwidth]{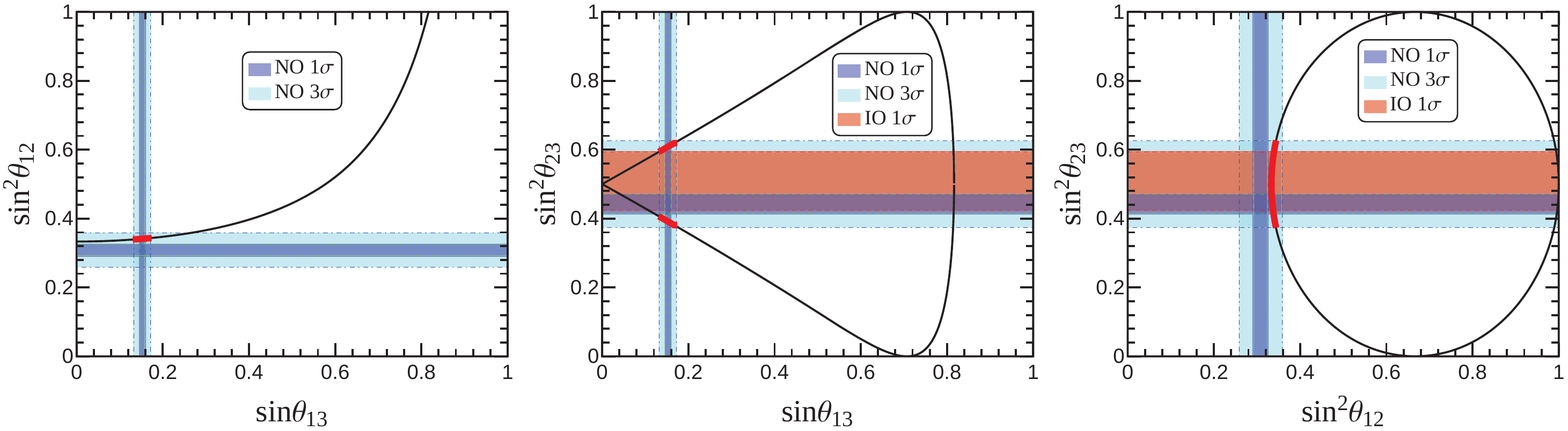}
\caption{\label{fig:caseXIII}The correlations among mixing parameters in case IV. The red filled regions denote the allowed values of the mixing parameters if we take the parameters $\varphi_6$ and $\varphi_7$ to be continuous (which is equivalent to taking the limit $n\rightarrow\infty $), where $\theta_{12}$ and $\theta_{13}$ are required to lie in their $3\sigma$ ranges. The $1\sigma$ and $3\sigma$ bounds of the mixing parameters are taken from Ref.~\cite{Capozzi:2013csa}. }
\end{center}
\end{figure}

\end{description}

\section{\label{sec:Z2xCP_charged_lepton}Lepton mixing from ``semi-direct'' approach with residual symmetry $Z_2\times CP$ in the charged lepton sector}
\cleqn

In the previous section, we assumed a $Z_2\times CP$ remnant symmetry in the neutrino sector and that an abelian subgroup of $\Delta(6n^2)$ is preserved in the charged lepton sector. In this section, we shall investigate another scenario: the remnant symmetry $Z_2\times CP$ is preserved in the charged lepton sector and the full symmetry $\Delta(6n^2)\rtimes H_{CP}$ is broken down to $K_{4}\rtimes H^{\nu}_{CP}$ in the neutrino sector. The phenomenological consequences of this scenario have been analyzed for the simple flavor symmetry group $\Delta(24)=S_4$ in Ref.~\cite{Li:2014eia}.
All the $Z_2$ subgroups of $\Delta(6n^2)$ have been listed in Eq.~\eqref{eq:z2_1} and Eq.~\eqref{eq:z2_2}. The $K_4$ subgroups of $\Delta(6n^2)$ are classified as follows:
\begin{eqnarray}
\nonumber &&K^{(c^{n/2},d^{n/2})}_4\equiv\left\{1, c^{n/2}, d^{n/2},
c^{n/2}d^{n/2}\right\},\\
\nonumber&& K^{(c^{n/2},abc^{y})}_4\equiv\left\{1, c^{n/2}, abc^{y},
abc^{y+n/2}\right\},\\
\nonumber&& K^{(d^{n/2},a^2bd^{z})}_4\equiv\left\{1, d^{n/2}, a^2bd^{z},
a^2bd^{z+n/2}\right\},\\
\label{eq:K4_subgroups}&&K^{(c^{n/2}d^{n/2},bc^{x}d^{x})}_4\equiv\left\{1, c^{n/2}d^{n/2}, bc^{x}d^{x},
bc^{x+n/2}d^{x+n/2}\right\}\,,
\end{eqnarray}
where $K^{(c^{n/2},d^{n/2})}_4$ is a normal subgroup of $\Delta(6n^2)$, and the remaining three $K_4$ subgroups are conjugate to each other,
Obviously this scenario is only possible if $n$ is divisible by 2. Because of the conjugate relations shown in Eq.~\eqref{eq:neutrino_conjugate_1} and Eq.~\eqref{eq:neutrino_conjugate_4}, we only need to consider the representative cases of $G_{l}=Z^{bc^{x}d^{x}}_2, Z^{c^{n/2}}_2$ and
$G_{\nu}=K^{(c^{n/2},d^{n/2})}_4$, $K^{(c^{n/2},abc^{y})}_4$,
$K^{(d^{n/2},a^2bd^{z})}_4$ and $K^{(c^{n/2}d^{n/2},bc^{x}d^{x})}_4$. Other possible choices of $G_{l}$ and $G_{\nu}$ are related to these representative residual symmetry by similarity transformations, and therefore the same lepton mixing matrices are generated.

Following the same procedure demonstrated in section~\ref{sec:Z2xCP_neutrino}, the hermitian combination $m^{\dagger}_{l}m_{l}$ of the charged lepton mass matrix and its diagonalization matrix can be straightforwardly calculated from the invariance under the remnant symmetry although the involved algebraic calculations are somewhat lengthy and tedious. To avoid distractions by too many details, we have left the calculations for this to Appendix~\ref{sec:charged_lepton_Z2CP}. Comparing with the scenario of $Z_2\times H^{\nu}_{CP}$ preserved in the neutrino sector which has been studied in section~\ref{subsec:neutrino_one_column}, we find that the unitary transformation $U_{l}$ is of the same form as $U_{\nu}$ listed in section~\ref{subsec:neutrino_one_column} if the both residual flavor symmetry and residual CP symmetry in the two occasions are identical. For the sake of presentation in the following, we briefly recapitulate the predicted form of $U_{l}$ for $G_{l}=Z^{bc^{x}d^{x}}_2$ and $Z^{c^{n/2}}_2$ here.

In case of $G_{l}=Z^{bc^xd^x}_2$, $X_{l\mathbf{r}}=\left\{\rho_{\mathbf{r}}(c^{\gamma}d^{-2x-\gamma}), \rho_{\mathbf{r}}(bc^{x+\gamma}d^{-x-\gamma})\right\}$, as shown in Appendix~\ref{sec:charged_lepton_Z2CP}, the charged lepton mass matrix is diagonalized by
\begin{equation}
\label{eq:ul_bcxdx_2re}U_{l}=\frac{1}{\sqrt{2}}
\left(
\begin{array}{ccc}
 e^{i\pi\frac{\gamma}{n}} & -e^{i\pi\frac{\gamma}{n}}\sin\theta &
 e^{i\pi\frac{\gamma}{n}}\cos\theta \\
 0 &  e^{-2i\pi\frac{x+\gamma}{n}}
   \sqrt{2}\cos\theta~ & ~e^{-2i\pi\frac{x+\gamma}{n}} \sqrt{2}\sin\theta
   \\
 -e^{i\pi\frac{2x+\gamma}{n}} & -e^{i\pi\frac{2x+\gamma}{n}}\sin\theta &
 e^{i\pi\frac{2x+\gamma}{n}}\cos\theta
\end{array}
\right)\,.
\end{equation}
For the residual symmetry $G_{l}=Z^{c^{n/2}}_2$, $X_{l\mathbf{r}}=\rho_{\mathbf{r}}(c^{\gamma}d^{\delta})$, the unitary matrix $U_{l}$ is of the form
\begin{equation}
U_{l}=\left(
\begin{array}{ccc}
e^{i\pi\frac{\gamma}{n}}\cos\theta &  e^{i\pi\frac{\gamma}{n}}\sin\theta
& 0 \\
-e^{i\pi\frac{\delta-\gamma}{n}}\sin\theta~ &
~e^{i\pi\frac{\delta-\gamma}{n}}\cos\theta & 0 \\
 0 & 0 & e^{-i\pi\frac{\delta}{n}}
\end{array}
\right)\,.
\end{equation}
The remnant symmetry could also be $G_{l}=Z^{c^{n/2}}_2$, $X_{l\mathbf{r}}=\rho_{\mathbf{r}}(abc^{\gamma}d^{\delta})$. The non-degeneracy of the charged lepton masses requires  $\delta=2\gamma~\text{mod}~n$, and then the diagonalization matrix $U_{l}$ is determined to be
\begin{equation}
U_{l}=\frac{1}{\sqrt{2}}\left(
\begin{array}{ccc}
 e^{i \phi } &~ e^{i \phi } & 0 \\
 -1 &~ 1 & 0 \\
 0 &~ 0 & \sqrt{2}
\end{array}
\right)\,,
\end{equation}
where $\phi$ is a free real parameter.

\subsection{\label{subsec:neutrino_one_row}Neutrino sector}

In this section, we shall assume that the $\Delta(6n^2)$ flavor symmetry is broken down to $K_4$ in the neutrino sector. Hence the neutrino diagonalization matrix $U_{\nu}$ is entirely fixed by the remnant $K_4$, and the residual CP symmetry allows us to further determine the three leptonic CP violating phases up to $\pi$. The residual CP symmetry $H^{\nu}_{CP}$ in the neutrino sector must be compatible with the remnant $K_4$ symmetry, and the consistency condition should be satisfied,
\begin{equation}
X_{\nu
\mathbf{r}}\rho^{*}_{\mathbf{r}}(g)X^{-1}_{\nu\mathbf{r}}=\rho_{\mathbf{r}}(g^{\prime}),\qquad
g, g^{\prime}\in K_4\,.
\end{equation}
Solving this equation, we can find the consistent remnant CP symmetries for different $K_4$ subgroups are as follows:
\begin{itemize}[labelindent=-0.7em, leftmargin=1.6em]

\item{$K^{(c^{n/2},d^{n/2})}_4$}
\begin{equation}
X_{\nu\mathbf{r}}=\rho_{\mathbf{r}}(h),\qquad h\in\Delta(6n^2)\,.
\end{equation}

\item{$K^{(c^{n/2},abc^{y})}_4$, $y=0,1,\ldots n-1$}
\begin{equation}
X_{\nu\mathbf{r}}=\rho_{\mathbf{r}}(c^{\gamma}d^{2y+2\gamma}),
\rho_{\mathbf{r}}(c^{\gamma}d^{2y+2\gamma+n/2}),
\rho_{\mathbf{r}}(abc^{\gamma}d^{2\gamma}),
\rho_{\mathbf{r}}(abc^{\gamma}d^{2\gamma+n/2})\,,
\end{equation}
with $\gamma=0,1,\ldots n-1$.

\item{$K^{(d^{n/2},a^2bd^{z})}_4$, $z=0,1,\ldots n-1$}
\begin{equation}
X_{\nu\mathbf{r}}=\rho_{\mathbf{r}}(c^{2z+2\delta}d^{\delta}),\rho_{\mathbf{r}}(c^{2z+2\delta+n/2}d^{\delta}),\rho_{\mathbf{r}}(a^2bc^{2\delta}d^{\delta}),\rho_{\mathbf{r}}(a^2bc^{2\delta+n/2}d^{\delta})\,,
\end{equation}
where $\delta=0,1,\ldots n-1$.

\item{$K^{(c^{n/2}d^{n/2},bc^{x}d^{x})}_4$, $x=0,1,\ldots n-1$}
\begin{equation}
X_{\nu\mathbf{r}}=\rho_{\mathbf{r}}(c^{\gamma}d^{-2x-\gamma}),\rho_{\mathbf{r}}(c^{\gamma}d^{-2x-\gamma+n/2}),\rho_{\mathbf{r}}(bc^{\gamma}d^{-\gamma}),\rho_{\mathbf{r}}(bc^{\gamma}d^{-\gamma+n/2})\,,
\end{equation}
with $\gamma=0,1,\ldots n-1$.

\end{itemize}

In this case, the light neutrino mass matrix must be subject to the constraints from the remnant family symmetry $K_4$ and the residual CP symmetry $H^{\nu}_{CP}$:
\begin{eqnarray}
\nonumber&&\rho^{T}_{\mathbf{3}}(g_{\nu})m_{\nu}\rho_{\mathbf{3}}(g_{\nu})=m_{\nu},\quad
g_{\nu}\in K_4\,,\\
&&X^{T}_{\nu\mathbf{3}}m_{\nu}X_{\nu\mathbf{3}}=m^{*}_{\nu},\quad X_{\nu}\in
H^{\nu}_{CP}\,.
\end{eqnarray}

\begin{description}[labelindent=-0.8em, leftmargin=0.3em]
  \item[~~(\romannumeral1)] {$G_{\nu}=K^{(c^{n/2},d^{n/2})}_4$,
      $X_{\nu\mathbf{r}}=\left\{\rho_{\mathbf{r}}(c^{\gamma}d^{\delta})\right\}$}

Since the representation matrices of both $c^{n/2}$ and $d^{n/2}$ are diagonal, the light neutrino mass matrix is constrained to be diagonal as well. Including the remnant CP symmetry, we find
\begin{equation}
m_{\nu}=\left(
\begin{array}{ccc}
 m_{11}e^{-2i\pi\frac{\gamma}{n}}  & 0 & 0 \\
 0 & m_{22}e^{2i\pi\frac{\gamma-\delta}{n}}  & 0   \\
 0 & 0 & m_{33}e^{2i\pi\frac{\delta}{n}}  \\
\end{array}
\right)\,,
\end{equation}
where $m_{11}$, $m_{22}$ and $m_{33}$ are real parameters. The neutrino
diagonalization matrix can be easily read out
\begin{equation}
U_{\nu}=\mathrm{diag}\left( e^{i\pi\frac{\gamma}{n}}, e^{-i\pi\frac{
\gamma-\delta}{n}}, e^{-i\pi\frac{\delta}{n}}\right)K_{\nu}\,,
\end{equation}
where $K_{\nu}$ is a diagonal matrix with element $\pm1$ or $\pm i$ to set
the light neutrino masses being positive. The light neutrino masses are
\begin{equation}
m_1=\left|m_{11}\right|,\qquad m_2=\left|m_{22}\right|,\qquad
m_3=\left|m_{33}\right|\,.
\end{equation}
We see that the light neutrino masses depend on only three real parameters, and we would like to stress again that the order of the light neutrino masses can not be fixed here, and therefore $U_{\nu}$ here and henceforth is determined up to column permutations. For other residual CP symmetries $X_{\nu\mathbf{r}}=\rho_{\mathbf{r}}(bc^{\gamma}d^{\delta})$,
$\rho_{\mathbf{r}}(ac^{\gamma}d^{\delta})$,
$\rho_{\mathbf{r}}(a^2c^{\gamma}d^{\delta})$,
$\rho_{\mathbf{r}}(abc^{\gamma}d^{\delta})$ and
$\rho_{\mathbf{r}}(a^2bc^{\gamma}d^{\delta})$ with $\gamma,
\delta=0,1,\ldots n-1$, the light neutrino masses are partially degenerate
such that they are not viable.

\item[~~(\romannumeral2)] {$G_{\nu}=K^{(c^{n/2},abc^{y})}_4$,
    $X_{\nu\mathbf{r}}=\left\{\rho_{\mathbf{r}}(c^{\gamma}d^{2y+2\gamma}),
    \rho_{\mathbf{r}}(abc^{y+\gamma}d^{2y+2\gamma})\right\}$}

In this case, the light neutrino mass matrix takes the form
\begin{equation}
m_{\nu}=\left(
\begin{array}{ccc}
 m_{11}e^{-2i\pi\frac{\gamma}{n}}  & m_{12}e^{-2i\pi\frac{y+\gamma}{n}} &
 0 \\
 m_{12}e^{-2i\pi\frac{y+\gamma}{n}}  & m_{11}e^{-2i\pi\frac{2y+\gamma}{n}}
 & 0 \\
 0 & 0 & m_{33}e^{4i\pi\frac{y+\gamma}{n}}
\end{array}
\right)\,,
\end{equation}
where $m_{11}$, $m_{12}$ and $m_{33}$ are real. It is diagonalized by the
unitary matrix $U_{\nu}$ with
\begin{equation}
\label{eq:unu_abcy_2}U_{\nu}=\frac{1}{\sqrt{2}}\left(
\begin{array}{ccc}
 e^{i\pi\frac{\gamma}{n}}   & e^{i\pi\frac{\gamma}{n}} & 0 \\
 -e^{i\pi\frac{2y+\gamma}{n}} & e^{i\pi\frac{2y+\gamma}{n}} & 0 \\
 0 & 0 & \sqrt{2}e^{-2i\pi\frac{y+\gamma}{n}}
\end{array}
\right)\,.
\end{equation}
The light neutrino masses are given by
\begin{equation}
m_1=\left|m_{11}-m_{12}\right|,\qquad
m_2=\left|m_{11}+m_{12}\right|,\qquad m_3=\left|m_{33}\right|\,.
\end{equation}
For the case of
$X_{\nu\mathbf{r}}=\left\{\rho_{\mathbf{r}}(c^{\gamma}d^{2y+2\gamma+n/2}),
\rho_{\mathbf{r}}(abc^{y+\gamma}d^{2y+2\gamma+n/2})\right\}$, the light
neutrino masses are degenerate, and therefore are not discussed here.

\item[~~(\romannumeral3)] {$G_{\nu}=K^{(d^{n/2},a^2bd^{z})}_4$,
    $X_{\nu\mathbf{r}}=\left\{\rho_{\mathbf{r}}(c^{2z+2\delta}d^{\delta}),
    \rho_{\mathbf{r}}(a^2bc^{2z+2\delta}d^{z+\delta})\right\}$}

The light neutrino mass matrix, which is invariant under both residual flavor and residual CP symmetry, is determined to be
\begin{equation}
m_{\nu}=\left(
\begin{array}{ccc}
 m_{11}e^{-4i\pi\frac{z+\delta}{n}}  & 0 & 0 \\
 0 & m_{22}e^{2i\pi\frac{2z+\delta}{n}}  &
 m_{23}e^{2i\pi\frac{z+\delta}{n}}  \\
 0 & m_{23}e^{2i\pi\frac{z+\delta}{n}} &
   m_{22}e^{2i\pi\frac{\delta}{n}}
\end{array}
\right)\,,
\end{equation}
where $m_{11}$, $m_{22}$ and $m_{23}$ are real. The unitary matrix $U_{\nu}$ is
\begin{equation}
U_{\nu}=\frac{1}{\sqrt{2}}\left(
\begin{array}{ccc}
 \sqrt{2} e^{2i\pi\frac{z+\delta}{n}} & 0 & 0 \\
 0 & e^{-i\pi\frac{2z+\delta}{n}} &   e^{-i\pi\frac{2z+\delta}{n}} \\
 0 & -e^{-i\pi\frac{\delta}{n}} & e^{-i\pi\frac{\delta}{n}}
\end{array}
\right)\,.
\end{equation}
The light neutrino mass eigenvalues are given by
\begin{equation}
m_1=\left|m_{11}\right|,\qquad  m_2=\left|m_{22}-m_{23}\right|,\qquad
m_3=\left|m_{22}+m_{23}\right|\,.
\end{equation}
For the value of
$X_{\nu\mathbf{r}}=\left\{\rho_{\mathbf{r}}(c^{2z+2\delta+n/2}d^{\delta}),
\rho_{\mathbf{r}}(a^2bc^{2z+2\delta+n/2}d^{z+\delta})\right\}$, the
neutrino masses are degenerate.

\item[~~(\romannumeral4)] {$G_{\nu}=K^{(c^{n/2}d^{n/2},bc^{x}d^{x})}_4$,
    $X_{\nu\mathbf{r}}=\left\{\rho_{\mathbf{r}}(c^{\gamma}d^{-2x-\gamma}),
    \rho_{\mathbf{r}}(bc^{x+\gamma}d^{-x-\gamma})\right\}$}

In this case, we find the light neutrino mass matrix is of the form
\begin{equation}
m_{\nu}=\left(
\begin{array}{ccc}
 m_{11}e^{-2i\pi\frac{\gamma}{n}}  & 0 &
 m_{13}e^{-2i\pi\frac{x+\gamma}{n}}  \\
 0 & m_{22}e^{4i\pi\frac{x+\gamma}{n}}  & 0   \\
 m_{13}e^{-2i\pi\frac{x+\gamma}{n}}  & 0 &
    m_{11}e^{-2i\pi\frac{2x+\gamma}{n}}
\end{array}
\right)\,,
\end{equation}
where $m_{11}$, $m_{13}$ and $m_{22}$ are real. The unitary matrix
$U_{\nu}$ diagonalizing this neutrino mass matrix is
\begin{equation}
U_{\nu}=\frac{1}{\sqrt{2}}\left(
\begin{array}{ccc}
 e^{i\pi\frac{\gamma}{n}}  & 0 & e^{i\pi\frac{\gamma}{n}} \\
 0 & \sqrt{2} e^{-2i\pi\frac{x+\gamma}{n}} & 0   \\
 -e^{i\pi\frac{2x+\gamma}{n}} & 0 & e^{i\pi\frac{2x+\gamma}{n}}
\end{array}
\right)\,.
\end{equation}
Finally the neutrino masses are
\begin{equation}
m_1=\left|m_{11}-m_{13}\right|,\qquad  m_2=\left|m_{22}\right|,\qquad
m_3=\left|m_{11}+m_{13}\right|\,.
\end{equation}
For the remaining value of
$X_{\nu\mathbf{r}}=\left\{\rho_{\mathbf{r}}(c^{\gamma}d^{-2x-\gamma+n/2}),
\rho_{\mathbf{r}}(bc^{x+\gamma}d^{-x-\gamma+n/2})\right\}$, the light
neutrino masses are degenerate.

\end{description}

\subsection{\label{subsec:PMNS_one_row}Predictions for lepton flavor mixing}

\begin{table}[t!]
\renewcommand{\tabcolsep}{2.0mm}
\centering
\begin{tabular}{|c||c|c|}
\hline \hline
 &  &     \\ [-0.16in]
 &  $G_{l}=Z^{bc^{x^{\prime}}d^{x^{\prime}}}_2$  &  $G_{l}=Z^{c^{n/2}}_2$  \\

  &   &      \\ [-0.16in]\hline
 &   &       \\ [-0.16in]

$G_{\nu}=K^{(c^{n/2},d^{n/2})}_4$  &  $\frac{1}{\sqrt{2}}\left(\begin{array}{c}
1\\
-1\\
0
\end{array}
\right)^{T}$\quad \xmark  &  $\left(\begin{array}{c}
1\\
0\\
0
\end{array}\right)^{T}$\quad \xmark   \\
 &   &         \\ [-0.16in]\hline
 &   &        \\ [-0.16in]

$G_{\nu}=K^{(c^{n/2},abc^{y})}_4$  & $\frac{1}{2}\left(\begin{array}{c}
1\\
1\\
-\sqrt{2}
\end{array}
\right)^{T}$\quad \cmark  & $\left(\begin{array}{c}
1\\
0\\
0
\end{array}
\right)^{T}$\quad \xmark   \\
 &   &         \\ [-0.16in]\hline
 &   &        \\ [-0.16in]

$G_{\nu}=K^{(d^{n/2},a^2bd^{z})}_4$  &  $\frac{1}{2}\left(\begin{array}{c}
1\\
1\\
-\sqrt{2}
\end{array}
\right)^{T}$\quad \cmark  & $\frac{1}{\sqrt{2}}\left(\begin{array}{c}
1\\
-1\\
0
\end{array}
\right)^{T}$\quad \xmark    \\
 &   &         \\ [-0.16in]\hline
 &   &        \\ [-0.16in]

$G_{\nu}=K^{(c^{n/2}d^{n/2},bc^{x}d^{x})}_4$  & $\left(\begin{array}{c}
\cos\left(\frac{x-x^{\prime}}{n}\pi\right)\\
-i\sin\left(\frac{x-x^{\prime}}{n}\pi\right) \\
0
\end{array}\right)^{T}$ \quad \xmark  &  $\frac{1}{\sqrt{2}}\left(\begin{array}{c}
1\\
-1\\
0
\end{array}
\right)^{T}$ \quad \xmark  \\

 &   &      \\ [-0.16in]\hline\hline

\end{tabular}
\caption{\label{tab:PMNS_row} The determined form of one row of the PMNS matrix for different remnant symmetries $G_{\nu}$ and $G_{l}$ which are $K_4$ and $Z_2$ subgroups of $\Delta(6n^2)$ family symmetry group respectively. The superscript ``$T$'' means transpose. The symbol ``\xmark'' denotes that the resulting lepton mixing is ruled out since there is at least one zero element in the fixed row, and the symbol ``\cmark'' denote that the resulting mixing is viable. Note that for $G_{l}=Z^{bc^{x^{\prime}}d^{x^{\prime}}}_2$, the cases of $G_{\nu}=K^{(c^{n/2},abc^{y})}_4$ and $G_{\nu}=K^{(d^{n/2},a^2bd^{z})}_4$  are equivalent because the remnant symmetries are related by group conjugation as $b(bc^{x^{\prime}}d^{x^{\prime}})b=bc^{-x^{\prime}}d^{-x^{\prime}}$,
$bd^{n/2}b=c^{n/2}$ and $b(a^2bd^{z})b=abc^{-z}$. }
\end{table}

As the different residual symmetries related by group conjugation lead to the same predictions for the lepton mixing matrix, we only need to consider the cases of $G_{l}=Z^{bc^{x}d^{x}}_2, Z^{c^{n/2}}_2$ and
$G_{\nu}=K^{(c^{n/2},d^{n/2})}_4$, $K^{(c^{n/2},abc^{y})}_4$,
$K^{(d^{n/2},a^2bd^{z})}_4$ and $K^{(c^{n/2}d^{n/2},bc^{x}d^{x})}_4$. Compared with section~\ref{sec:Z2xCP_neutrino}, one row instead of one column of the PMNS matrix is fixed by remnant flavor symmetry in this scenario. The explicit form of this row vector for different remnant symmetry is summarized in Table~\ref{tab:PMNS_row}. We see that only one independent case is viable. Taking into account the compatible remnant CP symmetry, we can predict both mixing angles and CP phases in terms of one free parameter.

\begin{description}[labelindent=-0.5em, leftmargin=0.1em]

\item[~~(\uppercase\expandafter{\romannumeral5})]

$G_{l}=\left\{1,bc^{x^{\prime}}d^{x^{\prime}}\right\}$,
$X_{l\mathbf{r}}=\left\{\rho_{\mathbf{r}}(c^{\gamma^{\prime}}d^{-2x^{\prime}-\gamma^{\prime}}),
\rho_{\mathbf{r}}(bc^{x^{\prime}+\gamma^{\prime}}d^{-x^{\prime}-\gamma^{\prime}})\right\}$,
$G_{\nu}=K^{(c^{n/2},abc^{y})}_4$ and
$X_{\nu\mathbf{r}}=\left\{\rho_{\mathbf{r}}(c^{\gamma}d^{2y+2\gamma}),
\rho_{\mathbf{r}}(abc^{y+\gamma}d^{2y+2\gamma})\right\}$

Combining the unitary transformation $U_{l}$ in Eq.~\eqref{eq:ul_bcxdx_2re} and $U_{\nu}$ in Eq.~\eqref{eq:unu_abcy_2}, we can pin down the lepton flavor mixing matrix as follows:
\begin{equation}
\label{eq:PMNS_case_XVII}U^{V}_{PMNS}=\frac{1}{2}\left(
\begin{array}{ccc}
 \sin\theta+\sqrt{2}e^{i\varphi_8}\cos\theta ~&~
 \sin\theta-\sqrt{2}e^{i\varphi_8}\cos\theta ~&~
 \sqrt{2}e^{i\varphi_9}\sin\theta   \\
  1 ~&~ 1 ~&~ -\sqrt{2}e^{i\varphi_9} \\
 \cos\theta-\sqrt{2}e^{i\varphi_8}\sin\theta   ~&~
 \cos\theta+\sqrt{2}e^{i\varphi_8}\sin\theta ~&~
 \sqrt{2}e^{i\varphi_9}\cos\theta
\end{array}
\right)\,,
\end{equation}
with
\begin{equation}
\varphi_8=\frac{3\gamma^{\prime}+2x^{\prime}+2y}{n}\pi,\qquad
\varphi_9=-\frac{3\gamma+2x^{\prime}+2y}{n}\pi\,.
\end{equation}
Here $\varphi_8$ and $\varphi_9$ are independent, they are determined by the remnant symmetry, and they can take the values,
\begin{equation}
\varphi_8, \varphi_9~~\textrm{mod}~~2\pi=0, \frac{1}{n}\pi, \frac{2}{n}\pi,\ldots, \frac{2n-1}{n}\pi\,.
\end{equation}
In order to be in accordance with the present neutrino oscillation data,
$\left(1/2, 1/2, -e^{i\varphi_9}/\sqrt{2}\right)$ can only be the second
or the third row. Note that as usual permutation of the second and the third rows of $U^{V}_{PMNS}$ is also viable. We can read out the lepton mixing parameters,
\begin{eqnarray}
\nonumber&&\hskip-0.3in\sin^2\theta_{13}=\frac{1}{2}\sin^2\theta,\qquad
\sin^2\theta_{12}=\frac{1}{2}-\frac{\sqrt{2}\sin2\theta\cos\varphi_8}{3+\cos2\theta},\quad \sin^2\theta_{23}=\frac{2}{3+\cos2\theta}\,,\\
\nonumber&&\hskip-0.3in\left|\tan\delta_{CP}\right|=\left|\frac{(3+\cos2\theta)\tan\varphi_8}{1+3\cos2\theta}\right|,\qquad
\left|J_{CP}\right|=\frac{1}{8\sqrt{2}}\left|\sin2\theta\sin\varphi_8\right|,\\
\nonumber&&\hskip-0.3in\left|\tan\alpha_{21}\right|=\left|\frac{8\sqrt{2}(1+3\cos2\theta)\sin2\theta\sin\varphi_8}
{7+12\cos2\theta+13\cos4\theta+8\sin^22\theta\cos2\varphi_8}\right|,\\
\label{eq:mixing_parameters_case_XVII}&&\hskip-0.3in\left|\tan\alpha^{\prime}_{31}\right|=\left|\frac{\sin^2\theta\sin2\varphi_9+\sqrt{2}\sin2\theta\sin(2\varphi_9-\varphi_8)+2\cos^2\theta\sin(2\varphi_9-2\varphi_8)}
{\sin^2\theta\cos2\varphi_9+\sqrt{2}\sin2\theta\cos(2\varphi_9-\varphi_8)+2\cos^2\theta\cos(2\varphi_9-2\varphi_8)}\right|\,.
\end{eqnarray}
We see that all mixing parameters depend on $\theta$ and $\varphi_8$ except $\left|\tan\alpha^{\prime}_{31}\right|$ which involves $\varphi_9$ additionally. It is interesting that the mixing angles $\theta_{13}$ and $\theta_{23}$ are related as follows
\begin{equation}
\label{eq:correlation_13_23_XVII}2\cos^2\theta_{13}\sin^2\theta_{23}=1,\quad \textrm{or}\quad 2\cos^2\theta_{13}\sin^2\theta_{23}=1-2\sin^2\theta_{13}\,,
\end{equation}
where the second relation is for the PMNS matrix obtained by exchanging the second and the third rows of $U^{V}_{PMNS}$. Moreover, $\theta_{12}$ and $\theta_{13}$ are related by
\begin{equation}
\cos^2\theta_{13}\cos2\theta_{12}=\pm2\sin\theta_{13}\sqrt{\cos2\theta_{13}}\cos\varphi_8\,,
\end{equation}
which is relevant to the parameter $\varphi_8$. The $3\sigma$ bound of $\sin^2\theta_{13}$ gives the limit on $\theta$:
\begin{equation}
\theta\in\left[0.060\pi,0.078\pi\right]\cup\left[0.922\pi,0.940\pi\right]\,.
\end{equation}
The equation for $\sin^2\theta_{12}$ in Eq.~\eqref{eq:mixing_parameters_case_XVII} leads to
\begin{equation}
\frac{1}{2}\left(1-\left|\cos\varphi_8\right|\right)\leq\sin^2\theta_{12}\leq\frac{1}{2}\left(1+\left|\cos\varphi_8\right|\right)\,,
\end{equation}
Hence $\varphi_8$ is constrained to lie in the region
\begin{equation}
\varphi_8\in\left[0,0.409\pi\right]\cup\left[0.591\pi,1.409\pi\right]\cup\left[1.591\pi,2\pi\right]\,.
\end{equation}
The numerical results are displayed in Fig.~\ref{fig:caseXVII_angles_ranges} and Fig.~\ref{fig:caseXVII_phases_ranges}. Note that conserved CP corresponding to $\varphi_8=0, \pi$ is always viable. If we require that all three mixing angles are in their $3\sigma$ intervals, we find that $0.141\leq\sin\theta_{13}\leq0.172$, $0.328\leq\sin^2\theta_{12}\leq0.359$, and $\sin^2\theta_{23}$ is around 0.488 and 0.512 due to the correlation shown in Eq.~\eqref{eq:correlation_13_23_XVII}. Note that $\theta_{23}$ is very close to maximal mixing. Therefore precisely measuring the lepton mixing angles at JUNO or long baseline neutrino experiments can test this mixing pattern directly. For the CP phases, $\delta_{CP}$ and $\alpha_{21}$ are predicted to be in the intervals of $\left|\sin\delta_{CP}\right|\leq0.586$ and $\left|\sin\alpha_{21}\right|\leq0.396$ while $\alpha^{\prime}_{31}$ can be any value for sufficient large $n$. The correlations between different mixing parameters are shown in Fig.~\ref{fig:caseXVII}.

\begin{figure}[hptb!]
\begin{center}
\includegraphics[width=0.99\textwidth]{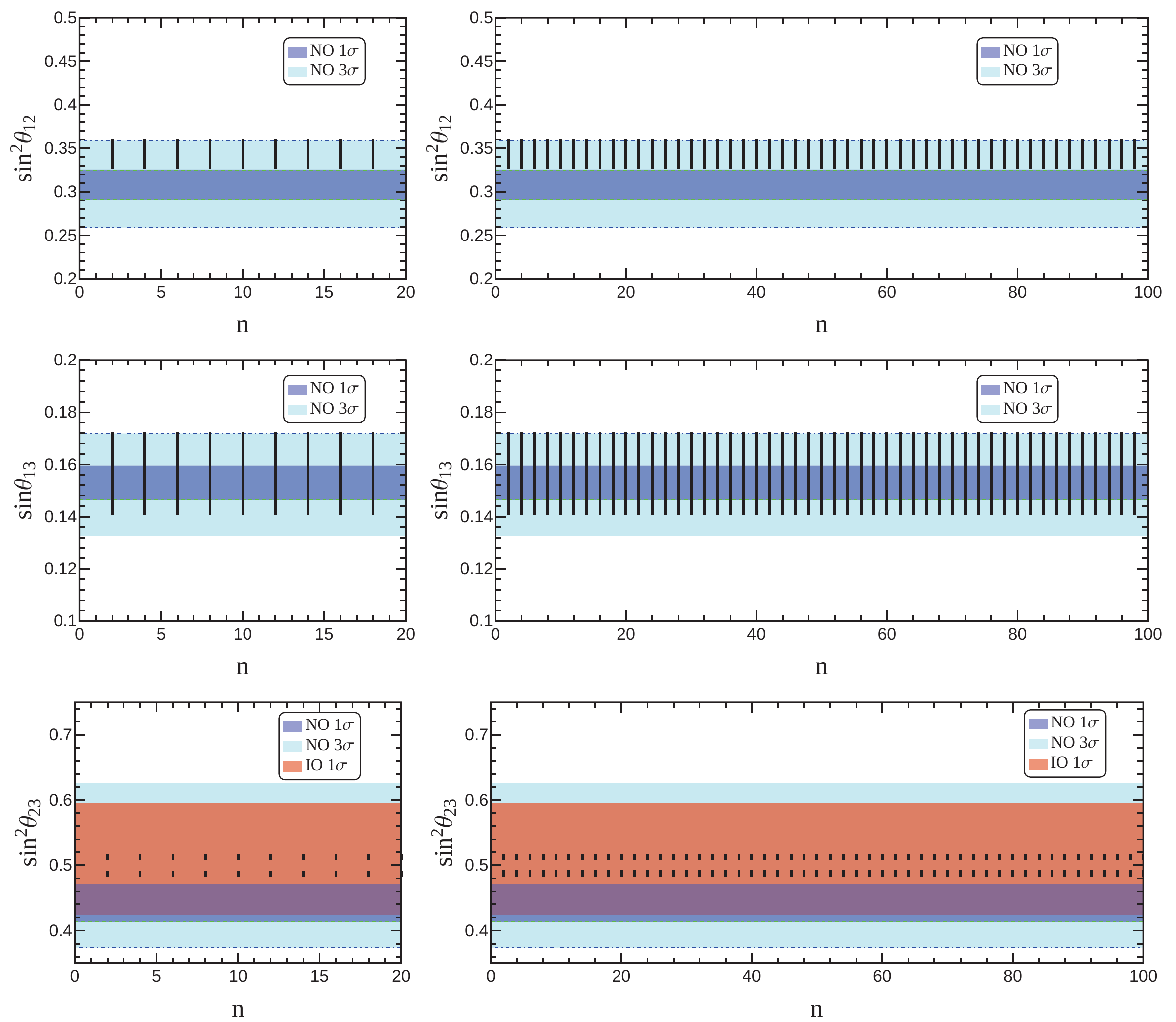}
\caption{\label{fig:caseXVII_angles_ranges}Numerical results in case V: the allowed ranges of $\sin^2\theta_{12}$, $\sin\theta_{13}$ and $\sin^2\theta_{23}$ for different $n$, where the three lepton mixing angles are required to lie in the $3\sigma$ regions. The $1\sigma$ and $3\sigma$ bounds of the mixing angles are taken from Ref.~\cite{Capozzi:2013csa}. Note that $n$ should be divisible by 2 in this case.}
\end{center}
\end{figure}

\begin{figure}[hptb!]
\begin{center}
\includegraphics[width=0.99\textwidth]{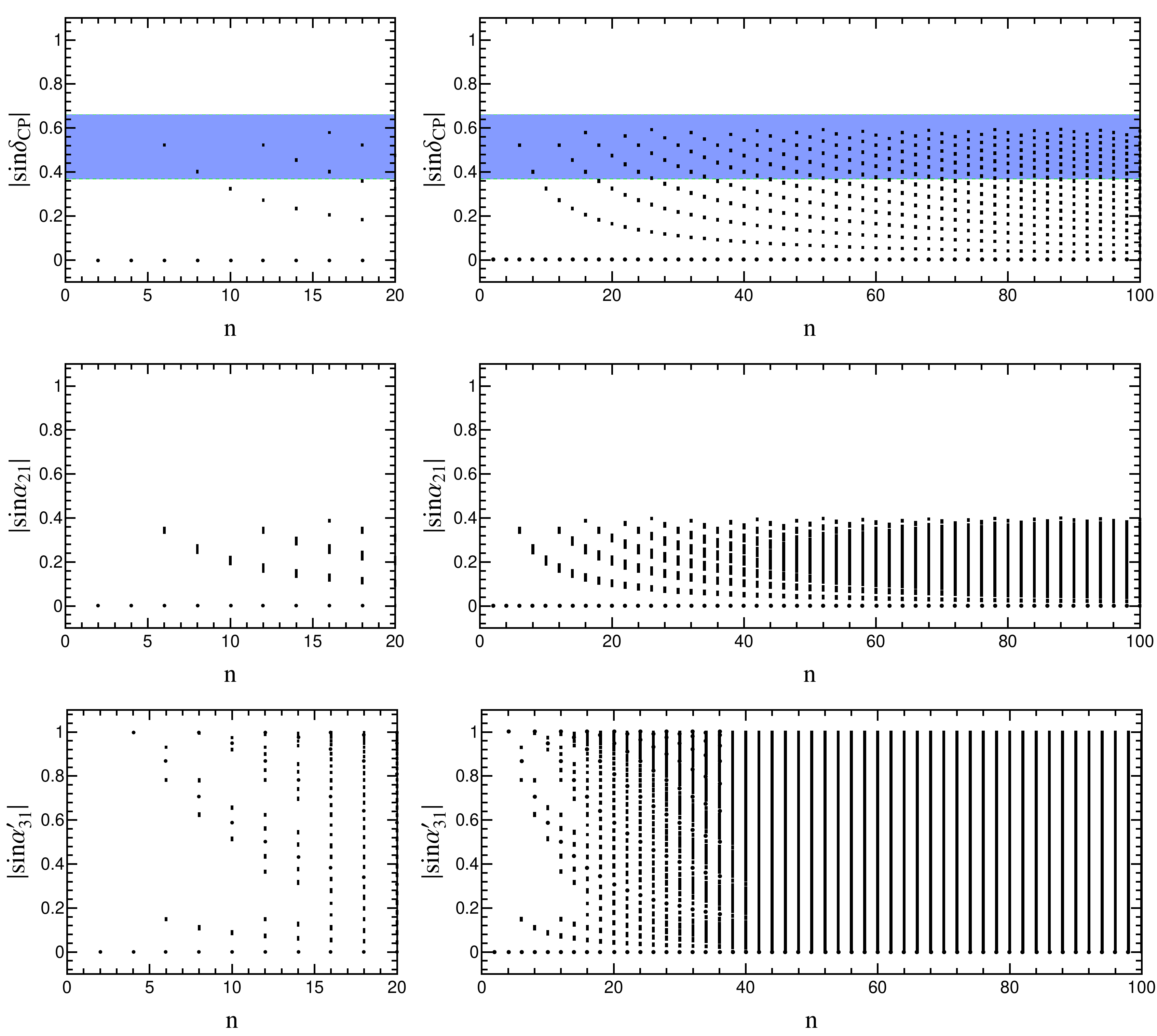}
\caption{\label{fig:caseXVII_phases_ranges}Numerical results in case V: the allowed ranges of $\left|\sin\delta_{CP}\right|$, $\left|\sin\alpha_{21}\right|$ and $\left|\sin\alpha^{\prime}_{31}\right|$ for different $n$, where the three lepton mixing angles are required to lie in the $3\sigma$ regions. The $1\sigma$ and $3\sigma$ bounds of the mixing angles are taken from Ref.~\cite{Capozzi:2013csa}. Note that $n$ should be divisible by 2 in this case. }
\end{center}
\end{figure}

\begin{figure}[hptb!]
\begin{center}
\includegraphics[width=0.90\textwidth]{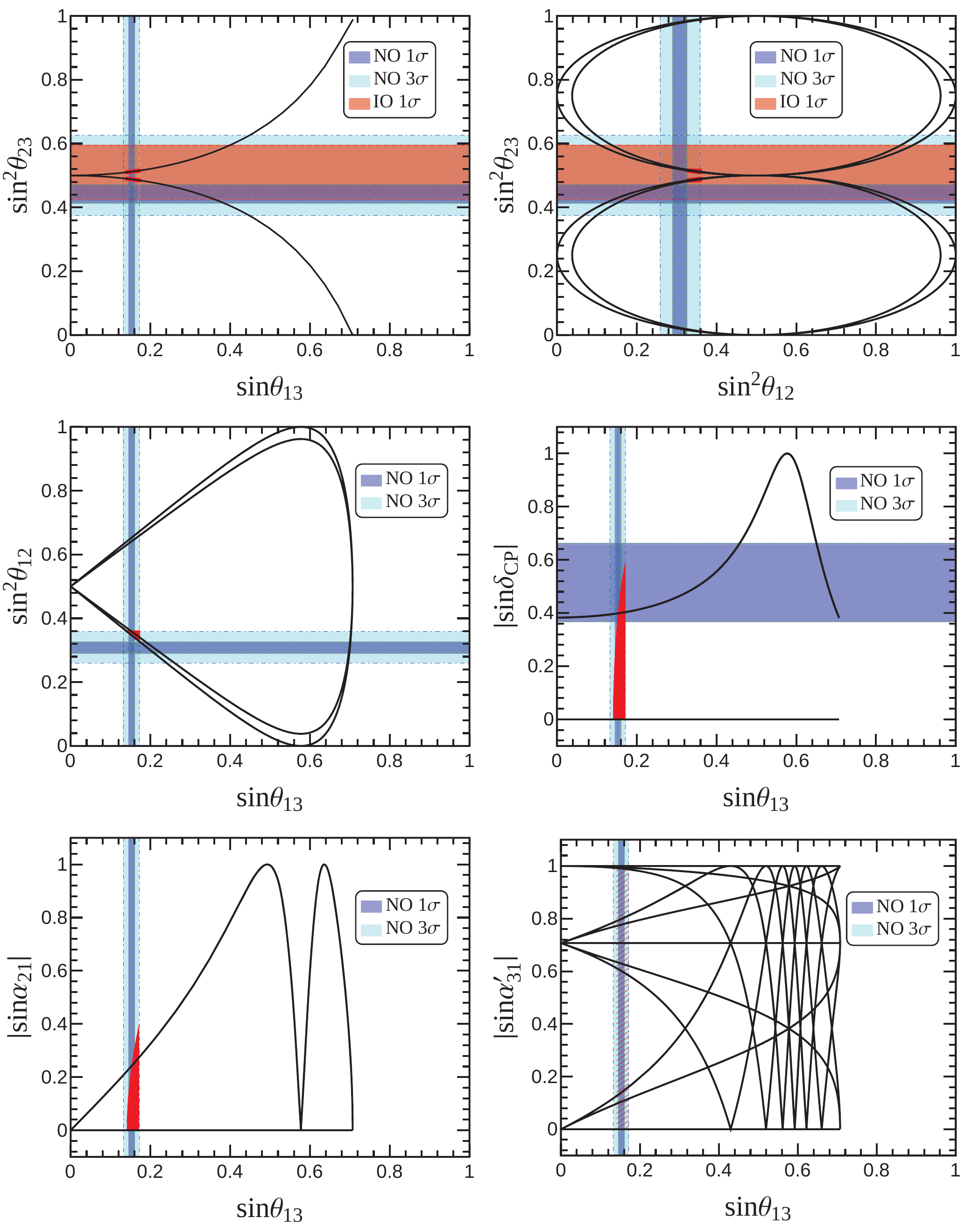}
\caption{\label{fig:caseXVII}The correlations among mixing parameters in case V. The red filled regions denote the allowed values of the mixing parameters if we take the parameters $\varphi_8$ and $\varphi_9$ to be continuous (which is equivalent to taking the limit $n\rightarrow\infty $) and the three mixing angles are required to lie in their $3\sigma$ ranges. Note that the Majorana phase $\alpha^{\prime}_{31}$ is not constrained in this limit. The black curves represent the phenomenologically viable correlations for $n=8$. The $1\sigma$ and $3\sigma$ bounds of the mixing parameters are taken from Ref.~\cite{Capozzi:2013csa}.}
\end{center}
\end{figure}

\end{description}

\section{\label{sec:nubeta}Neutrinoless Double-Beta Decay}

\begin{figure}
\begin{center}
\begin{tabular}{cc}
\includegraphics[width=0.45\linewidth]{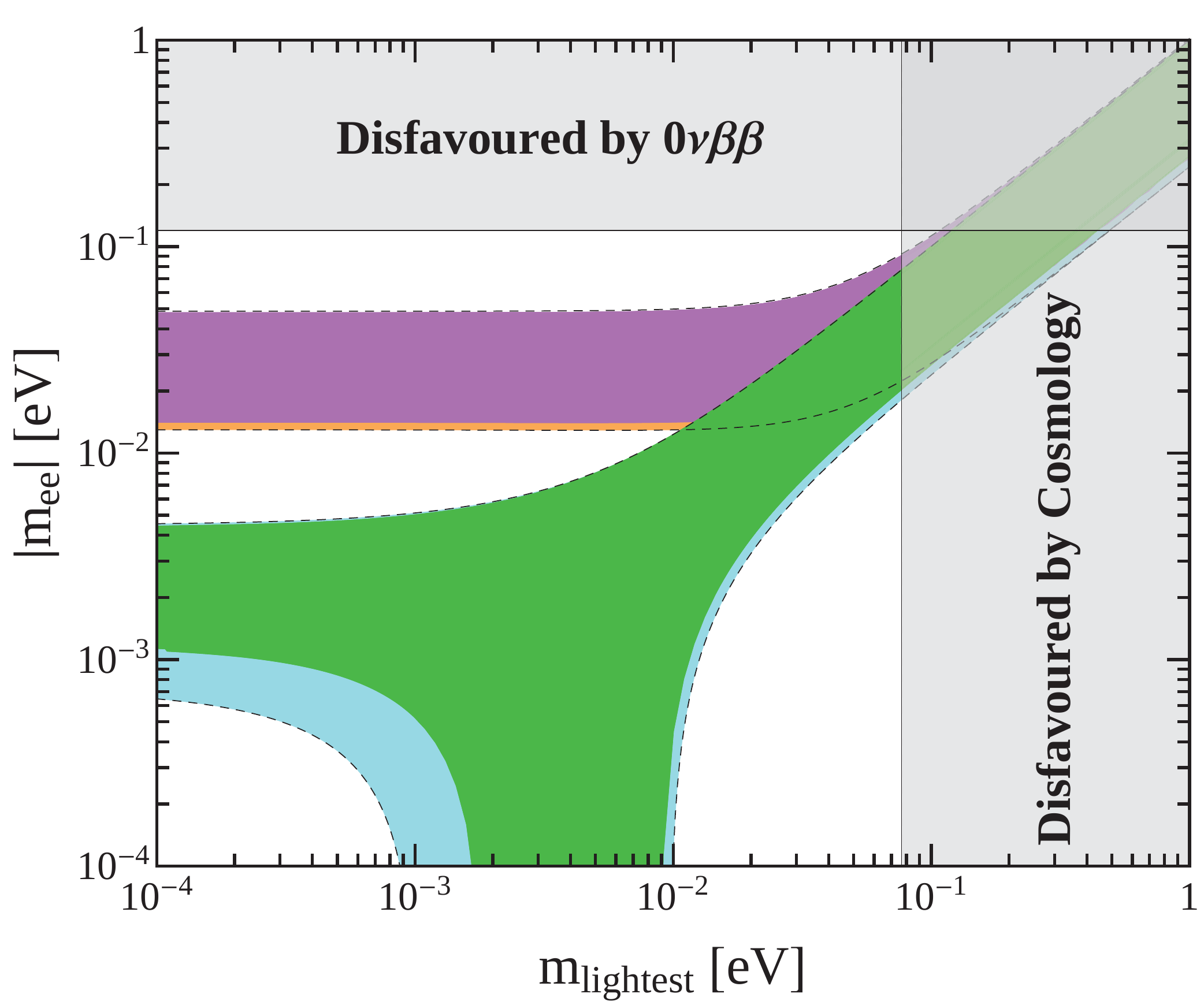} &
\includegraphics[width=0.45\linewidth]{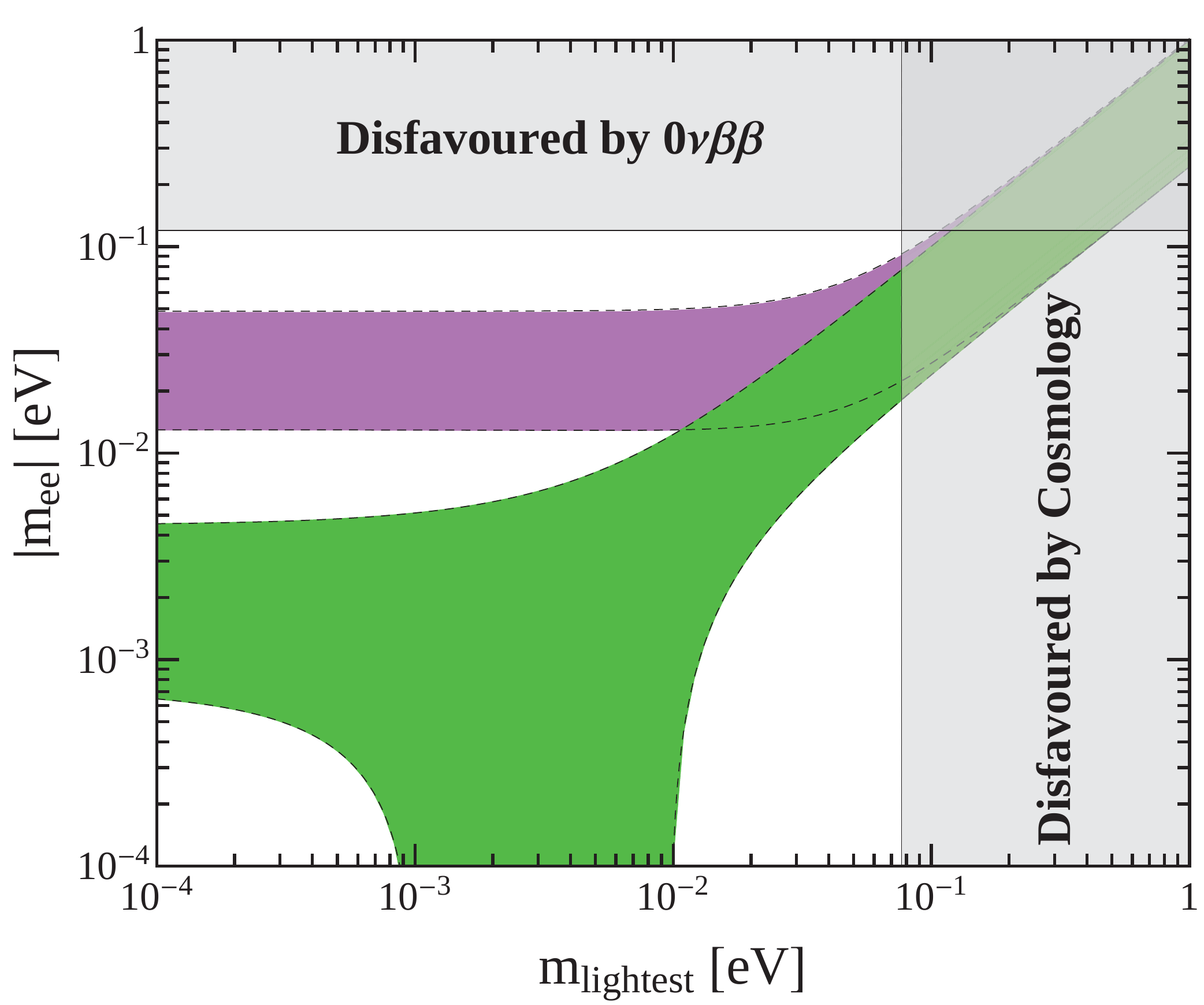} \\
\includegraphics[width=0.45\linewidth]{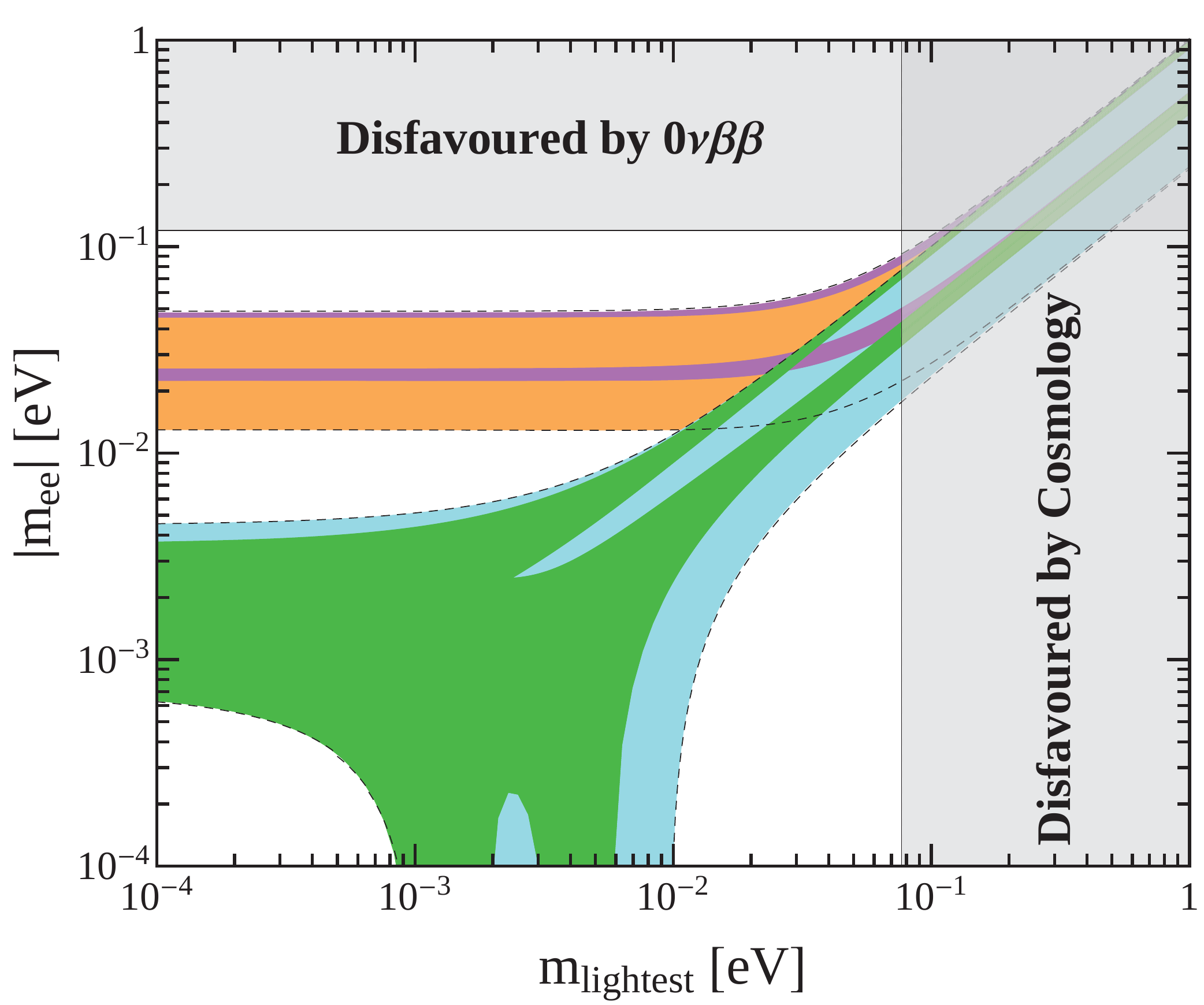} &
\includegraphics[width=0.45\linewidth]{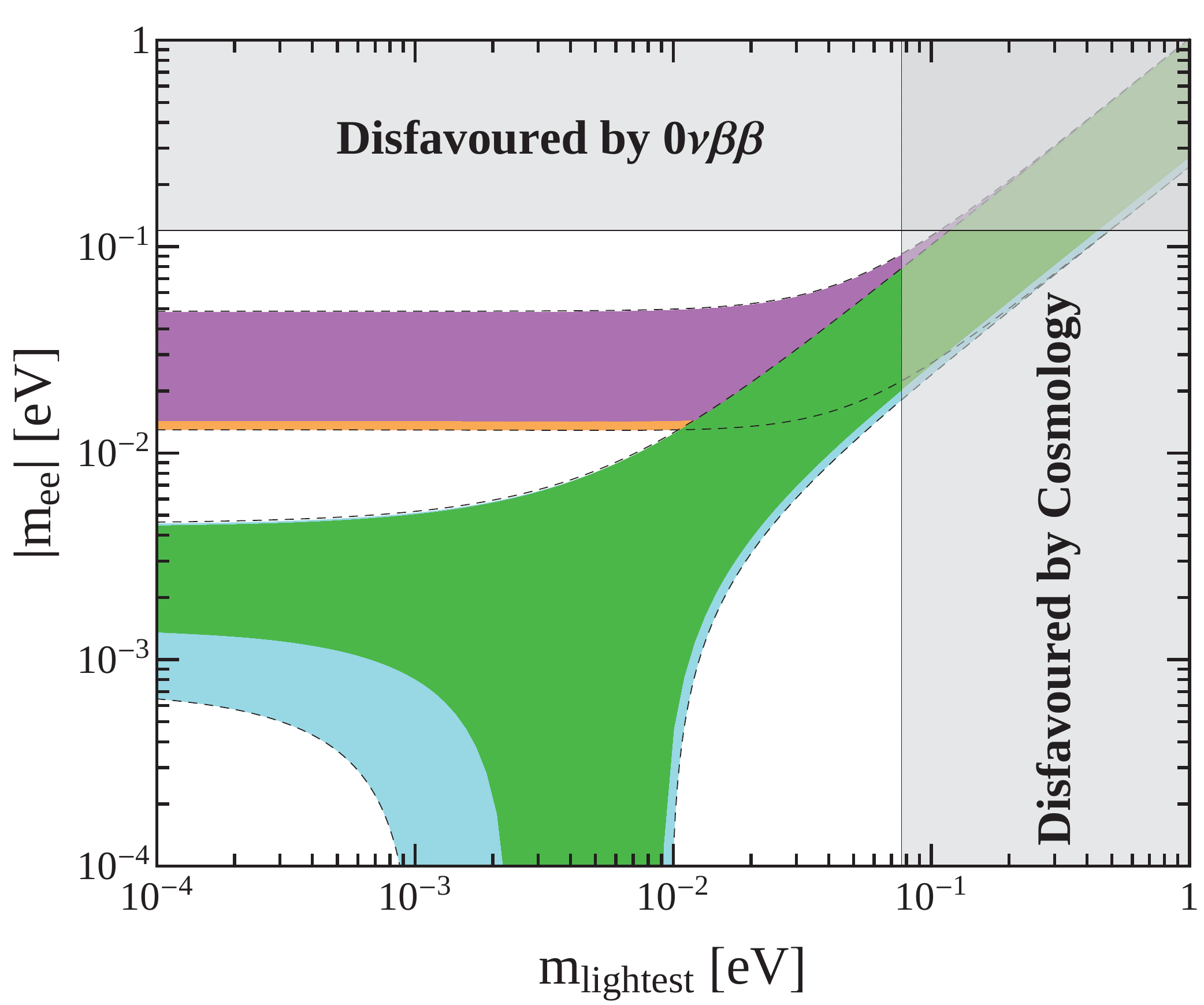} \\
\includegraphics[width=0.45\linewidth]{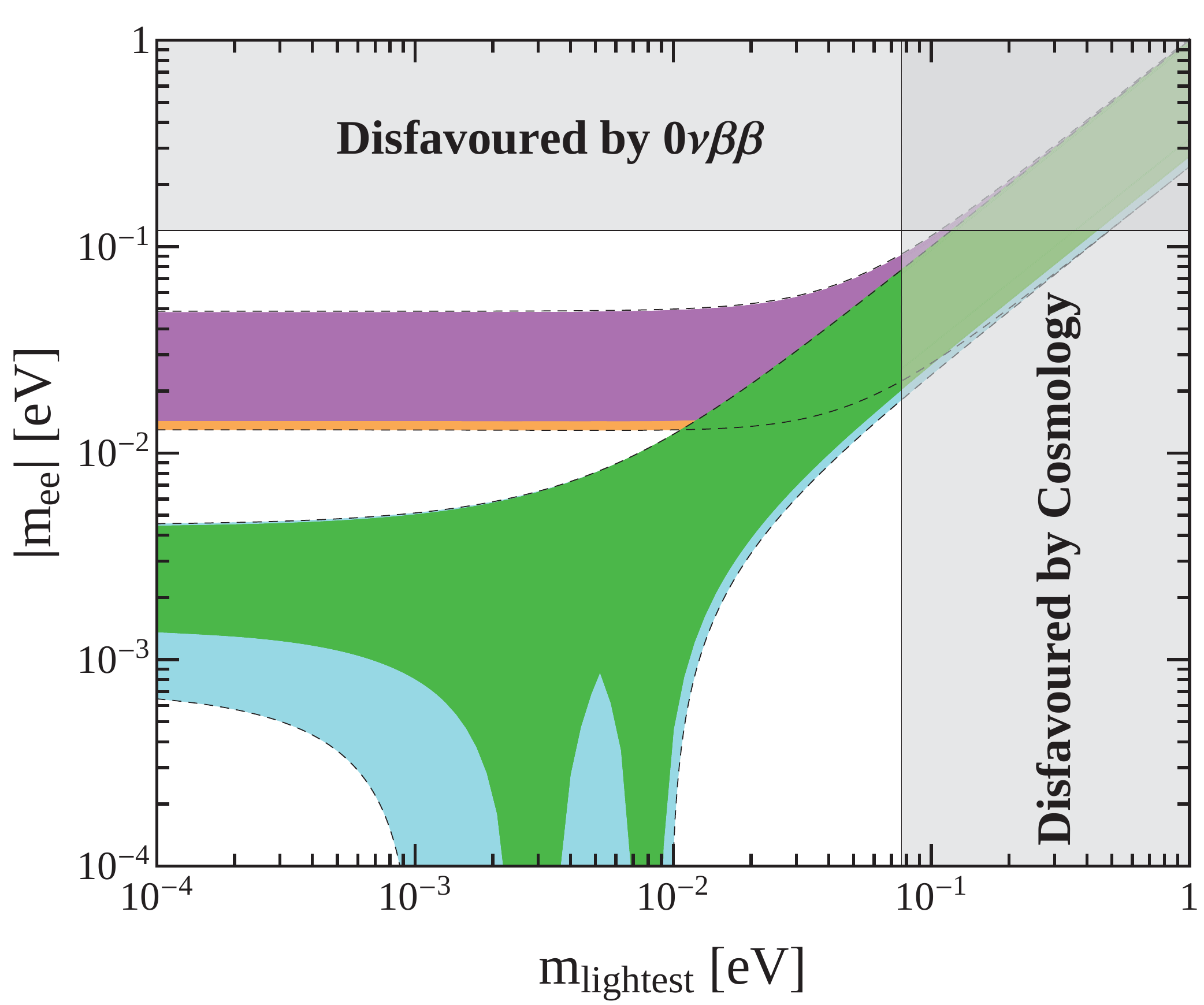} &
\includegraphics[width=0.45\linewidth]{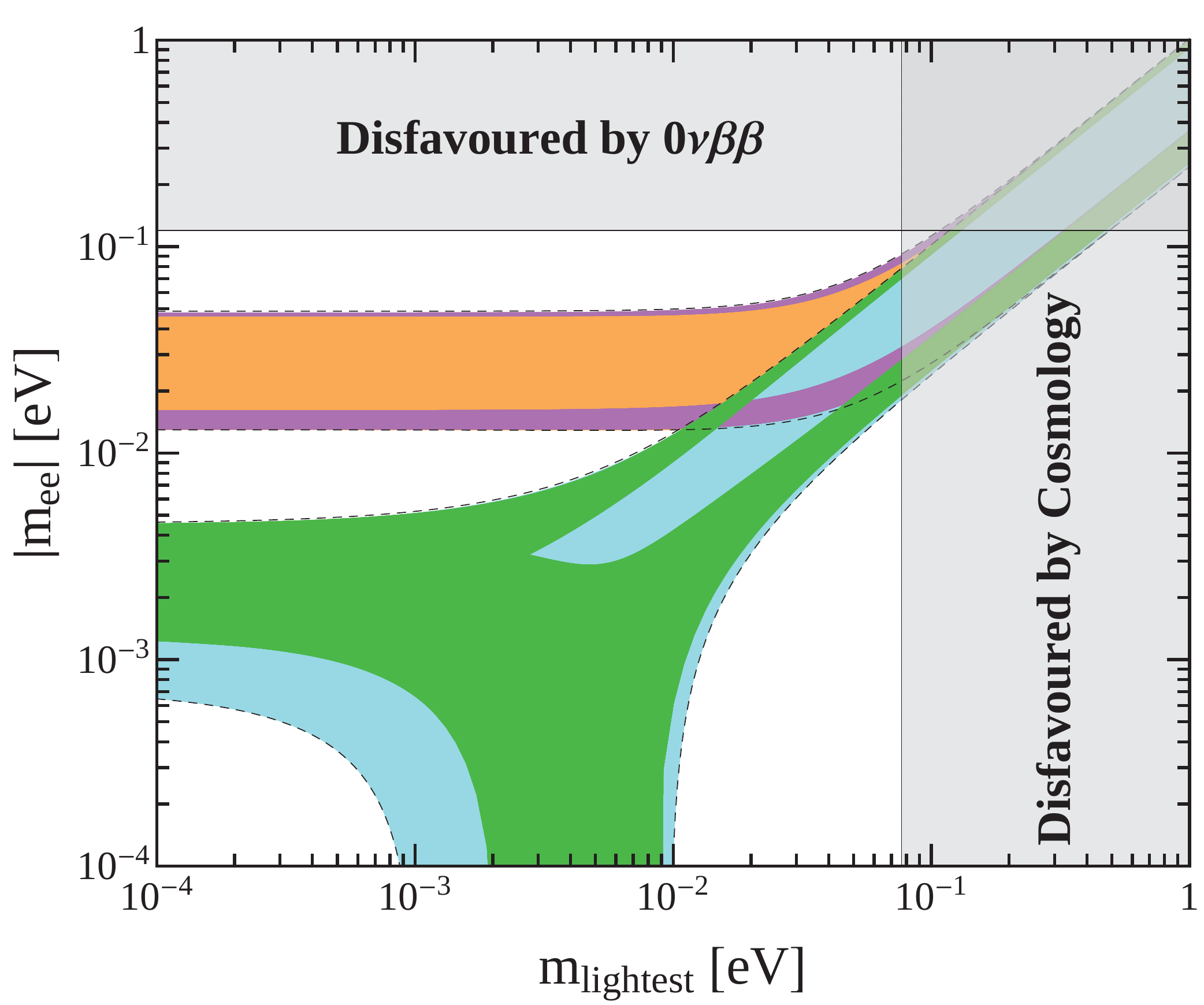}
\end{tabular}
\caption{\label{fig:meeplots1}The allowed ranges of the effective mass for neutrinoless double-beta decay for all viable cases of lepton mixing in semidirect models with a $\Delta(6n^2)$ flavour group in the limit of $n\rightarrow\infty$. The top row corresponds to case I, with 1st-3rd ordering on the left and 7th to 9th ordering on the right, the middle row contains case II and III, and the bottom row case IV and V. Light blue and yellow areas indicate the currently allowed three sigma region for normal and inverted hierarchy, respectively. Purple regions correspond to predictions assuming inverted hierarchy, green regions to normal hierarchy.
The upper bound $|m_{ee}|<0.120$ eV is given by measurements by the EXO-200~\cite{AugerAR,Albert:2014awa} and KamLAND-ZEN experiments~\cite{Gando:2012zm}. Planck data in combination with other CMB and BAO measurements~\cite{AdeZUV} provides a limit on the sum of neutrino masses of $m_1+m_2+m_3<0.230$ eV from which the upper limit on the mass of the lightest neutrino can be derived.}
\end{center}
\end{figure}

\begin{figure}
\begin{center}
\begin{tabular}{cc}
\includegraphics[width=0.44\linewidth]{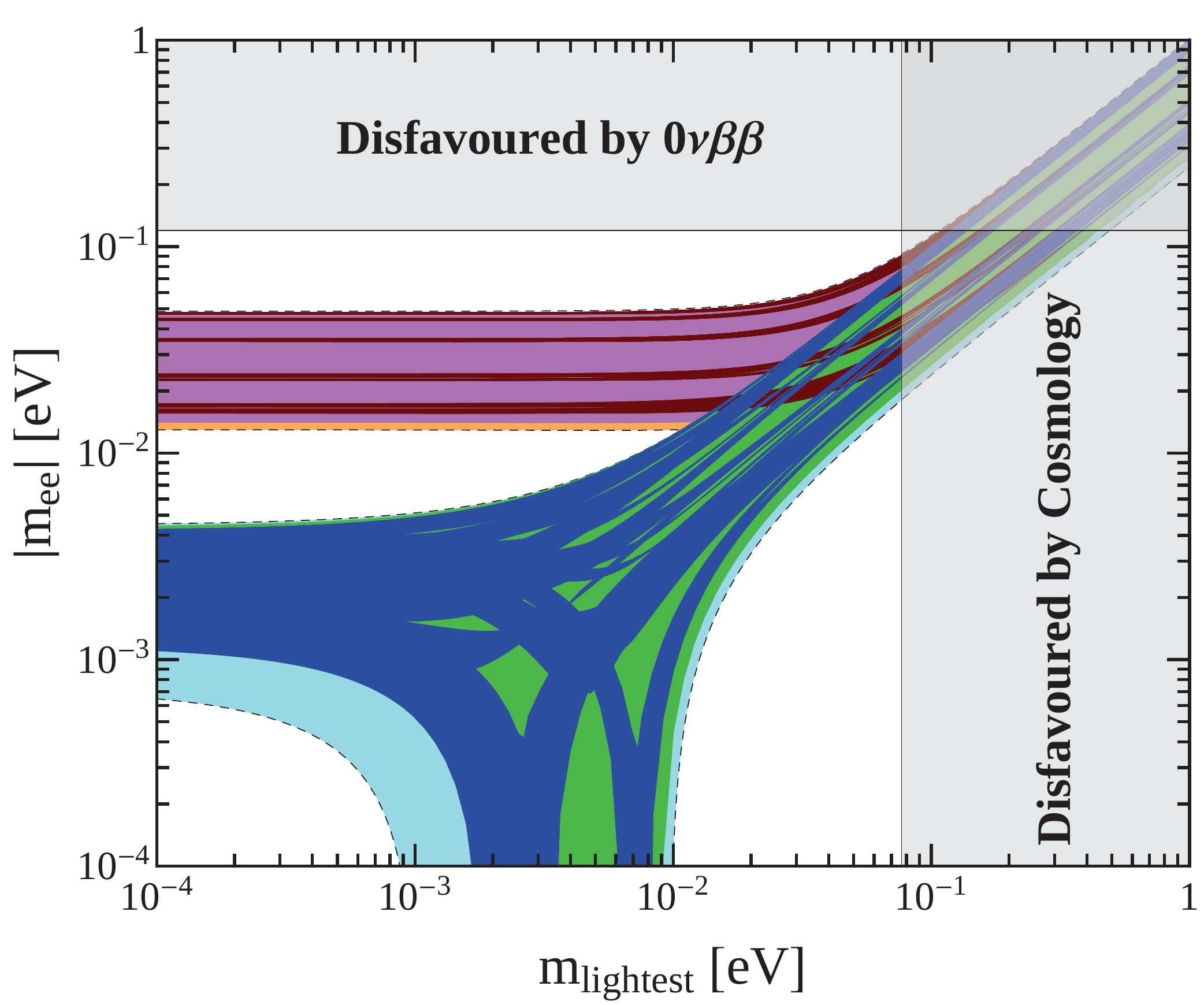} &
\includegraphics[width=0.44\linewidth]{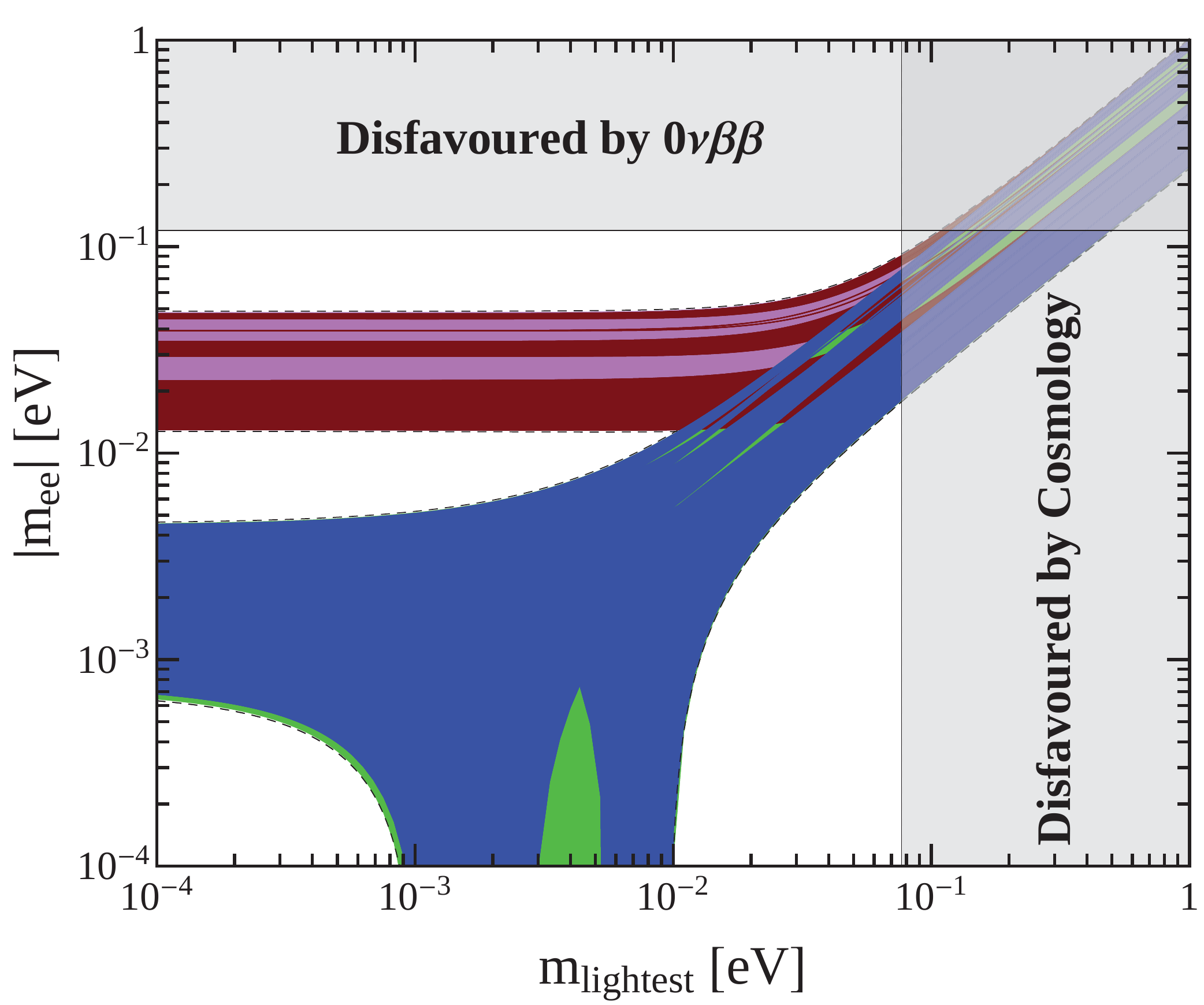} \\
\includegraphics[width=0.44\linewidth]{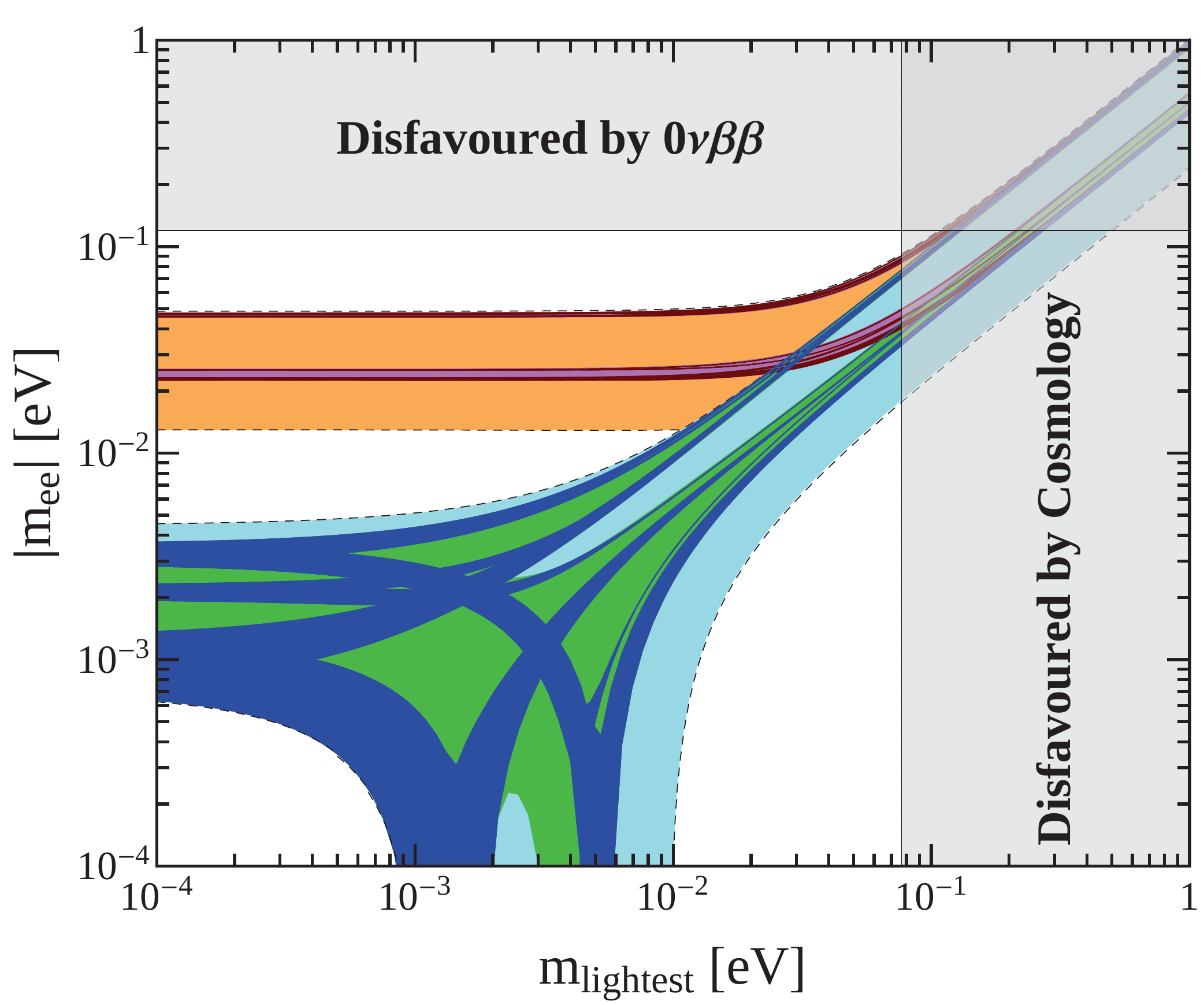} &
\includegraphics[width=0.44\linewidth]{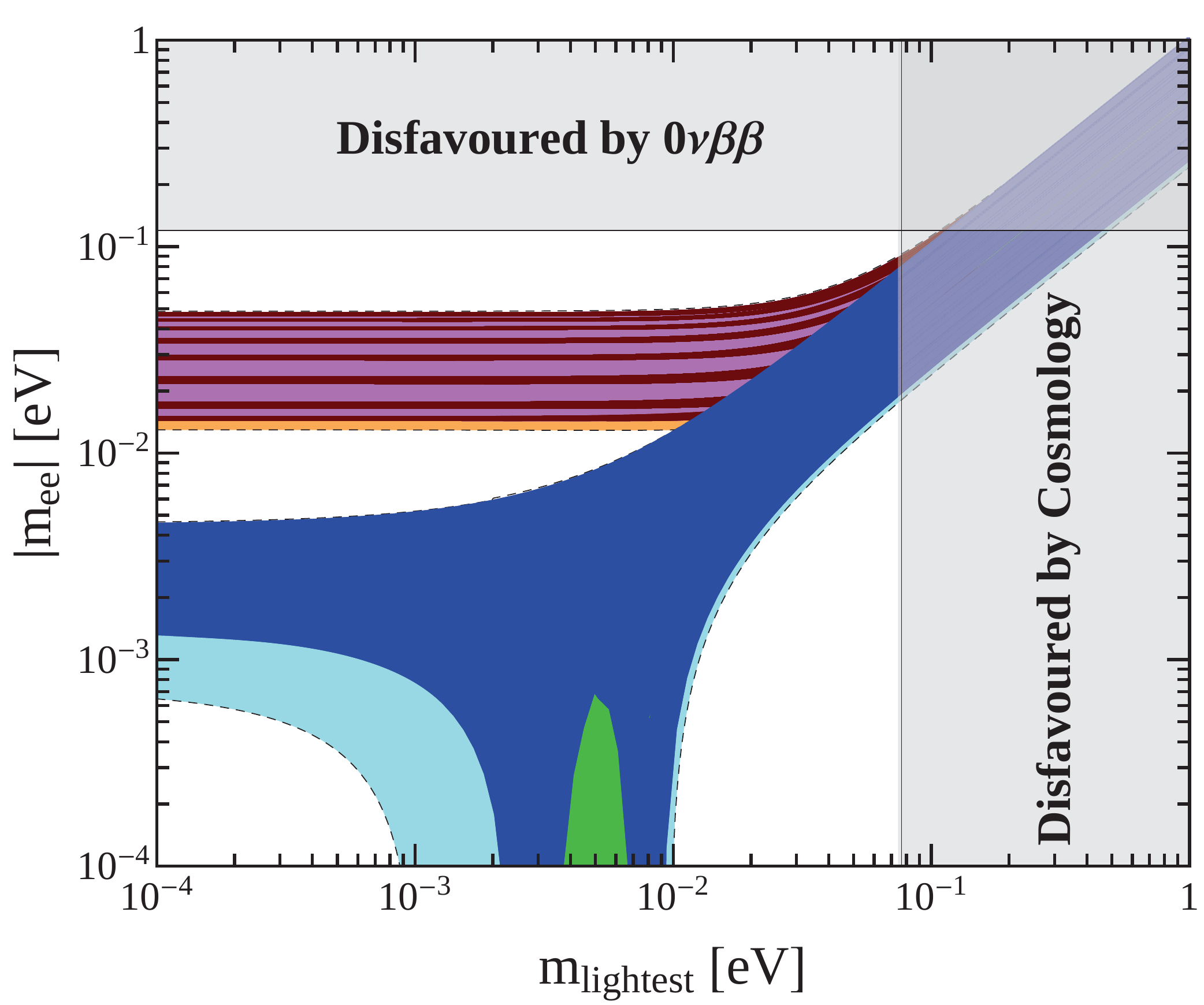} \\
\includegraphics[width=0.44\linewidth]{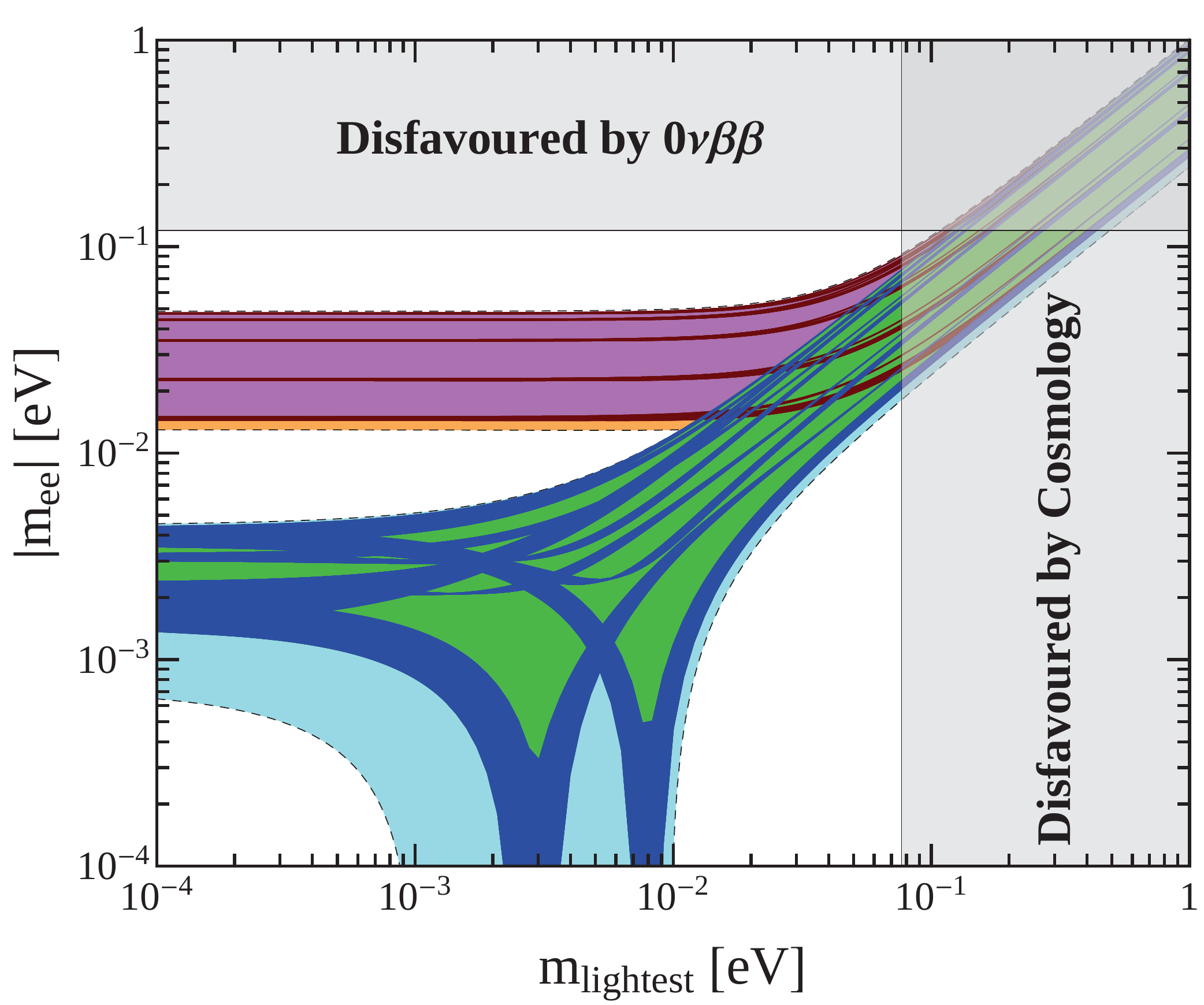} &
\includegraphics[width=0.44\linewidth]{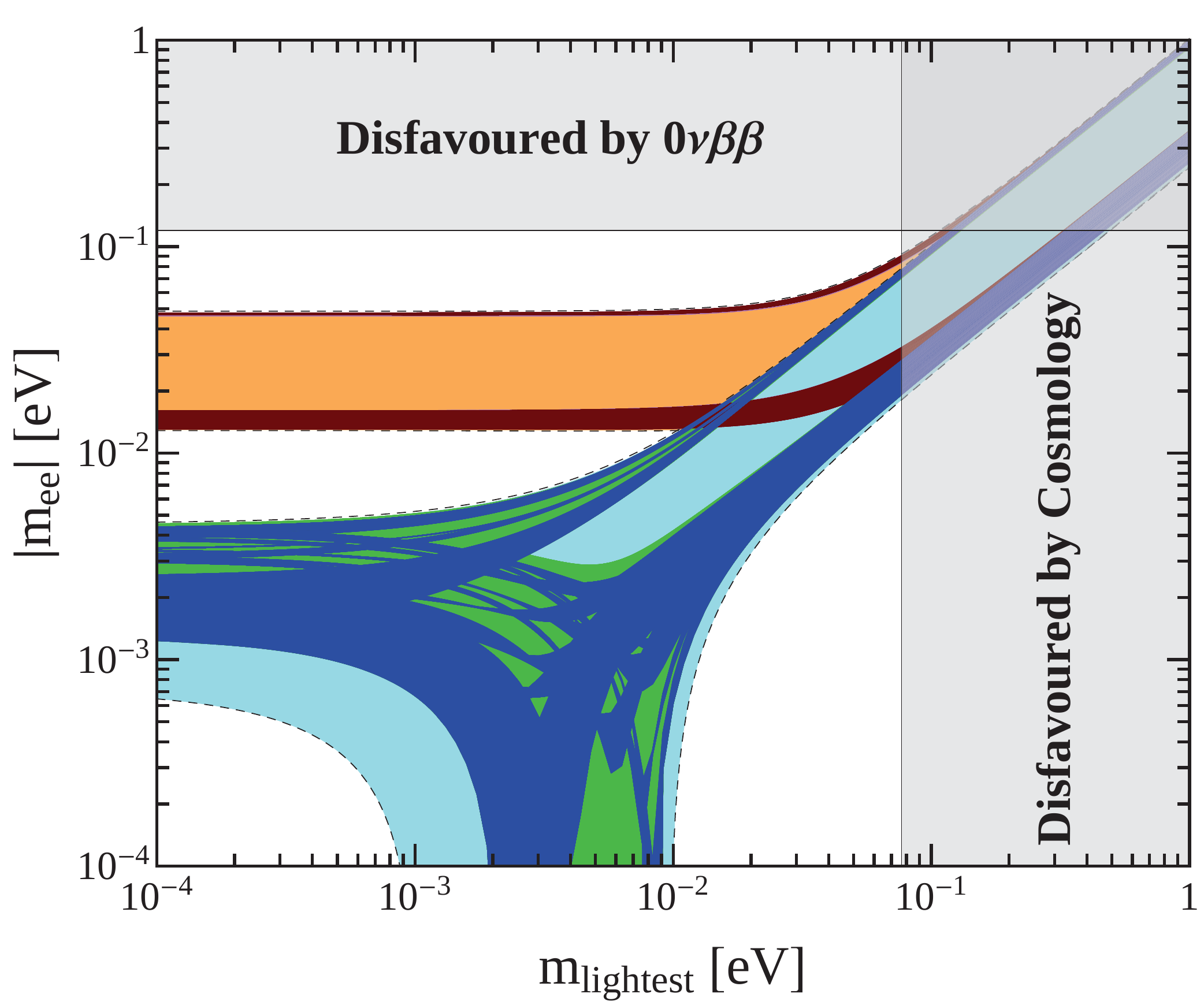}
\end{tabular}
\caption{\label{fig:meeplots2}The allowed ranges of the effective mass for neutrinoless double-beta decay for all viable cases of lepton mixing in semidirect models with a $\Delta(6n^2)$ flavour group. The top row corresponds to case I, with 1st-3rd ordering on the left and 7th to 9th ordering on the right, the middle row contains case II and III, and the bottom row case IV and V. Light blue and yellow areas indicate the currently allowed three sigma region for normal and inverted hierarchy, respectively. Purple regions correspond to predictions assuming inverted hierarchy, green regions to normal hierarchy in the limit of $n\rightarrow\infty$. Blue and red regions represent predictions for normal and inverted hierarchy for the value $n=8$ (in the top-right panel, we choose $n=5$ which is the smallest viable value of $n$ in that case). The upper bound $|m_{ee}|<0.120$ eV is given by measurements by the EXO-200~\cite{AugerAR,Albert:2014awa} and KamLAND-ZEN experiments~\cite{Gando:2012zm}. Planck data in combination with other CMB and BAO measurements~\cite{AdeZUV} provides a limit on the sum of neutrino masses of $m_1+m_2+m_3<0.230$ eV from which the upper limit on the mass of the lightest neutrino can be derived.}
\end{center}
\end{figure}

The rare process, neutrinoless double-beta decay ($0\nu2\beta$),is an important probe for the Majorana nature of neutrino and lepton number violation, a sizable number of new experiments are currently running, under construction, or in the planing phase. The effective mass of neutrinoless double-beta decay is~\cite{pdg}
\begin{equation}
\left|m_{ee}\right|=\left|(m_1 c_{12}^2+m_2s^2_{12}e^{i \alpha_{21}})c_{13}^2+m_3 s_{13}^2 e^{i \alpha_{31}'}\right|
\label{mee}
\end{equation}
For normal hierarchy, the masses are
\begin{equation}
 m_1=m_l,\quad  m_2=\sqrt{m_l^2+\delta m^2},  \quad m_3=\sqrt{m_l^2+\Delta m^2+\delta m^2/2}\,,
\end{equation}
and for inverted hierarchy
\begin{equation}
m_1=\sqrt{m_l^2-\Delta m^2-\delta m^2/2},\quad m_2=\sqrt{m_l^2-\Delta m^2+\delta m^2/2},\quad m_3=m_l\,,
\end{equation}
where $m_{l}$ denotes the lightest neutrino masses, and $\delta m^2\equiv m^2_2-m^2_1$ and $\Delta m^2\equiv m^2_3-(m^2_1+m^2_2)/2$ defined in Ref.~\cite{Capozzi:2013csa}. The experimental error on the neutrino mass splitting is not taken into account during the analysis, instead the best fit values from~\cite{Capozzi:2013csa} are used:
\begin{equation}
\delta m^2=7.54\times10^{-5}\textrm{eV}^2,\qquad \Delta m^2=2.43\times 10^{-3}(-2.38\times 10^{-3})\textrm{eV}^2\,,
\end{equation}
for normal (inverted) hierarchy. In the following, the properties of the effective mass are examined for all viable cases of lepton mixing discussed in this paper. In Fig.~\ref{fig:meeplots1} the allowed ranges of the effective mass are shown for each case in the limit of $n\rightarrow\infty$, where the three mixing angles are required to lie in the measured $3\sigma$ intervals~\cite{Capozzi:2013csa}(As previously mentioned, the $3\sigma$ lower bound of $\sin^2\theta_{12}$ is chosen to be 0.254 instead of 0.259 in case II). Furthermore, the predictions for the representative value $n=8$ ($n=5$ in case I for the 7th-9th ordering) are plotted in Fig.~\ref{fig:meeplots2} in  order to be read easily. The results for any finite value of $n$ must be part of the shown one corresponding to $n\rightarrow\infty$. Moreover, the plotting would change only a little bit if the experimental errors on $\delta m^2$ and $\Delta m^2$ are taken into account. Note that only one distinct prediction for the effective mass arises except in case I. One reason for this is that, as discussed before, many of the possible permutations of the mixing matrix can be identified with each other. Furthermore, permuting the second and third row has no effect on the effective mass as $\theta_{23}$ does not appear in Eq.~\eqref{mee}.

As shown in Fig.~\ref{fig:meeplots1}, for inverted hierarchy neutrino mass spectrum, almost all the $3\sigma$ range values of the effective masses $\left|m_{ee}\right|$ can be reproduced in the limit $n\rightarrow\infty$ in case I, case III and case IV. However, the predictions for $\left|m_{ee}\right|$ are around the upper bound (about 0.05eV) or lower bound (about 0.013 eV) in case V. The reason is that the solar mixing angle is in a narrow region $0.328\leq\sin^2\theta_{12}\leq0.359$ and the Majorana phase $\alpha_{21}$ is constrained to be $\left|\sin\alpha_{21}\right|\leq0.586$ in this case, as displayed in Fig.~\ref{fig:caseXVII_angles_ranges} and Fig.~\ref{fig:caseXVII_phases_ranges}. Similarly $\left|m_{ee}\right|$ is near the upper bound and 0.025 eV in case II. Therefore if the effective mass is measured to be far from 0.013 eV, 0.025 eV and 0.05 eV for inverted hierarchy by future experiments, the mixing patterns in cases II and V could be ruled out.

For normal hierarchy neutrino mass spectrum, a sizable part of the experimentally allowed $3\sigma$ region of $\left|m_{ee}\right|$ can be generated in all cases, and the effective mass could be rather small. In particular, the prediction in case I with 7th to 9th ordering approximately coincides with the present $3\sigma$ region. 
Unfortunately the predictions for normal hierarchy are still out of reach of projected experiments known to the author. As a result, it is should be generally difficult to test the $\Delta(6n^2)$ family symmetry and generalized CP symmetry through neutrinoless double beta decay experiments in the case of normal ordering spectrum.

\section{\label{sec:conclusions}Conclusions}
\cleqn

We have performed a detailed analysis of $\Delta (6n^2)$ family symmetry combined with the generalised CP symmetry $H_{\rm{CP}}$ in the lepton sector. We have investigated the lepton mixing parameters which can be obtained from the original symmetry $\Delta (6n^2)\rtimes H_{\rm{CP}}$ breaking to different remnant symmetries in the neutrino and charged lepton sectors, namely $G_{\nu}$ and $G_l$ subgroups in the neutrino and the charged lepton sector respectively, while the remnant CP symmetries from the breaking of $H_{\rm{CP}}$ are $H^{\nu}_{\rm{CP}}$ and $H^{l}_{\rm{CP}}$, respectively.

We have assumed a preserved symmetry smaller than the full Klein symmetry, as in the semi-direct approach, leading to predictions which depend on a single undetermined real parameter, which mainly controls the reactor angle. We have discussed the resulting mass and mixing predictions for all possible cases where the $\Delta (6n^2)$ family symmetry with generalised CP is broken to $G_{\nu}=Z_2$ with $G_{l}=K_4,Z_p,p>2$ and $G_{\nu}=K_4$ with $G_{l}=Z_2$. We have focused on five phenomenologically allowed cases and have presented the resulting predictions for the PMNS parameters as a function of $n$, as well as the predictions for neutrinoless double beta decay.

It is remarkable that the CP phases are predicted to take irregular values rather than 0, $\pi$ or $\pm\pi/2$. In particular, compared to $\Delta (6n^2)$ in the direct approach where the full Klein symmetry is identified as a subgroup, the result $|\sin \delta_{\rm{CP}}|=0$ corresponding to the Dirac CP phase being either zero or $\pm \pi$ is relaxed in the indirect approach followed here. Compared to the indirect approach to $S_4$ (which corresponds in our notation to $n=2$) where $|\sin \delta_{\rm{CP}}|=1$ is the only available possibility for non-zero CP violation, for larger values of $n$ we find many different possibilities for the oscillation phase, where the number of possible predictions grows rapidly as $n$ is increased. However it is interesting that, even for very large $n$, the predictions for the PMNS parameters is bounded within certain ranges, providing non-trivial tests of the semi-direct approach followed here. It remains to be seen if any of these possibilities will closely correspond to the observed future precise determination of leptonic mixing angles and CP
violating parameters in the future.

\vskip0.4in

\section*{Acknowledgements}

The research was partially supported by the National Natural Science
Foundation of China under Grant Nos. 11275188 and 11179007. SK acknowledges
support from the STFC Consolidated ST/J000396/1 grant. TN and SK acknowledge support from the European Union FP7 ITN-INVISIBLES (Marie Curie Actions, PITN- GA-2011- 289442). One of the author (G.J.D.) is grateful to Chang-Yuan Yao for his kind help on plotting the figures in this paper.

\appendix

\cleqn

\section{\label{sec:Appd_group_theory}The group theory of $\Delta(6n^2)$}
\cleqn
\indent

$\Delta(6n^2)$ is non-abelian finite subgroup of $SU(3)$. The  $\Delta(6n^2)$ is
isomorphic to $(Z_n\times Z_n)\rtimes S_3$, where $S_3$ is isomorphic to
$Z_3\rtimes Z_2$, and it can be conveniently defined by four generators $a$,
$b$, $c$ and $d$ obeying the relations~\cite{Escobar:2008vc}:
\begin{eqnarray}
\nonumber&a^3=b^2=(ab)^2=1,\\
\nonumber&c^n=d^n=1,\qquad cd=dc\,,\\
\label{eq:abcd_generators}&aca^{-1}=c^{-1}d^{-1},\quad ada^{-1}=c,\quad bcb^{-1}=d^{-1}, \quad bdb^{-1}=c^{-1}\,.
\end{eqnarray}
The elements $a$ and $b$ are the generators of $S_3$ while
$c$ and $d$ generate $Z_n\times Z_n$, and the last line defines the
semidirect product $``\rtimes"$. Note that the generator $d=bc^{-1}b^{-1}$
is not independent. All the group elements can be written into the form
\begin{equation}
g=a^{\alpha}b^{\beta}c^{\gamma}d^{\delta}\,,
\end{equation}
where $\alpha=0,1,2$, $\beta=0,1$, $\gamma, \delta=0,1,2,\ldots n-1$. In the following we list the elements of $\Delta(6n^2)$ by order of the generated cyclic subgroup.
\begin{itemize}

 \item Elements of order 2, if $n$ even:
 \begin{equation}
 c^{n/2},d^{n/2},c^{n/2}d^{n/2}
 \end{equation}
\item Elements of order 2, always:
\begin{equation}
bc^\epsilon d^\epsilon, abc^\gamma, a^2 b d^\delta
\end{equation}
with $\epsilon, \gamma, \delta=0,\ldots,n-1$.

\item Elements of order 3, if 3 divides $n$:
\begin{equation}
 c^{n/3},d^{n/3},\ldots
\end{equation}
where the dots indicate all possible products and powers of the two first elements.

\item Elements of order 3, always:
\begin{equation}
ac^\gamma d^\delta, a^2 c^\gamma d^\delta
\end{equation}
with $\gamma,\delta=0,\ldots,n-1$

\item Elements of order $m$ where $m$ divides $n$, if $m$ and $n$ are even:
\begin{equation}
bc^{\delta+2kn/m}d^\delta,abc^\gamma d^{2kn/m},a^2 b c^{2kn/m} d^\delta
\label{orderm1}
\end{equation}
with $\gamma,\delta=0,\ldots,n-1$ and $0\leq k \leq m/2$

\item Elements of order $m$ where $m$ divides $n$, always:
\begin{equation}
c^{kn/m}d^{ln/m}
\end{equation}
with $k,l=0,\ldots,n-1$.
\end{itemize}

The $\Delta(6n^2)$ group have been thoroughly studied in Ref.~\cite{Escobar:2008vc}. In the following, we shall review the basic aspects, which is relevant to our present work. The conjugacy classes of $\Delta(6n^2)$ group are of the following forms:
\begin{itemize}
\item{$n\neq3\, \mathbb{Z}$}

\small\begin{subequations}
\begin{eqnarray}
 1 & :& 1C_1=\left\{1\right\},\\
 n-1 & :&  3C_{1}^{(\rho)}=
\left\{c^{\rho}d^{-\rho}, c^{-2\rho}d^{-\rho},
c^{\rho}d^{2\rho}
\right \},
~~~~ \rho=1,2,...,n-1,
\\
\label{eq:conjugcy_class_rs1}\dfrac{n^2-3n+2}{6}& :&
6C_{1}^{(\rho,\sigma)}=\left \{
c^{\rho}d^{\sigma}, c^{\sigma-\rho}d^{-\rho},
c^{-\sigma}d^{\rho-\sigma},
c^{-\sigma}d^{-\rho},
c^{\sigma-\rho}d^{\sigma},
c^{\rho}d^{\rho-\sigma}
\right \},\\
1 &:& 2n^2C_{2}= \left\{ac^{z}d^{y},~
a^2c^{-y}d^{-z}|z,y=0,1,\ldots,n-1\right\},\\
 &&\hskip-1.4in n:
 3nC_{3}^{(\rho)}=\left\{bc^{\rho+x}d^{x},~a^2bc^{-\rho}d^{-x-\rho},
 ~abc^{-x}d^{\rho}|x=0,1,\ldots,n-1\right\}, \rho=0,1,\ldots,n-1.
\end{eqnarray}
\end{subequations}
The convention used here is that the quantity left of the
colon is the number of classes of the kind on the right of
the colon. In Eq.~\eqref{eq:conjugcy_class_rs1}, the parameter
$\rho,\sigma =0,1,...,n-1$, but excluding possibilities given by
\begin{equation}\label{eq:cond}
\rho+\sigma=~0~ \textrm{mod}~ n, \quad
2\rho-\sigma=~0~ \textrm{mod}~ n, \quad
\rho-2\sigma=~0~ \textrm{mod}~ n \,.
\end{equation}
\item{$n=3\, \mathbb{Z}$}

\small\begin{subequations}
\begin{eqnarray}
 1 & :& 1C_1=\left\{1\right\},\\
 2 &:& 1C_{1}^{(\nu)} =
\left \{
c^{\nu}d^{2\nu}
\right \},
\quad \nu=\mbox{$\frac{n}{3},\frac{2n}{3}$},
\\
 n-3  &:&  3C_{1}^{(\rho)}=
\left \{
c^{\rho}d^{-\rho}, c^{-2\rho}d^{-\rho},
c^{\rho}d^{2\rho}
\right \},
~~
\rho\neq
\mbox{$\frac{n}{3},\frac{2n}{3}$},
\\
\label{eq:conjugcy_class_rs2} \dfrac{n^2-3n+6}{6}&:&
6C_{1}^{(\rho,\sigma)}=
\left \{
c^{\rho}d^{\sigma}, c^{\sigma-\rho}d^{-\rho},
c^{-\sigma}d^{\rho-\sigma},
c^{-\sigma}d^{-\rho},
c^{\sigma-\rho}d^{\sigma},
c^{\rho}d^{\rho-\sigma}
\right \},\\
\nonumber &&\hskip-1.0in 3: \frac{2n^2}{3}C_{2}^{(\tau)}=\{ac^{\tau-y-3x}
d^{y},
  a^2c^{-y}d^{y+3x-\tau}|y=0,1,...,n-1,x=0,1,\ldots,\mbox{$\frac{n-3}{3}$}\},
  \tau=0,1,2\,,\\
 &&\hskip-1.0in n:
 3nC_{3}^{(\rho)}=\left\{bc^{\rho+x}d^{x},a^2bc^{-\rho}d^{-x-\rho},
 abc^{-x}d^{\rho}|x=0,1,\ldots,n-1\right\}, \rho=0,1,\ldots,n-1.
\end{eqnarray}
\end{subequations}
In Eq.~\eqref{eq:conjugcy_class_rs2}, $\rho,\sigma =0,1,...,n-1$, again
excluding possibilities given by Eq.~\eqref{eq:cond}.

\end{itemize}

The irreducible representations and their representation matrices of the $\Delta(6n^2)$ group are as follows~\cite{Escobar:2008vc}:
\begin{description}
  \item[\textbf{(i)}] {$n\neq3\, \mathbb{Z}$}

  \begin{itemize}[labelindent=-0.7em, leftmargin=0.1em]
   \item{One-dimensional representations}
  \begin{subequations}
  \begin{eqnarray}
  \mathbf{1_{1}} &:& a=b=c=d=1,\\
  \mathbf{1_{2}} &:& a=c=d=1, ~b=-1,
 \end{eqnarray}
  \end{subequations}

  \item{Two-dimensional representation}

 \begin{equation}\label{eq:twodimrep1}
 \mathbf{2}: a=\frac{1}{2}\begin{pmatrix}-1 & ~-\sqrt{3} \\
    \sqrt{3} &  ~-1~ \end{pmatrix},~~
 b=\begin{pmatrix}1 & 0\\ 0 &
           ~-1~ \end{pmatrix},~~
 c=d= \begin{pmatrix}1 & 0 \\ 0 &
           ~1~ \end{pmatrix}\,,
\end{equation}
which is related to the basis chosen in Ref.~\cite{Escobar:2008vc} by
a unitary transformation $U$ with
\begin{equation}
\label{eq:EL2our}U=\frac{1}{\sqrt{2}}\begin{pmatrix}
1  &  i  \\
1  & -i
\end{pmatrix}\,.
\end{equation}
In our new basis, all the Clebsch-Gordan (CG) coefficients are real, as is shown in the Appendix~\ref{sec:appendix_CG_coefficients}. Hence our basis is the so-called the ``CP'' basis. The conventional CP transformation $\varphi\rightarrow \varphi^{*}$ can be consistently imposed onto the theory in our basis, and all the coupling constant would be constrained to be real.

\item{Three-dimensional representations}

\begin{subequations}
\begin{eqnarray}
&&\hskip-0.45in \mathbf{3}_{1,k}: a=\begin{pmatrix}0 &1 &0 \\ 0&0&1 \\
   1&0&0\end{pmatrix},~
   b=\begin{pmatrix} 0 &0 &1 \\ 0&1&0 \\
   1&0&0\end{pmatrix},~
   c=\begin{pmatrix} \eta^{k}&0 &0 \\ 0&\eta^{-k}&0 \\
   0&0&1\end{pmatrix},~
   d=\begin{pmatrix}1 &0 &0 \\ 0&\eta^{k}&0 \\
   0&0&\eta^{-k}\end{pmatrix}, \\
 &&\hskip-0.45in\mathbf{3}_{2,k}: a=\begin{pmatrix}0 &1 &0 \\ 0&0&1 \\
   1&0&0\end{pmatrix},~
   b=-\begin{pmatrix} 0 &0 &1 \\ 0&1&0 \\
   1&0&0\end{pmatrix},~
   c=\begin{pmatrix} \eta^{k}&0 &0 \\ 0&\eta^{-k}&0 \\
   0&0&1\end{pmatrix},~
   d=\begin{pmatrix}1 &0 &0 \\ 0&\eta^{k}&0 \\
   0&0&\eta^{-k}\end{pmatrix},
\end{eqnarray}
\end{subequations}
where $\eta\equiv e^{2\pi i/n}$ and $k=1,2,\ldots n-1$.

\item{Six-dimensional representations}

\begin{equation}\label{eq:6drep}
\mathbf{6}_{\widetilde{(k,l)}} ~:~ a = \begin{pmatrix} a_{1}& 0 \\  0 &
a_{2} \end{pmatrix},~~~~
b = \begin{pmatrix} 0    & {\mathbb{1}_3} \\  {\mathbb{1}_3}  & 0
\end{pmatrix},~~~~
c = \begin{pmatrix} c_{1} & 0 \\ 0 & c_{2}
\end{pmatrix},~~~~
d = \begin{pmatrix} d_{1} & 0 \\ 0 & d_{2}
\end{pmatrix}\ ,
\end{equation}
with
\begin{eqnarray}\label{eq:6drep2}
 a_{1} =
    \begin{pmatrix}
    0 & 1 & 0 \\  0 & 0 & 1 \\ 1 & 0 & 0
    \end{pmatrix},~~~~
    &
    a_{2} =
    \begin{pmatrix}
    0 & 0 & 1 \\  1 & 0 & 0 \\ 0 & 1 & 0
    \end{pmatrix},~~~~
    \\
    c_{1} = d_{2}^{-1}=
    \begin{pmatrix}
    \eta^{l}& 0 & 0 \\ 0 & \eta^{k} & 0 \\ 0 & 0 & \eta^{-l-k}
    \end{pmatrix},~~~~
    &
    c_{2} = d_{1}^{-1}=
    \begin{pmatrix}
    \eta^{l+k} & 0 & 0 \\ 0 & \eta^{-l} & 0 \\ 0 & 0 & \eta^{-k}
    \end{pmatrix}~
    \ .
\end{eqnarray}
Here $\widetilde{\phantom{wa}}$ denotes the mapping
\begin{equation}
\widetilde{\begin{pmatrix} k \\ l \end{pmatrix}} ~\longmapsto
~\mathrm{either} ~
\begin{pmatrix} k \\ l \end{pmatrix}, ~~
\begin{pmatrix} -k-l \\ k \end{pmatrix}, ~~
\begin{pmatrix} l \\ -k-l \end{pmatrix}, ~~
\begin{pmatrix} -l \\ -k \end{pmatrix}, ~~
\begin{pmatrix} k+l \\ -l \end{pmatrix},  ~\mathrm{or} ~~
\begin{pmatrix} -k \\ k+l \end{pmatrix}\label{mapping}
\,,
\end{equation}
$k,l=0,1,\ldots n-1$, and the following cases are forbidden.
\begin{equation}
 l=0,\quad k=0,\quad k+l=0~ \mathrm{mod}~n\,.
\end{equation}

\end{itemize}

\item[\textbf{(ii)}]{$n=3\, \mathbb{Z}$}

   \begin{itemize}[labelindent=-0.7em, leftmargin=0.1em]
   \item{One-dimensional representations}
  \begin{subequations}
  \begin{eqnarray}
  \mathbf{1_{1}} &:& a=b=c=d=1,\\
  \mathbf{1_{2}} &:& a=c=d=1, ~b=-1,
  \end{eqnarray}
  \end{subequations}

  \item{Two-dimensional representation}

\begin{subequations}
\begin{eqnarray}\label{eq:2dimb}
&&\hskip-0.3in\mathbf{2_1}: a=\frac{1}{2}\begin{pmatrix}-1 & ~-\sqrt{3} \\
    \sqrt{3} &  ~-1~ \end{pmatrix},~~
b=\begin{pmatrix}1 & 0\\ 0 &
           ~-1~ \end{pmatrix},~~c=d=\begin{pmatrix}1 & ~0 \\
   0 &  ~1~ \end{pmatrix},\\
&&\hskip-0.3in\mathbf{2_2}: a=\frac{1}{2}\begin{pmatrix}-1 & ~-\sqrt{3} \\
    \sqrt{3} &  ~-1~ \end{pmatrix},~~
b=\begin{pmatrix}1 & 0\\ 0 &
           ~-1~ \end{pmatrix},~~c=d=\frac{1}{2}\begin{pmatrix}-1 &
           ~\sqrt{3} \\
    -\sqrt{3} &  ~-1~ \end{pmatrix},\\
&&\hskip-0.3in \mathbf{2_3}: a =\frac{1}{2}\begin{pmatrix}-1 & ~-\sqrt{3} \\
    \sqrt{3} &  ~-1~ \end{pmatrix},~~b=\begin{pmatrix}1 & 0\\ 0 &
           ~-1~ \end{pmatrix},~~c=d=\frac{1}{2}\begin{pmatrix}-1 &
           ~-\sqrt{3} \\
    \sqrt{3} &  ~-1~ \end{pmatrix},\\
&&\hskip-0.3in \mathbf{2_4}:  a =\begin{pmatrix}1 & 0\\ 0 &
           ~1~ \end{pmatrix},~~
~b=\begin{pmatrix}1 & 0\\ 0 &
           ~-1~ \end{pmatrix},~~c=d=\frac{1}{2}\begin{pmatrix}-1 &
           ~-\sqrt{3} \\
    \sqrt{3} &  ~-1~ \end{pmatrix}.
\end{eqnarray}
\end{subequations}
They are related to the representation matrices of
Ref.~\cite{Escobar:2008vc} by the unitary transformation $U$ in
Eq.~\eqref{eq:EL2our}.

\item{Three-dimensional representations}

\begin{subequations}
\begin{eqnarray}
&&\hskip-0.50in\mathbf{3}_{1,k}: a=\begin{pmatrix}0 &1 &0 \\ 0&0&1 \\
   1&0&0\end{pmatrix},~
   b=\begin{pmatrix} 0 &0 &1 \\ 0&1&0 \\
   1&0&0\end{pmatrix},~
   c=\begin{pmatrix} \eta^{k}&0 &0 \\ 0&\eta^{-k}&0 \\
   0&0&1\end{pmatrix},~
   d=\begin{pmatrix}1 &0 &0 \\ 0&\eta^{k}&0 \\
   0&0&\eta^{-k}\end{pmatrix}, \\
&&\hskip-0.50in \mathbf{3}_{2,k}: a=\begin{pmatrix}0 &1 &0 \\ 0&0&1 \\
   1&0&0\end{pmatrix},~
   b=-\begin{pmatrix} 0 &0 &1 \\ 0&1&0 \\
   1&0&0\end{pmatrix},~
   c=\begin{pmatrix} \eta^{k}&0 &0 \\ 0&\eta^{-k}&0 \\
   0&0&1\end{pmatrix},~
   d=\begin{pmatrix}1 &0 &0 \\ 0&\eta^{k}&0 \\
   0&0&\eta^{-k}\end{pmatrix},
\end{eqnarray}
\end{subequations}
where $k=1,2,\ldots n-1$.

\item{Six-dimensional representations}

\begin{equation}\label{eq:6drep}
\mathbf{6}_{\widetilde{(k,l)}} ~:~ a = \begin{pmatrix} a_{1}& 0 \\  0 &
a_{2} \end{pmatrix},~~~~
b = \begin{pmatrix} 0    & {\mathbb{1}_3} \\  {\mathbb{1}_3}  & 0
\end{pmatrix},~~~~
c = \begin{pmatrix} c_{1} & 0 \\ 0 & c_{2}
\end{pmatrix},~~~~
d = \begin{pmatrix} d_{1} & 0 \\ 0 & d_{2}
\end{pmatrix}\,.
\end{equation}
The $3\times3$ unitary matrices $a_{1,2}$, $c_{1,2}$ and $d_{1,2}$ are
given in Eq.~\eqref{eq:6drep2}. Again the following values are prohibited:
\begin{equation}
l=0,\quad k=0,\quad k=l=n/3,\quad k=l=2n/3, \quad k+l=0 ~\mathrm{mod} ~n\,.
\end{equation}

\end{itemize}

\end{description}

\section{\label{sec:appendix_CG_coefficients}Clebsch-Gordan coefficients for
$\Delta(6n^2)$ group with $n\neq3\mathbb{Z}$}
\cleqn

In the following, we shall present all the CG coefficients in the form of
$x\otimes y$ in our chosen basis, $x_i$ denotes the element of the left base
vectors $x$, and $y_i$ is the element of the right base vectors $y$. We
shall see explicitly that all the CG coefficients are real.

\begin{itemize}

\item[$\bullet$]{${\mathbf2} \otimes {\mathbf 2}={\mathbf
	  1_1}\oplus {\mathbf 1_2} \oplus{\mathbf 2}$}

\begin{equation}
\mathbf{2}\sim\left(\begin{array}{c}
x_2y_2-x_1y_1 \\
x_1y_2+x_2y_1
\end{array}
\right),\quad \mathbf{1_1}\sim x_1y_1+x_2y_2,\quad \mathbf{1_2}\sim
x_1y_2-x_2y_1\,.
\end{equation}

\item[$\bullet$]{$\mathbf{2} \otimes \mathbf{3}_{1,k}=\mathbf{
    3}_{1,k}\oplus\mathbf{3}_{2,k}$}

\begin{equation}
\mathbf{3}_{1,k}\sim\left(\begin{array}{c}
\left(x_1-\sqrt{3}\,x_2\right)y_1\\
-2x_1y_2\\
\left(x_1+\sqrt{3}\,x_2\right)y_3
\end{array}
\right),\quad \mathbf{3}_{2,k}\sim\left(\begin{array}{c}
\left(\sqrt{3}\,x_1+x_2\right)y_1\\
-2x_2y_2\\
\left(-\sqrt{3}\,x_1+x_2\right)y_3
\end{array}
\right)\,.
\end{equation}

\item[$\bullet$]{$\mathbf{2} \otimes
    \mathbf{3}_{2,k}=\mathbf{3}_{1,k}\oplus\mathbf{3}_{2,k}$}

\begin{equation}
\mathbf{3}_{1,k}\sim\left(\begin{array}{c}
\left(\sqrt{3}\,x_1+x_2\right)y_1\\
-2x_2y_2\\
\left(-\sqrt{3}\,x_1+x_2\right)y_3
\end{array}
\right),\quad \mathbf{3}_{2,k}\sim\left(\begin{array}{c}
\left(x_1-\sqrt{3}\,x_2\right)y_1\\
-2x_1y_2\\
\left(x_1+\sqrt{3}\,x_2\right)y_3
\end{array}
\right)\,.
\end{equation}

\item[$\bullet$]{$\mathbf{2} \otimes
    \mathbf{6}_{(k,l)}=\mathbf{6}_{(k,l)}\oplus\mathbf{6}_{(k,l)}$}

\begin{equation}
\mathbf{6}_{(k,l)}\sim\left(\begin{array}{c}
\left(\sqrt{3}\,x_1+x_2\right)y_1\\
-2x_2y_2\\
\left(-\sqrt{3}\,x_1+x_2\right)y_3\\
\left(\sqrt{3}\,x_1-x_2\right)y_4\\
2x_2y_5\\
-\left(\sqrt{3}\,x_1+x_2\right)y_6
\end{array}
\right),\quad \mathbf{6}_{(k,l)}\sim\left(\begin{array}{c}
2x_2y_1\\
\left(\sqrt{3}\,x_1-x_2\right)y_2\\
-\left(\sqrt{3}\,x_1+x_2\right)y_3\\
-2x_2y_4\\
\left(\sqrt{3}\,x_1+x_2\right)y_5\\
\left(-\sqrt{3}\,x_1+x_2\right)y_6
\end{array}
\right)\,.
\end{equation}

\item[$\bullet$]{$\mathbf{3}_{1,l} \otimes
    \mathbf{3}_{1,l'}=\mathbf{3}_{1,l+l'}\oplus\mathbf{6}_{\widetilde{(l,-l')}}$}
	
\begin{equation}
\mathbf{3}_{1,l+l'}\sim
	 \begin{pmatrix}
	 x_1 y_1 \\
	 x_2 y_2 \\
	 x_3 y_3 \\
	 \end{pmatrix},\quad
	 \mathbf{6}_{(-l,l-l')}\sim
	 \begin{pmatrix}
	 x_1 y_2 \\
	 x_2 y_3 \\
	 x_3 y_1 \\
	 x_3 y_2 \\
	 x_2 y_1 \\
	 x_1 y_3
	 \end{pmatrix},
	 \end{equation}

\item[$\bullet$]{$\mathbf{3}_{1,l} \otimes
    \mathbf{3}_{2,l'}=\mathbf{3}_{2,l+l'}\oplus\mathbf{6}_{\widetilde{(l,-l')}}$}

\begin{equation}
\mathbf{3}_{2,l+l'}\sim\begin{pmatrix}
	 x_1 y_1 \\
	 x_2 y_2 \\
	 x_3 y_3 \\
	 \end{pmatrix},\quad \mathbf{6}_{(-l,l-l')}\sim	 \begin{pmatrix}
 x_1 y_2 \\
 x_2 y_3 \\
 x_3 y_1 \\
 -x_3 y_2 \\
 -x_2 y_1 \\
 -x_1 y_3
\end{pmatrix},
\end{equation}

\item[$\bullet$]{$\mathbf{3}_{1,l} \otimes \mathbf{6}_{(k',l')}=
\mathbf{6}_{\widetilde{\tiny{\begin{pmatrix}k'\\l'-l\end{pmatrix}}}}
\oplus
\mathbf{6}_{\tiny{\widetilde{\begin{pmatrix}k'-l\\l'+l\end{pmatrix}}}}
\oplus
\mathbf{6}_{\tiny{\widetilde{\begin{pmatrix}l+k'\\l'\end{pmatrix}}}}$}

\begin{equation}
\mathbf{6}_{\tiny{{\begin{pmatrix}l'-l\\l-k'-l'\end{pmatrix}}}}\sim
\begin{pmatrix}
 x_1 y_3 \\
 x_2 y_1 \\
 x_3 y_2 \\
 x_3 y_6 \\
 x_2 y_4 \\
 x_1 y_5
\end{pmatrix},\quad
\mathbf{6}_{{\tiny{\begin{pmatrix}k'-l\\l'+l\end{pmatrix}}}}\sim
\begin{pmatrix}
 x_1 y_1 \\
 x_2 y_2 \\
 x_3 y_3 \\
 x_3 y_4 \\
 x_2 y_5 \\
 x_1 y_6 \\
\end{pmatrix},\quad
\mathbf{6}_{\tiny{{\begin{pmatrix}-l-k'-l'\\l+k'\end{pmatrix}}}}\sim
\begin{pmatrix}
 x_1 y_2 \\
 x_2 y_3 \\
 x_3 y_1 \\
 x_3 y_5 \\
 x_2 y_6 \\
 x_1 y_4 \\
\end{pmatrix}\,.
\end{equation}

\item[$\bullet$]{$\mathbf{3}_{2,l}\otimes\mathbf{3}_{2,l'}=\mathbf{3}_{1,l+l'}\oplus\mathbf{6}_{\widetilde{(l,-l')}}$}

\begin{equation}
\mathbf{3}_{1,l+l'}\sim
\begin{pmatrix}
 x_1 y_1 \\
 x_2 y_2 \\
 x_3 y_3 \\
\end{pmatrix},\quad
\mathbf{6}_{(-l,l-l')}\sim
\begin{pmatrix}
 x_1 y_2 \\
 x_2 y_3 \\
 x_3 y_1 \\
 x_3 y_2 \\
 x_2 y_1 \\
 x_1 y_3
\end{pmatrix}\,.
\end{equation}

\item[$\bullet$]{$\mathbf{3}_{2,l}\otimes\mathbf{6}_{(k',l')}=\mathbf{6}_{\widetilde{\tiny{\begin{pmatrix}k'\\l'-l\end{pmatrix}}}}
\oplus\mathbf{6}_{\tiny{\widetilde{\begin{pmatrix}k'-l\\l'+l\end{pmatrix}}}}\oplus\mathbf{6}_{\tiny{\widetilde{\begin{pmatrix}l+k'\\l'\end{pmatrix}}}}$}

\begin{equation}
\mathbf{6}_{\tiny{{\begin{pmatrix}l'-l\\l-k'-l'\end{pmatrix}}}}\sim
\begin{pmatrix}
 x_1 y_3 \\
 x_2 y_1 \\
 x_3 y_2 \\
 -x_3 y_6 \\
 -x_2 y_4 \\
 -x_1 y_5
\end{pmatrix},\quad
\mathbf{6}_{{\tiny{\begin{pmatrix}k'-l\\l'+l\end{pmatrix}}}}\sim
\begin{pmatrix}
 x_1 y_1 \\
 x_2 y_2 \\
 x_3 y_3 \\
 -x_3 y_4 \\
 -x_2 y_5 \\
 -x_1 y_6 \\
\end{pmatrix},\quad
\mathbf{6}_{\tiny{{\begin{pmatrix}-l-k'-l'\\l+k'\end{pmatrix}}}}\sim
\begin{pmatrix}
 x_1 y_2 \\
 x_2 y_3 \\
 x_3 y_1 \\
 -x_3 y_5 \\
 -x_2 y_6 \\
 -x_1 y_4 \\
\end{pmatrix}\,.
\end{equation}

\item[$\bullet$]{$\mathbf{6}_{(k,l)}\otimes\mathbf{6}_{(k',l')}=\sum_{p,s}\mathbf{6}_{\widetilde{\tiny{\left(
\begin{pmatrix}k\\l\end{pmatrix}+M^p_s
\begin{pmatrix}k'\\l'\end{pmatrix}\right )}}}$}

\begin{eqnarray}
\nonumber&&
\mathbf{6}_{\tiny{{\begin{pmatrix}k+k'\\l+l'\end{pmatrix}}}}\sim
\begin{pmatrix}
 x_1 y_1 \\
 x_2 y_2 \\
 x_3 y_3 \\
 x_4 y_4 \\
 x_5 y_5 \\
 x_6 y_6
\end{pmatrix},\quad
\mathbf{6}_{{\tiny{\begin{pmatrix}k-k'-l'\\l+k'\end{pmatrix}}}}\sim
\begin{pmatrix}
 x_1 y_2 \\
 x_2 y_3 \\
 x_3 y_1 \\
 x_4 y_5 \\
 x_5 y_6 \\
 x_6 y_4 \\
\end{pmatrix},\quad
\mathbf{6}_{\tiny{{\begin{pmatrix}k+l'\\l-l'-k'\end{pmatrix}}}}\sim
\begin{pmatrix}
 x_1 y_3 \\
 x_2 y_1 \\
 x_3 y_2 \\
 x_4 y_6 \\
 x_5 y_4 \\
 x_6 y_5 \\
\end{pmatrix}, \\
&&\hskip-0.15in
\mathbf{6}_{\tiny{{\begin{pmatrix}k-k'\\l+k'+l'\end{pmatrix}}}}\sim
\begin{pmatrix}
 x_1 y_4 \\
 x_2 y_6 \\
 x_3 y_5 \\
 x_4 y_1 \\
 x_5 y_3 \\
 x_6 y_2
\end{pmatrix},\quad
\mathbf{6}_{{\tiny{\begin{pmatrix}k+k'+l'\\l-l'\end{pmatrix}}}}\sim
\begin{pmatrix}
 x_1 y_5 \\
 x_2 y_4 \\
 x_3 y_6 \\
 x_4 y_2 \\
 x_5 y_1 \\
 x_6 y_3 \\
\end{pmatrix},\quad
\mathbf{6}_{\tiny{{\begin{pmatrix}k-l'\\l-k'\end{pmatrix}}}}\sim
\begin{pmatrix}
 x_1 y_6 \\
 x_2 y_5 \\
 x_3 y_4 \\
 x_4 y_3 \\
 x_5 y_2 \\
 x_6 y_1 \\
\end{pmatrix}\,.
\end{eqnarray}

\end{itemize}

For the case of $n=3\mathbb{Z}$, the CG-coefficients can be
straightforwardly calculated although it is somewhat lengthy.
Part of the CG coefficients are complex numbers in our chosen basis, the explicit form would not be reported here since generalized CP transformations can not be consistently defined in generic settings based on such groups unless the doublet representations $\mathbf{2_2}$, $\mathbf{2_3}$ and $\mathbf{2_4}$ are not introduced in a specific model.

\section{\label{sec:charged_lepton_Z2CP}Phenomenological implication of $Z_2\times CP$ in the charged lepton sector}

The full symmetry $\Delta(6n^2)\rtimes H_{CP}$ is broken down to $Z_2\times H^{l}_{CP}$ in the charged lepton sector. Similar to section~\ref{sec:Z2xCP_neutrino}, the hermitian combination $m^{\dagger}_{l}m_{l}$ of the charged lepton mass matrix can be constructed from its invariance under the remnant family symmetry $Z_2$ and the remnant CP symmetry $H^{l}_{CP}$,
\begin{eqnarray}
\nonumber&&\rho^{\dagger}_{\mathbf{3}}(g_{l})m^{\dagger}_{l}m_{l}\rho_{\mathbf{3}}(g_{l})=m^{\dagger}_{l}m_{l},\quad
g_{l}\in Z_2\,,\\
\label{eq:invariant_charged_lepton}&&X^{\dagger}_{l\mathbf{3}}m^{\dagger}_{l}m_{l}X_{l\mathbf{3}}=\left(m^{\dagger}_{l}m_{l}\right)^{*},\quad
X_{l}\in H^{l}_{CP}\,.
\end{eqnarray}

\begin{description}[labelindent=-0.8em, leftmargin=0.3em]
  \item[~~(\romannumeral1)] {$G_{l}=Z^{bc^xd^x}_2$,
      $X_{l\mathbf{r}}=\left\{\rho_{\mathbf{r}}(c^{\gamma}d^{-2x-\gamma}),
      \rho_{\mathbf{r}}(bc^{x+\gamma}d^{-x-\gamma})\right\}$ }

In this case, $m^{\dagger}_{l}m_{l}$ is determined to be of the form
\begin{equation}
m^{\dagger}_{l}m_{l}=\left(
\begin{array}{ccc}
 \widetilde{m}_{11} &  \widetilde{m}_{12}e^{i\pi\frac{2x+3 \gamma}{n}} &
 \widetilde{m}_{13}e^{-2i\pi\frac{x}{n}}   \\
 \widetilde{m}_{12}e^{-i\pi\frac{2x+3\gamma}{n}}  &  \widetilde{m}_{22} &
 \widetilde{m}_{12}e^{-i\pi\frac{4x+3\gamma}{n}}   \\
 \widetilde{m}_{13}e^{2i\pi\frac{x}{n}}  &
 \widetilde{m}_{12}e^{i\pi\frac{4 x+3 \gamma}{n}}  &   \widetilde{m}_{11}
 \\
\end{array}
\right)\,,
\end{equation}
where $\widetilde{m}_{11}$, $\widetilde{m}_{12}$, $\widetilde{m}_{13}$ and $\widetilde{m}_{22}$ are real parameters, and they have mass dimension of 2. This charged lepton mass matrix
is diagonalized by a unitary transformation $U_{l}$ via
\begin{equation}
U^{\dagger}_{l}m^{\dagger}_{l}m_{l}U_{l}=\text{diag}\left(m^2_{l_1},m^2_{l_2},m^2_{l_3}\right)\,,
\end{equation}
with
\begin{equation}
\label{eq:ul_bcxdx_2}U_{l}=\frac{1}{\sqrt{2}}
\left(
\begin{array}{ccc}
 e^{i\pi\frac{\gamma}{n}} & -e^{i\pi\frac{\gamma}{n}}\sin\theta &
 e^{i\pi\frac{\gamma}{n}}\cos\theta \\
 0 &  e^{-2i\pi\frac{x+\gamma}{n}}
   \sqrt{2}\cos\theta~ & ~e^{-2i\pi\frac{x+\gamma}{n}} \sqrt{2}\sin\theta
   \\
 -e^{i\pi\frac{2x+\gamma}{n}} & -e^{i\pi\frac{2x+\gamma}{n}}\sin\theta &
 e^{i\pi\frac{2x+\gamma}{n}}\cos\theta
\end{array}
\right)\,,
\end{equation}
where the angle $\theta$ is
\begin{equation}
\tan2\theta=\frac{2\sqrt{2}\widetilde{m}_{12}}{\widetilde{m}_{11}+\widetilde{m}_{13}-\widetilde{m}_{22}}\,.
\end{equation}
It is remarkable that the unitary transformation $U_{l}$ in Eq.~\eqref{eq:ul_bcxdx_2} coincides with $U_{\nu}$ in Eq.~\eqref{eq:unu_bcxdx}. The reason is that the two cases share the same residual symmetry. The charged lepton masses are given by
\begin{eqnarray}
\nonumber&&m^2_{l_1}=\widetilde{m}_{11}-\widetilde{m}_{13},\\
\nonumber&&m^2_{l_2}=\frac{1}{2}\left[\widetilde{m}_{11}+\widetilde{m}_{13}+\widetilde{m}_{22}-\text{sign}\left((\widetilde{m}_{11}+\widetilde{m}_{13}-\widetilde{m}_{22})\cos2\theta\right)\sqrt{(\widetilde{m}_{11}+\widetilde{m}_{13}-\widetilde{m}_{22})^2+8\widetilde{m}^2_{12}}\right],\\
\nonumber&&m^2_{l_3}=\frac{1}{2}\left[\widetilde{m}_{11}+\widetilde{m}_{13}+\widetilde{m}_{22}+\text{sign}\left((\widetilde{m}_{11}+\widetilde{m}_{13}-\widetilde{m}_{22})\cos2\theta\right)\sqrt{(\widetilde{m}_{11}+\widetilde{m}_{13}-\widetilde{m}_{22})^2+8\widetilde{m}^2_{12}}\right]\,.
\end{eqnarray}
In the present framework, we can not determine the order of $m^2_{l_1}$, $m^2_{l_2}$ and $m^2_{l_3}$, i.e. we don't know which one of $m^2_{l_1}$, $m^2_{l_2}$, $m^2_{l_3}$ is electron (muon or tau) mass squared. As a result, the diagonalization matrix $U_{l}$ in Eq.~\eqref{eq:ul_bcxdx_2} is also determined up to rephasing and permutations of its column vectors. The same holds true for the following cases.

\item[~~(\romannumeral2)] {$G_{l}=Z^{abc^y}_2$,
    $X_{l\mathbf{r}}=\left\{\rho_{\mathbf{r}}(c^{\gamma}d^{2y+2\gamma}),~~
    \rho_{\mathbf{r}}(abc^{y+\gamma}d^{2y+2\gamma})\right\}$}

The charged lepton mass matrix satisfying the invariant conditions of Eq.~\eqref{eq:invariant_charged_lepton} takes the form
\begin{equation}
m^{\dagger}_{l}m_{l}=\left(
\begin{array}{ccc}
 \widetilde{m}_{11} & \widetilde{m}_{12}e^{-2i\pi\frac{y}{n}}  &
 \widetilde{m}_{13}e^{i \pi\frac{2 y+3\gamma}{n}} \\
\widetilde{m}_{12}  e^{2i\pi\frac{y}{n}} & \widetilde{m}_{11} &
\widetilde{m}_{13} e^{i\pi\frac{4y+3\gamma}{n}} \\
\widetilde{m}_{13} e^{-i\pi\frac{2y+3\gamma}{n}}  &
\widetilde{m}_{13}e^{-i\pi\frac{4y+3\gamma}{n}} &  \widetilde{m}_{33} \\
\end{array}
\right)\,,
\end{equation}
where $\widetilde{m}_{11}$, $\widetilde{m}_{12}$, $\widetilde{m}_{13}$ and $\widetilde{m}_{33}$ are real. The charged lepton diagonalization matrix $U_{l}$ is given by
\begin{equation}
U_{l}=\frac{1}{\sqrt{2}}
\left(
\begin{array}{ccc}
e^{i\pi\frac{\gamma}{n}} & e^{i\pi\frac{\gamma}{n}}\cos\theta &
e^{i\pi\frac{\gamma}{n}}\sin\theta \\
-e^{i\pi\frac{2y+\gamma}{n}} & e^{i\pi\frac{2y+\gamma}{n}} \cos\theta &
e^{i\pi\frac{2y+\gamma}{n}}\sin\theta \\
0 & -e^{-2i\pi\frac{y+\gamma}{n}}\sqrt{2}\sin\theta~ &
~e^{-2i\pi\frac{y+\gamma}{n}}\sqrt{2}\cos\theta \\
\end{array}
\right)\,,
\end{equation}
with
\begin{equation}
\tan2\theta=\frac{2\sqrt{2}\,\widetilde{m}_{13}}{\widetilde{m}_{33}-\widetilde{m}_{11}-\widetilde{m}_{12}}\,.
\end{equation}
The charged lepton masses are determined to be
\begin{eqnarray}
\nonumber&&m^2_{l_1}=\widetilde{m}_{11}-\widetilde{m}_{12},\\
\nonumber&&m^2_{l_2}=\frac{1}{2}\left[\widetilde{m}_{11}+\widetilde{m}_{12}+\widetilde{m}_{33}+\text{sign}\left((\widetilde{m}_{11}+\widetilde{m}_{12}-\widetilde{m}_{33})\cos2\theta\right)\sqrt{(\widetilde{m}_{11}+\widetilde{m}_{12}-\widetilde{m}_{33})^2+8m^2_{13}}\right],\\
\nonumber&&m^2_{l_3}=\frac{1}{2}\left[\widetilde{m}_{11}+\widetilde{m}_{12}+\widetilde{m}_{33}-\text{sign}\left((\widetilde{m}_{11}+\widetilde{m}_{12}-\widetilde{m}_{33})\cos2\theta\right)\sqrt{(\widetilde{m}_{11}+\widetilde{m}_{12}-\widetilde{m}_{33})^2+8m^2_{13}}\right]\,.
\end{eqnarray}

\item[~~(\romannumeral3)]{$G_{l}=Z^{a^2bd^z}_2$,
    $X_{l\mathbf{r}}=\left\{\rho_{\mathbf{r}}(c^{2z+2\delta}d^{\delta}),
    \rho_{\mathbf{r}}(a^2bc^{2z+2\delta}d^{z+\delta})\right\}$}

The charged lepton mass matrix invariant under both residual flavor and residual CP symmetries is
\begin{equation}
m^{\dagger}_{l}m_{l}=\left(
\begin{array}{ccc}
 \widetilde{m}_{11} ~&~ \widetilde{m}_{12}e^{i\pi\frac{4z+3\delta}{n}} ~
 &~  \widetilde{m}_{12}e^{i\pi\frac{2z+3\delta}{n}}  \\
 \widetilde{m}_{12} e^{-i\pi\frac{4 z+3\delta}{n}} ~&~ \widetilde{m}_{22}
 ~&~    \widetilde{m}_{23}e^{-2i\pi\frac{z}{n}}  \\
 \widetilde{m}_{12}e^{-i\pi\frac{2z+3\delta}{n}}  ~&~
 \widetilde{m}_{23}e^{2i\pi\frac{z}{n}}  ~&~ \widetilde{m}_{22}   \\
\end{array}
\right)\,,
\end{equation}
where $\widetilde{m}_{11}$, $\widetilde{m}_{12}$, $\widetilde{m}_{22}$ and $\widetilde{m}_{23}$ are real. The unitary transformation $U_{l}$ follows immediately,
\begin{equation}
U_{l}=\frac{1}{\sqrt{2}}
\left(
\begin{array}{ccc}
 0 & ~-e^{2i\pi\frac{z+\delta}{n}}\sqrt{2}\sin\theta~ &
    ~e^{2i\pi\frac{z+\delta}{n}}\sqrt{2}\cos\theta \\
 e^{-i\pi\frac{2z+\delta}{n}} & e^{-i\pi\frac{2z+\delta}{n}}\cos\theta &
 e^{-i\pi\frac{2z+\delta}{n}}\sin\theta \\
 -e^{-i\pi\frac{\delta}{n}} & e^{-i\pi\frac{\delta}{n}}\cos\theta &
 e^{-i\pi\frac{\delta}{n}}\sin\theta
\end{array}
\right)\,,
\end{equation}
with the angle $\theta$ specified by
\begin{equation}
\tan2\theta=\frac{2\sqrt{2}\widetilde{m}_{12}}{\widetilde{m}_{11}-\widetilde{m}_{22}-\widetilde{m}_{23}}\,.
\end{equation}
Finally the charged lepton mass eigenvalues are
\begin{eqnarray}
\nonumber&&m^2_{l_1}=\widetilde{m}_{22}-\widetilde{m}_{23},\\
\nonumber&&m^2_{l_2}=\frac{1}{2}\left[\widetilde{m}_{11}+\widetilde{m}_{22}+\widetilde{m}_{23}-\text{sign}\left((\widetilde{m}_{11}-\widetilde{m}_{22}-\widetilde{m}_{23})\cos2\theta\right)\sqrt{(\widetilde{m}_{11}-\widetilde{m}_{22}-\widetilde{m}_{23})^2+8m^2_{12}}\right],\\
\nonumber&&m^2_{l_3}=\frac{1}{2}\left[\widetilde{m}_{11}+\widetilde{m}_{22}+\widetilde{m}_{23}+\text{sign}\left((\widetilde{m}_{11}-\widetilde{m}_{22}-\widetilde{m}_{23})\cos2\theta\right)\sqrt{(\widetilde{m}_{11}-\widetilde{m}_{22}-\widetilde{m}_{23})^2+8m^2_{12}}\right]\,.
\end{eqnarray}

\item[~~(\romannumeral4)]{$G_{l}=Z^{c^{n/2}}_2=\left\{1,c^{n/2}\right\}$,
    $X_{l\mathbf{r}}=\left\{\rho_{\mathbf{r}}(c^{\gamma}d^{\delta}),
    \rho_{\mathbf{r}}(abc^{\gamma}d^{\delta})\right\}$}

\begin{itemize}

\item{$X_{l\mathbf{r}}=\rho_{\mathbf{r}}(c^{\gamma}d^{\delta})$}

The charged lepton mass matrix is constrained to be of the following form
\begin{equation}
m^{\dagger}_{l}m_{l}=\left(
\begin{array}{ccc}
 \widetilde{m}_{11} & \widetilde{m}_{12}e^{i\pi\frac{2\gamma-\delta}{n}}
 & 0 \\
  \widetilde{m}_{12}e^{-i\pi\frac{2\gamma-\delta}{n}} &
  \widetilde{m}_{22} & 0 \\
 0 & 0 & \widetilde{m}_{33}
\end{array}
\right)\,,
\end{equation}
where $\widetilde{m}_{11}$, $\widetilde{m}_{12}$, $\widetilde{m}_{22}$ and $\widetilde{m}_{33}$ are real. It is diagonalized by the unitary matrix $U_{l}$ with
\begin{equation}
U_{l}=\left(
\begin{array}{ccc}
e^{i\pi\frac{\gamma}{n}}\cos\theta &  e^{i\pi\frac{\gamma}{n}}\sin\theta
& 0 \\
-e^{i\pi\frac{\delta-\gamma}{n}}\sin\theta~ &
~e^{i\pi\frac{\delta-\gamma}{n}}\cos\theta & 0 \\
 0 & 0 & e^{-i\pi\frac{\delta}{n}}
\end{array}
\right)\,,
\end{equation}
where
\begin{equation}
\tan2\theta=\frac{2\widetilde{m}_{12}}{\widetilde{m}_{22}-\widetilde{m}_{11}}\,.
\end{equation}
The charged lepton masses are determined to be
\begin{eqnarray}
\nonumber&&
m^2_{l_1}=\frac{1}{2}\left[\widetilde{m}_{11}+\widetilde{m}_{22}-\text{sign}\left((\widetilde{m}_{22}-\widetilde{m}_{11})\cos2\theta\right)\sqrt{(\widetilde{m}_{22}-\widetilde{m}_{11})^2+4m^2_{12}}\right],\\
\nonumber&&
m^2_{l_2}=\frac{1}{2}\left[\widetilde{m}_{11}+\widetilde{m}_{22}+\text{sign}\left((\widetilde{m}_{22}-\widetilde{m}_{11})\cos2\theta\right)\sqrt{(\widetilde{m}_{22}-\widetilde{m}_{11})^2+4m^2_{12}}\right],\\
&&m^2_{l_3}=\widetilde{m}_{33}\,.
\end{eqnarray}

\item{$X_{l\mathbf{r}}=\rho_{\mathbf{r}}(abc^{\gamma}d^{\delta})$}

Similar to the discussed situation that $Z_2\times CP$ is preserved in the neutrino sector, the CP transformation should be symmetric as well otherwise the charged lepton masses would be at least partially degenerate~\footnote{From the remnant symmetry invariant conditions in Eq.~\eqref{eq:invariant_charged_lepton}, we can derive that $U^{\dagger}_{l}X_{l\mathbf{3}}U^{*}_{l}$ should be a diagonal matrix. As a consequence, the CP transformation $X_{l\mathbf{3}}$ is symmetric.}. Therefore we shall focus on the case of $\delta=2\gamma~\text{mod}~ n$ in the following. Then the charged lepton mass matrix is fixed to be
\begin{equation}
m^{\dagger}_{l}m_{l}=\left(
\begin{array}{ccc}
 \widetilde{m}_{11} &  \widetilde{m}_{12}e^{i \phi} & 0 \\
 \widetilde{m}_{12}e^{-i\phi }  & \widetilde{m}_{11} & 0 \\
 0 & 0 & \widetilde{m}_{33}
\end{array}
\right)\,,
\end{equation}
where $\widetilde{m}_{11}$, $\widetilde{m}_{12}$, $\widetilde{m}_{33}$ and $\phi$ are free real parameters. Notice that $m^{\dagger}_{l}m_{l}$ is independent of the parameter $\gamma$. The unitary matrix $U_{l}$ is of the form
\begin{equation}
U_{l}=\frac{1}{\sqrt{2}}\left(
\begin{array}{ccc}
 e^{i \phi } &~ e^{i \phi } & 0 \\
 -1 &~ 1 & 0 \\
 0 &~ 0 & \sqrt{2}
\end{array}
\right)\,.
\end{equation}
The charged lepton masses are given by
\begin{eqnarray}
\nonumber&&m^2_{l_1}=\widetilde{m}_{11}-\widetilde{m}_{12},\\
\nonumber&&m^2_{l_2}=\widetilde{m}_{11}+\widetilde{m}_{12},\\
&&m^2_{l_2}=\widetilde{m}_{33}\,.
\end{eqnarray}

\end{itemize}

\item[~~(\romannumeral5)]{$G_{l}=Z^{d^{n/2}}_2=\left\{1,d^{n/2}\right\}$,
    $X_{l\mathbf{r}}=\left\{\rho_{\mathbf{r}}(c^{\gamma}d^{\delta}),
    \rho_{\mathbf{r}}(a^2bc^{\gamma}d^{\delta})\right\}$}

\begin{itemize}

\item{$X_{l\mathbf{r}}=\rho_{\mathbf{r}}(c^{\gamma}d^{\delta})$}

In this case, the charged lepton mass matrix takes the form
\begin{equation}
m^{\dagger}_{l}m_{l}=\left(
\begin{array}{ccc}
 \widetilde{m}_{11} & 0 & 0 \\
 0 & \widetilde{m}_{22} &
 \widetilde{m}_{23}e^{-i\pi\frac{\gamma-2\delta}{n}}  \\
 0 & \widetilde{m}_{23}e^{i\pi\frac{\gamma -2\delta}{n}}  &
 \widetilde{m}_{33}
\end{array}
\right)\,,
\end{equation}
where $\widetilde{m}_{11}$, $\widetilde{m}_{22}$, $\widetilde{m}_{23}$ and $\widetilde{m}_{33}$ are real. The charged lepton diagonalization matrix is
\begin{equation}
U_{l}=\left(
\begin{array}{ccc}
 e^{i\pi\frac{\gamma}{n}}   & 0 & 0 \\
 0 & e^{i\pi\frac{\delta-\gamma}{n}}\cos\theta~ &
 ~e^{i\pi\frac{\delta-\gamma}{n}}\sin\theta \\
 0 & -e^{-i\pi\frac{\delta}{n}}\sin\theta~ &
 ~e^{-i\pi\frac{\delta}{n}}\cos\theta
\end{array}
\right)\,,
\end{equation}
with
\begin{equation}
\tan2\theta=\frac{2\widetilde{m}_{23}}{\widetilde{m}_{33}-\widetilde{m}_{22}}\,.
\end{equation}
The mass eigenvalues of the charged lepton are found to be
\begin{eqnarray}
\nonumber&&\hskip-0.3in m^2_{l_1}=\widetilde{m}_{11},\\
\nonumber&&\hskip-0.3in
m^2_{l_2}=\frac{1}{2}\left[\widetilde{m}_{22}+\widetilde{m}_{33}-\text{sign}\left((\widetilde{m}_{33}-\widetilde{m}_{22})\cos2\theta\right)\sqrt{(\widetilde{m}_{33}-\widetilde{m}_{22})^2+4m^2_{23}}\right],\\
&&\hskip-0.3in
m^2_{l_3}=\frac{1}{2}\left[\widetilde{m}_{22}+\widetilde{m}_{33}+\text{sign}\left((\widetilde{m}_{33}-\widetilde{m}_{22})\cos2\theta\right)\sqrt{(\widetilde{m}_{33}-\widetilde{m}_{22})^2+4m^2_{23}}\right]\,.
\end{eqnarray}

\item{$X_{l\mathbf{r}}=\rho_{\mathbf{r}}(a^2bc^{\gamma}d^{\delta})$}

This generalized CP transformation is symmetric only if $\gamma=2\delta~\textrm{mod}~n$. One can easily find that the charged lepton mass matrix is constrained to be of the form
\begin{equation}
m^{\dagger}_{l}m_{l}=\left(
\begin{array}{ccc}
 \widetilde{m}_{11} & 0 & 0 \\
 0 & \widetilde{m}_{22} & \widetilde{m}_{23} e^{i \phi }  \\
 0 & \widetilde{m}_{23}e^{-i \phi }  & \widetilde{m}_{22}
\end{array}
\right)\,,
\end{equation}
where $\widetilde{m}_{11}$, $\widetilde{m}_{22}$, $\widetilde{m}_{23}$ and $\phi$ are real. It is diagonalized by the unitary matrix
\begin{equation}
U_{l}=\frac{1}{\sqrt{2}}
\left(
\begin{array}{ccc}
 \sqrt{2} & 0 & 0 \\
 0 & e^{i \phi } & e^{i \phi } \\
 0 & -1 & 1
\end{array}
\right)\,.
\end{equation}
The charged lepton masses are
\begin{eqnarray}
\nonumber&&m^2_{l_1}=\widetilde{m}_{11},\\
\nonumber&&m^2_{l_2}=\widetilde{m}_{22}-\widetilde{m}_{23},\\
&&m^2_{l_3}=\widetilde{m}_{22}+\widetilde{m}_{23}\,.
\end{eqnarray}

\end{itemize}

\item[~~(\romannumeral6)]{$G_{l}=Z^{c^{n/2}d^{n/2}}_2=\left\{1,c^{n/2}d^{n/2}\right\}$,
    $X_{l\mathbf{r}}=\left\{\rho_{\mathbf{r}}(c^{\gamma}d^{\delta}),
    \rho_{\mathbf{r}}(bc^{\gamma}d^{\delta})\right\}$}

\begin{itemize}

\item{$X_{l\mathbf{r}}=\rho_{\mathbf{r}}(c^{\gamma}d^{\delta})$}

Remnant symmetry leads to the following charged lepton mass matrix
\begin{equation}
m^{\dagger}_{l}m_l=\left(
\begin{array}{ccc}
 \widetilde{m}_{11} & 0 &
 \widetilde{m}_{13}e^{i\pi\frac{\gamma+\delta}{n}}  \\
 0 & \widetilde{m}_{22} & 0 \\
 \widetilde{m}_{13}e^{-i\pi\frac{\gamma+\delta}{n}}   & 0 &
 \widetilde{m}_{33}
\end{array}
\right)\,,
\end{equation}
where $\widetilde{m}_{11}$, $\widetilde{m}_{13}$, $\widetilde{m}_{22}$ and $m_{33}$ are real parameters. The unitary transformation $U_{l}$ is of the form
\begin{equation}
U_{l}=\left(
\begin{array}{ccc}
 e^{i\pi\frac{\gamma}{n}}\cos\theta & 0 &
 e^{i\pi\frac{\gamma}{n}}\sin\theta \\
 0 & e^{i\pi\frac{\delta-\gamma}{n}} & 0   \\
 -e^{-i\pi\frac{\delta}{n}}\sin\theta & 0 &
   e^{-i\pi\frac{\delta}{n}}\cos\theta \\
\end{array}
\right)\,,
\end{equation}
with
\begin{equation}
\tan2\theta=\frac{2\widetilde{m}_{13}}{\widetilde{m}_{33}-\widetilde{m}_{11}}\,.
\end{equation}
The charged lepton mass eigenvalues are given by
\begin{eqnarray}
\nonumber&&\hskip-0.3in
m^2_{l_1}=\frac{1}{2}\left[\widetilde{m}_{11}+\widetilde{m}_{33}-\text{sign}\left((\widetilde{m}_{33}-\widetilde{m}_{11})\cos2\theta\right)\sqrt{(\widetilde{m}_{33}-\widetilde{m}_{11})^2+4\widetilde{m}^2_{13}}\right],\\
\nonumber&&\hskip-0.3in m^2_{l_2}=\widetilde{m}_{22},\\
&&\hskip-0.3in
m^2_{l_3}=\frac{1}{2}\left[\widetilde{m}_{11}+\widetilde{m}_{33}+\text{sign}\left((\widetilde{m}_{33}-\widetilde{m}_{11})\cos2\theta\right)\sqrt{(\widetilde{m}_{33}-\widetilde{m}_{11})^2+4\widetilde{m}^2_{13}}\right]\,.
\end{eqnarray}

\item{$X_{l\mathbf{r}}=\rho_{\mathbf{r}}(bc^{\gamma}d^{\delta})$}

The non-degeneracy of the charged lepton masses requires $\gamma+\delta=0~\text{mod}~n$ for which the generalized CP
transformation matrix $\rho_{\mathbf{r}}(bc^{\gamma}d^{\delta})$ is symmetric. The charged lepton mass matrix fulfilling the invariant condition in Eq.~\eqref{eq:invariant_charged_lepton} is of the form
\begin{equation}
m^{\dagger}_{l}m_{l}=\left(
\begin{array}{ccc}
\widetilde{m}_{11} & 0 &  \widetilde{m}_{13}e^{i\phi} \\
 0 & \widetilde{m}_{22} & 0 \\
 \widetilde{m}_{13}e^{-i\phi}  & 0 & \widetilde{m}_{11}
\end{array}
\right)\,,
\end{equation}
where $\widetilde{m}_{11}$, $\widetilde{m}_{13}$, $\widetilde{m}_{22}$
and $\phi$ are real. The charged lepton diagonalization matrix is
\begin{equation}
U_{l}=\frac{1}{\sqrt{2}}\left(
\begin{array}{ccc}
 e^{i \phi } & 0 & e^{i \phi } \\
 0 & \sqrt{2} & 0 \\
 -1 & 0 & 1
\end{array}
\right)\,.
\end{equation}
Finally the charged lepton masses are
\begin{eqnarray}
\nonumber&&m^2_{l_1}=\widetilde{m}_{11}-\widetilde{m}_{13},\\
\nonumber&&m^2_{l_2}=\widetilde{m}_{22},\\
&&m^2_{l_3}=\widetilde{m}_{11}+\widetilde{m}_{13}\,.
\end{eqnarray}

\end{itemize}

\end{description}
Comparing with phenomenological predictions of $Z_2\times CP$ in the neutrino sector analyzed in section~\ref{subsec:neutrino_one_column}, we see that the diagonalization matrix $U_{l}$ is of the same form as $U_{\nu}$ provided the remnant flavor and remnant CP symmetries are the same in the two occasions.


\begin{thebibliography}{}

\bibitem{An:2012eh}
  F.~P.~An {\it et al.}  [DAYA-BAY Collaboration],
   Phys.\ Rev.\ Lett.\  {\bf 108}, 171803 (2012)  [arXiv:1203.1669
   [hep-ex]];  
  Chin.\  Phys.\ C {\bf 37}, 011001 (2013)  [arXiv:1210.6327 [hep-ex]].


\bibitem{Ahn:2012nd}
  J.~K.~Ahn {\it et al.}  [RENO Collaboration],
  Phys.\ Rev.\ Lett.\  {\bf 108}, 191802 (2012)  [arXiv:1204.0626 [hep-ex]].


\bibitem{Abe:2011fz}
  Y.~Abe {\it et al.}  [DOUBLE-CHOOZ Collaboration],
  Phys.\ Rev.\ Lett.\  {\bf 108}, 131801 (2012)  [arXiv:1112.6353 [hep-ex]];
  Phys.\ Rev.\ D {\bf 86}, 052008 (2012)  [arXiv:1207.6632 [hep-ex]].


\bibitem{Daya_Bay_NOW_14}
Neutrino Oscillation Workshop, Conca Specchiulla (Otranto, Lecce, Italy),
September 7-14, 2014,
\url{http://www.ba.infn.it/~now/now2014/web-content/index.html}.



\bibitem{Tortola:2012te}
 D.~V.~Forero, M.~Tortola and J.~W.~F.~Valle,
  Phys.\ Rev.\ D {\bf 86}, 073012 (2012)
  [arXiv:1205.4018 [hep-ph]].


\bibitem{Capozzi:2013csa}
  F.~Capozzi, G.~L.~Fogli, E.~Lisi, A.~Marrone, D.~Montanino and A.~Palazzo,
  Phys.\ Rev.\ D {\bf 89}, 093018 (2014)
  [arXiv:1312.2878 [hep-ph]].


\bibitem{Gonzalez-Garcia:2014bfa}
  M.~C.~Gonzalez-Garcia, M.~Maltoni and T.~Schwetz,
  arXiv:1409.5439 [hep-ph].



\bibitem{Altarelli:2010gt}
  G.~Altarelli and F.~Feruglio,
  Rev.\ Mod.\ Phys.\  {\bf 82}, 2701 (2010)
  [arXiv:1002.0211 [hep-ph]];
  H.~Ishimori, T.~Kobayashi, H.~Ohki, Y.~Shimizu, H.~Okada and M.~Tanimoto,
  Prog.\ Theor.\ Phys.\ Suppl.\  {\bf 183}, 1 (2010)
  [arXiv:1003.3552 [hep-th]];
  W.~Grimus and P.~O.~Ludl,
  J.\ Phys.\ A {\bf 45}, 233001 (2012)
  [arXiv:1110.6376 [hep-ph]].


\bibitem{King:2013eh}
  S.~F.~King and C.~Luhn,
  Rept.\ Prog.\ Phys.\  {\bf 76} (2013) 056201
  [arXiv:1301.1340 [hep-ph]];
  S.~F.~King, A.~Merle, S.~Morisi, Y.~Shimizu and M.~Tanimoto,
  arXiv:1402.4271 [hep-ph].


\bibitem{King:2014iia}
  S.~F.~King,
  JHEP {\bf 1408} (2014) 130
  [arXiv:1406.7005 [hep-ph]].


\bibitem{Ecker:1981wv}
  G.~Ecker, W.~Grimus and W.~Konetschny,
  Nucl.\ Phys.\ B {\bf 191} (1981) 465;
  G.~Ecker, W.~Grimus and H.~Neufeld,
  Nucl.\ Phys.\ B {\bf 247} (1984) 70;
  G.~Ecker, W.~Grimus and H.~Neufeld,
  J.\ Phys.\ A {\bf 20} (1987) L807;
  H.~Neufeld, W.~Grimus and G.~Ecker,
  Int.\ J.\ Mod.\ Phys.\ A {\bf 3}, 603 (1988).

\bibitem{Grimus:1995zi}
  W.~Grimus and M.~N.~Rebelo,
  Phys.\ Rept.\  {\bf 281}, 239 (1997)
  [arXiv:9506272[hep-ph]].


\bibitem{Harrison:2002kp}
  P.~F.~Harrison and W.~G.~Scott,
  Phys.\ Lett.\ B {\bf 535}, 163 (2002)
  [hep-ph/0203209];
  P.~F.~Harrison and W.~G.~Scott,
  Phys.\ Lett.\ B {\bf 547}, 219 (2002)
  [hep-ph/0210197];
  P.~F.~Harrison and W.~G.~Scott,
  Phys.\ Lett.\ B {\bf 594}, 324 (2004)
  [hep-ph/0403278].


\bibitem{Grimus:2003yn}
  W.~Grimus and L.~Lavoura,
  Phys.\ Lett.\ B {\bf 579}, 113 (2004)
  [hep-ph/0305309];
 W.~Grimus and L.~Lavoura,
  arXiv:1207.1678;  
  P.~M.~Ferreira, W.~Grimus, L.~Lavoura and P.~O.~Ludl,
  JHEP {\bf 1209}, 128 (2012)
  [arXiv:1206.7072].


\bibitem{Farzan:2006vj}
  Y.~Farzan and A.~Y.~.Smirnov,
  JHEP {\bf 0701}, 059 (2007)
  [hep-ph/0610337].


\bibitem{Feruglio:2012cw}
  F.~Feruglio, C.~Hagedorn and R.~Ziegler,
  JHEP {\bf 1307}, 027 (2013)
  [arXiv:1211.5560 [hep-ph]].


\bibitem{Ding:2013hpa}
  G.~-J.~Ding, S.~F.~King, C.~Luhn and A.~J.~Stuart,
  JHEP {\bf 1305}, 084 (2013)
  [arXiv:1303.6180 [hep-ph]].


\bibitem{Li:2013jya}
  C.~-C.~Li and G.~-J.~Ding,
  Nucl.\ Phys.\ B {\bf 881}, 206 (2014)
  [arXiv:1312.4401 [hep-ph]].


\bibitem{Li:2014eia}
  C.~C.~Li and G.~J.~Ding,
  arXiv:1408.0785 [hep-ph].



\bibitem{Feruglio:2013hia}
  F.~Feruglio, C.~Hagedorn and R.~Ziegler,
  Eur.\ Phys.\ J.\ C {\bf 74}, 2753 (2014)
  [arXiv:1303.7178 [hep-ph]].


\bibitem{Luhn:2013lkn}
  C.~Luhn,
  Nucl.\ Phys.\ B {\bf 875}, 80 (2013)
  [arXiv:1306.2358 [hep-ph]].


\bibitem{Ding:2013bpa}
  G.~-J.~Ding, S.~F.~King and A.~J.~Stuart,
  JHEP {\bf 1312} (2013) 006
  [arXiv:1307.4212].


\bibitem{Krishnan:2012me}
  R.~Krishnan, P.~F.~Harrison and W.~G.~Scott,
  JHEP {\bf 1304}, 087 (2013)
  [arXiv:1211.2000 [hep-ph]].


\bibitem{Mohapatra:2012tb}
  R.~N.~Mohapatra and C.~C.~Nishi,
  Phys.\ Rev.\ D {\bf 86}, 073007 (2012)  [arXiv:1208.2875 [hep-ph]].


\bibitem{Nishi:2013jqa}
  C.~C.~Nishi,
  Phys.\ Rev.\ D {\bf 88}, 033010 (2013)
  [arXiv:1306.0877 [hep-ph]].


\bibitem{Ding:2013nsa}
  G.~-J.~Ding and Y.~-L.~Zhou,
  arXiv:1312.5222 [hep-ph];
  G.~J.~Ding and Y.~L.~Zhou,
  JHEP {\bf 1406}, 023 (2014)
  [arXiv:1404.0592 [hep-ph]].



\bibitem{Holthausen:2012dk}
  M.~Holthausen, M.~Lindner and M.~A.~Schmidt,
  JHEP {\bf 1304}, 122 (2013)
  [arXiv:1211.6953 [hep-ph]].


\bibitem{Chen:2014tpa}
  M.~C.~Chen, M.~Fallbacher, K.~T.~Mahanthappa, M.~Ratz and A.~Trautner,
  Nucl.\ Phys.\ B {\bf 883}, 267 (2014)
  [arXiv:1402.0507 [hep-ph]].



\bibitem{Branco:1983tn}
  G.~C.~Branco, J.~M.~Gerard and W.~Grimus,
  Phys.\ Lett.\ B {\bf 136}, 383 (1984);  
  I.~de Medeiros Varzielas and D.~Emmanuel-Costa,
  Phys.\ Rev.\ D {\bf 84}, 117901 (2011)  [arXiv:1106.5477 [hep-ph]]; %
  I.~de Medeiros Varzielas, D.~Emmanuel-Costa and P.~Leser,
  Phys.\ Lett.\ B {\bf 716}, 193 (2012)  [arXiv:1204.3633 [hep-ph]];
  I.~de Medeiros Varzielas,
  JHEP {\bf 1208}, 055 (2012)  [arXiv:1205.3780 [hep-ph]];
  G.~Bhattacharyya, I.~de Medeiros Varzielas and P.~Leser,
  Phys.\ Rev.\ Lett.\  {\bf 109}, 241603 (2012)  [arXiv:1210.0545 [hep-ph]];
  I.~d.~M.~Varzielas and D.~Pidt,
  arXiv:1307.0711 [hep-ph].


\bibitem{Chen:2009gf}
  M.~-C.~Chen and K.~T.~Mahanthappa,
  Phys.\ Lett.\ B {\bf 681}, 444 (2009)  [arXiv:0904.1721 [hep-ph]];
  A.~Meroni, S.~T.~Petcov and M.~Spinrath,
Phys.\ Rev.\ D {\bf 86}, 113003 (2012)  [arXiv:1205.5241 [hep-ph]];
  S.~Antusch, S.~F.~King and M.~Spinrath,
  Phys.\ Rev.\ D {\bf 87}, 096018 (2013)  [arXiv:1301.6764 [hep-ph]].


\bibitem{Antusch:2011sx}
  S.~Antusch, S.~F.~King, C.~Luhn and M.~Spinrath,
   Nucl.\ Phys.\ B {\bf 850}, 477 (2011)  [arXiv:1103.5930 [hep-ph]];
  S.~Antusch, M.~Holthausen, M.~A.~Schmidt and M.~Spinrath,
  Nucl.\ Phys.\ B {\bf 877}, 752 (2013)
  [arXiv:1307.0710 [hep-ph]].


\bibitem{Girardi:2013sza}
  I.~Girardi, A.~Meroni, S.~T.~Petcov and M.~Spinrath,
  JHEP {\bf 1402}, 050 (2014)
  [arXiv:1312.1966 [hep-ph]].


\bibitem{King:2014rwa}
  S.~F.~King and T.~Neder,
  arXiv:1403.1758 [hep-ph].


\bibitem{King:2013vna}
  S.~F.~King, T.~Neder and A.~J.~Stuart,
  Phys.\ Lett.\ B {\bf 726} (2013) 312
  [arXiv:1305.3200 [hep-ph]].


\bibitem{Ding:2014ssa}
  G.~J.~Ding and S.~F.~King,
  Phys.\ Rev.\ D {\bf 89} (2014) 093020
  [arXiv:1403.5846 [hep-ph]].


\bibitem{Hagedorn:2014wha}
  C.~Hagedorn, A.~Meroni and E.~Molinaro,
  arXiv:1408.7118 [hep-ph].


\bibitem{GAP4:2011}
The GAP~Group, {\em {GAP -- Groups, Algorithms, and Programming, Version
  4.4.12}}, 2008.
\newblock \url{http://www.gap-system.org}.

\bibitem{REPSN:2011}
V.~Dabbaghian, {\em {REPSN} - for constructing representations of finite
  groups, {GAP} package, Version 3.0.2}.
\newblock The GAP~Group, 2011.
\newblock \url{http://www.gap-system.org/Packages/repsn.html}.

\bibitem{SmallGroups:2011}
H.U.Besche, B.Eick, and E.O'Brien, {\em {SmallGroups} - library of all
'small'
  groups, {GAP} package, Version included in {GAP} 4.4.12}.
\newblock The GAP~Group, 2002.
\newblock \url{http://www.gap-system.org/Packages/sgl.html}.

\bibitem{SONATA:2003}
E.~Aichinger, F.~Binder, J.~Ecker, P.~Mayr, and C.~N{\"o}bauer, {\em
{SONATA} -
  system of near-rings and their applications, {GAP} package, Version 2},
  2003.
\newblock \url{http://www.algebra.uni-linz.ac.at/Sonata/}.



\bibitem{Toorop:2011jn}
  R.~d.~A.~Toorop, F.~Feruglio and C.~Hagedorn,
  Phys.\ Lett.\ B {\bf 703}, 447 (2011)
  [arXiv:1107.3486 [hep-ph]];
  R.~de Adelhart Toorop, F.~Feruglio and C.~Hagedorn,
  Nucl.\ Phys.\ B {\bf 858}, 437 (2012)
  [arXiv:1112.1340 [hep-ph]].

\bibitem{Ding:2012xx}
  G.~-J.~Ding,
Nucl.\ Phys.\ B {\bf 862}, 1 (2012)
  [arXiv:1201.3279 [hep-ph]].


\bibitem{King:2012in}
  S.~F.~King, C.~Luhn and A.~J.~Stuart,
  Nucl.\ Phys.\ B {\bf 867}, 203 (2013)
  [arXiv:1207.5741 [hep-ph]].


\bibitem{pdg}
  K.~A.~Olive {\it et al.}  [Particle Data Group Collaboration],
  Chin.\ Phys.\ C {\bf 38}, 090001 (2014).



\bibitem{JUNO}
Talk with professor Wei Wang. JUNO experiment, \url{http://english.ihep.cas.cn/rs/fs/juno0815/}.


\bibitem{Albright:2008rp}
  C.~H.~Albright and W.~Rodejohann,
  Eur.\ Phys.\ J.\ C {\bf 62}, 599 (2009)
  [arXiv:0812.0436 [hep-ph]].


\bibitem{KingPSA}
  S.~F.~King, A.~Merle and A.~J.~Stuart,
JHEP {\bf 1312}, 005 (2013).
[arXiv:1307.2901 [hep-ph]].


\bibitem{AugerAR}
  M.~Auger {\it et al.}  [EXO Collaboration],
Phys.\ Rev.\ Lett.\  {\bf 109}, 032505 (2012).
[arXiv:1205.5608 [hep-ex]].



\bibitem{Albert:2014awa}
  J.~B.~Albert {\it et al.}  [EXO-200 Collaboration],
  Nature {\bf 510} (2014) 229-234
  [arXiv:1402.6956 [nucl-ex]].


\bibitem{Gando:2012zm}
  A.~Gando {\it et al.}  [KamLAND-Zen Collaboration],
  Phys.\ Rev.\ Lett.\  {\bf 110} (2013) 062502
  [arXiv:1211.3863 [hep-ex]].



\bibitem{AdeZUV}
  P.~A.~R.~Ade {\it et al.}  [Planck Collaboration],
[arXiv:1303.5076 [astro-ph.CO]].


\bibitem{Escobar:2008vc}
  J.~A.~Escobar and C.~Luhn,
  J.\ Math.\ Phys.\  {\bf 50}, 013524 (2009)  [arXiv:0809.0639 [hep-th]].



\end{thebibliography}
\end{document}